\documentclass[11pt]{article}
\usepackage{caption}
\pdfoutput=1
\usepackage{jheppub}
\usepackage{dsfont}
\usepackage{upgreek}
\usepackage{float, extarrows, tikz-cd}
\usepackage{graphicx}
\usepackage{subfig}
\usepackage{slashed}
\usepackage{tabularx,ragged2e}
\usepackage{longtable}
\usepackage{amssymb}
\usepackage{amsmath,amssymb}
\usepackage{slashed}
\usepackage{xcolor}
\usepackage{dsfont}
\usepackage{verbatim}
\usepackage{mathtools,xcolor,ytableau,amsfonts,tikz}
\usepackage{physics}
\usepackage{simpler-wick}
\usepackage{microtype}
\usepackage[nobottomtitles*]{titlesec}
\usepackage{color,soul}
\usepackage{cancel}
\setulcolor{orange}
\newcolumntype{C}{>{\Centering\arraybackslash}X}

\usepackage{array}
\newcolumntype{P}[1]{>{\centering\arraybackslash}p{#1}}

\numberwithin{equation}{section}

\let\oldfootnote\footnote
\renewcommand{\footnote}[1]{%
    \begingroup%
    \linespread{1}%    % <- linespread for footnote: 1, 1.1, 1.2 etc
    \oldfootnote{#1}%
    \endgroup%
}

\newcommand{\mL}{\mathcal{L}}
\newcommand{\mO}{\mathcal{O}}
\newcommand{\pd}{\partial}
\newcommand{\eps}{\varepsilon}
\newcommand{\mR}{\mathcal{R}}
\newcommand{\Gb}{\overline{G}}

\newcommand\N{{\mathcal{N}}}

\newcommand\beq{\begin{equation}}
\newcommand\eeq{\end{equation}}

\author[a,b]{Gabriel Cuomo,}
\author[c]{Adar Sharon}

\affiliation[a]{SISSA, Via Bonomea 265, 34136 Trieste, Italy}
\affiliation[b]{INFN, Sezione di Trieste, Via Valerio 2, 34127 Trieste, Italy}
\affiliation[c]{Mani L. Bhaumik Institute for Theoretical Physics, Department of Physics and Astronomy,
University of California, Los Angeles, CA 90095}

\emailAdd{gcuomo@sissa.it}			 
\emailAdd{asharon@physics.ucla.edu}    

\title{Moduli spaces in CFT: vortex defects}

\abstract{We construct and study a class of codimension-two conformal defects in CFTs with spontaneously broken conformal symmetry, assuming that a continuous global symmetry is also broken on the moduli space. Our construction can be viewed as an analytic continuation of the saddle point describing large charge operators and generalizes known BPS defects in supersymmetric theories, such as Gukov-Witten surface operators in $\mathcal{N}=4$ SYM and vortex loops in ABJM theory, but does not rely on supersymmetry. These defects are characterized by continuous, unbounded parameters controlling the expectation values of operators parametrizing the moduli space, and interpolate between a quantum regime and a semiclassical regime in which the local physics approaches that of the symmetry-breaking vacuum. 
This connection allows us to derive exact results from the effective description of the broken phase. In particular, our results for Gukov-Witten surface operators agree with localization, and we compute the $g$-function for a broad class of $1/2$-BPS vortex loops in ABJM theory. Finally, we relate the spectrum of defect operators,  in a suitable scaling limit, to the spectrum of massive particles in the symmetry-breaking vacuum.}

\begin{document}
\maketitle

\section{Introduction}\label{Sec_Intro}

Moduli spaces of vacua on which conformal symmetry is spontaneously broken appear to be non-generic in conformal field theories (CFTs). Indeed, the existence of a flat direction requires a nontrivial fine tuning of the dilaton potential, which is naturally realized only in free theories and supersymmetric models. It is perhaps therefore not surprising that all known CFTs with moduli spaces are either free or supersymmetric, or possess only an approximate moduli space in certain parametric limits.

Given how restrictive this condition is, it is natural to ask how the existence of a moduli space is reflected in the CFT data of the theory in the conformal vacuum. Ideally, one would like both to identify sharp criteria on CFT data that diagnose the existence of a flat direction and to develop specialized tools for studying observables in such theories, without necessarily assuming supersymmetry (SUSY). 

We began addressing these questions in \cite{Cuomo:2024vfk,Cuomo:2024fuy} (see also \cite{Karananas:2017zrg,Hellerman:2017veg,Ivanovskiy:2024vel} for related work). The main goal of this work is to make further progress in understanding the physics of CFTs with moduli spaces by studying a special class of conformal defects whose properties are closely related to the physics on the moduli space, in a precise sense that will be clarified below. We argue that such defects arise generically in CFTs with non-conformal vacua whenever a continuous internal symmetry is also spontaneously broken in some region of the moduli space. As we explain below, the defects we study generalize known supersymmetric examples, such as Gukov-Witten surface operators in $\mathcal{N}=4$ super Yang-Mills (SYM)~~\cite{Gukov:2006jk,Gukov:2014gja} and vortex loops in ABJM theory~\cite{Drukker:2008jm}. Besides generalizing these constructions to other models, our methodology provides a new handle on these objects and allows us to derive new exact results both for the known examples and for the broader class of theories considered here.

Our work is motivated by the observation that there exists an intriguing parallel between the bootstrap problem for boundary or defect CFTs and that of CFTs in a vacuum with broken conformal symmetry, formerly discussed in \cite{Cuomo:2024vfk}. In both cases, the usual local CFT data are supplemented by additional observables associated with the choice of defect in the former case, and the choice of vacuum in the latter. For a conformal defect, the additional data are the spectrum and OPE coefficients of defect operators, and the bulk-to-defect OPE coefficients. On the moduli space, their analogues are the spectrum and scattering amplitudes of asymptotic states in the broken vacuum, together with form factors of local operators in these states. In both cases, these enlarged sets of data are tied together by nontrivial consistency conditions. 

A priori, the analogy only goes so far. Indeed, a $p$-dimensional conformal defect breaks the Euclidean conformal group $SO(d+1,1)$ to $SO(d-p)\times SO(p+1,1)$, while a non conformal vacuum preserves the $d$-dimensional Euclidean group $ISO(d)$. Additionally, while all known CFTs admit nontrivial conformal defects (and boundaries in the absence of a gravitational anomaly), spontaneous conformal symmetry breaking is rare, as emphasized above. 

Yet, the above parallel suggests a natural way to bridge defect CFT data with asymptotic observables on the moduli space. Suppose, for instance, that one could construct a $p$-dimensional conformal defect with an \emph{unbounded} marginal parameter $h>0$,\footnote{In the sense that the limit $h\to\infty$ lies at infinite distance in the Zamolodchikov metric.} such that, when $h\gg 1$, one or more of the operators providing local coordinates on the moduli space acquire an expectation value of the form
\begin{equation}\label{eq_would_be_intro}
    \langle \mO_{\Delta}(r)\rangle
    \simeq
    c_{\mO}\frac{h^{\Delta}}{r^{\Delta}}\,,
\end{equation}
where $r$ is the distance from the defect, $\Delta$ is the scaling dimension of $\mO_{\Delta}$ and $c_{\mO}$ an $O(1)$ coefficient. It is clear that  $h$ can be unbounded only if there exists a moduli space of vacua. Consider indeed the limit $h\rightarrow\infty$, while simultaneously moving far away from the defect, $r=\bar r+\delta r$ with $\bar r\rightarrow\infty$, keeping
\begin{equation}
    v=\frac{h}{\bar r}
\end{equation}
fixed. In a region of size $\delta r\ll \bar r$, which becomes infinitely large in the limit $\bar{r}\rightarrow \infty$, the expectation value is approximately constant
\begin{equation}
    \langle \mO_{\Delta}(\bar r+\delta r)\rangle
    =
    c_{\mO}v^{\Delta}\left[1+\mathcal{O}\left(\frac{\delta r}{\bar r}\right)\right]\,.
\end{equation}
The theory in this region is therefore indistinguishable from the CFT placed in a state characterized by $\langle\mO_{\Delta}\rangle=c_{\mO}v^{\Delta}$. Such a state provides a stable vacuum only if the CFT admits a moduli space of vacua.

The above argument becomes particularly transparent by performing a Weyl transformation to $\mathrm{AdS}_{p+1}\times S^{d-p-1}$, with common radius $R$. In this Weyl frame, the expectation value in \eqref{eq_would_be_intro} becomes constant,
\begin{equation}\label{eq_intro_would_be_AdS}
    \left.\langle \mO_{\Delta}\rangle\right|_{\mathrm{AdS}_{p+1}\times S^{d-p-1}}
    \simeq c_{\mO}
    \left(\frac{h}{R}\right)^{\Delta}
    \,.
\end{equation}
As $h\rightarrow\infty$, the scale set by the operator VEV becomes parametrically larger than the curvature scale $1/R$, and the physics at distances much smaller than $R$ approaches that of the flat-space theory in the corresponding vacuum on the moduli space. It also follows then that the defect bootstrap problem approaches the moduli space one in this limit, as we will explain in greater detail in the main text.

This perspective also makes immediately clear that defects realizing expectation values \emph{exactly} of the form \eqref{eq_would_be_intro} are, in fact, even more difficult to find than moduli spaces themselves. Indeed, as is well known, even when a theory admits a flat direction in flat space, curvature-dependent terms in the effective potential generally lift the moduli space in curved space. Consequently, states with arbitrarily large constant expectation values as in \eqref{eq_intro_would_be_AdS} are generically unstable in interacting theories.\footnote{As we will discuss in sec.~\ref{sec_source_defects}, exceptions arise in free theories.}

\begin{figure}[t]
\centering
\includegraphics[width=0.45\linewidth]{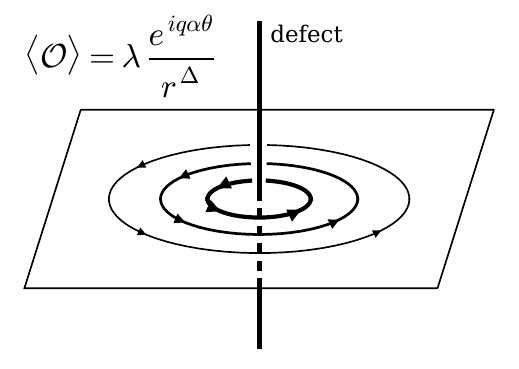}
\caption{A conformal vortex defect: for an operator $\langle\mathcal O\rangle$ of dimension $\Delta$ and charge $q$, its phase winds around the defect by $2\pi q \alpha$ with $\alpha$ the defect monodromy, while its magnitude decays as $|\lambda|/r^{\Delta}$.}
\label{fig:vortex_defect}
\end{figure}

In this paper, we show that the above construction can be modified to overcome this difficulty in theories in which a continuous global symmetry is also spontaneously broken on the moduli space. In this case, we will argue that there generically exist codimension-two defects for which the local operators parametrizing the moduli space acquire expectation values of the form
\begin{equation}\label{eq_1pt_intro}
    \langle \mO_{\Delta,\{q_a\}}(z,\bar{z})\rangle
    =
    \frac{\lambda_{\mO}}
    {z^{\frac{\Delta-\alpha^a(h)q_a}{2}}
    \bar{z}^{\frac{\Delta+\alpha^a(h)q_a}{2}}}\,,\qquad
    |\lambda_{\mO}|\stackrel{h\gg1}{\propto}h^{\Delta}\,,
\end{equation}
where $z=r e^{i\theta}$ is a complex transverse coordinate and $\{q_a\}$ denotes the charges of the operator under a set of commuting internal generators $\{Q_a\}$. Here $\lambda_{\mO}$ is a complex parameter whose absolute value grows as $h^{\Delta}$ with the unbounded modulus, while the monodromies $\alpha^a(h)$ are in general $h$-dependent and approach finite $O(1)$ values as $h\rightarrow\infty$. The phase of $\lambda_{\mO}$ is specified by additional marginal parameters, associated with the breaking of the internal symmetry. In particular, the $\alpha^a(h)$ need not be properly quantized, and in general our construction therefore defines a monodromy defect~\cite{Giombi:2021uae,Bianchi:2021snj}, attached to a codimension one topological operator implementing the corresponding internal symmetry transformation~\cite{Gaiotto:2014kfa}, see figure \ref{fig:vortex_defect}.

The expectation value \eqref{eq_1pt_intro} breaks transverse rotations and the internal symmetry separately, while preserving both the conformal group $SO(d-1,1)$ along the defect and the diagonal combination
\begin{equation}
    J+\alpha^a(h)Q_a\,.
\end{equation}
The defects we consider can therefore be viewed physically as conformal vortex configurations, in which the phase of the order parameter winds around a codimension-two locus with angular velocity set by the monodromies $\alpha^a(h)$.\footnote{Similar defects were recently studied in~\cite{Kravchuk:2025evf} in the $O(2N)$ model, although in that setup one-point functions cannot be made arbitrarily large due to the absence of a moduli space.} In the limit $h\rightarrow\infty$, the winding is locally negligible, and defect correlation functions approach those of the theory in the corresponding Poincar\'e invariant symmetry-breaking vacuum.

Known examples of vortex defects giving rise to expectation values of the form \eqref{eq_1pt_intro} include Gukov-Witten surface operators in $\mathcal{N}=4$ SYM~\cite{Gukov:2006jk,Gukov:2014gja} and vortex loops in ABJM theory~\cite{Drukker:2008jm}. In these examples, supersymmetry further constrains the winding by requiring the expectation values of chiral operators to be holomorphic. To establish the existence of analogous conformal vortices more generally, we follow the strategy of \cite{Cuomo:2024fuy} and show that configurations of the form \eqref{eq_1pt_intro} arise as self-consistent saddle points of the EFT describing the dilaton and the other light fields on the moduli space. In particular, the dilaton expectation value is controlled by the parameter $h$, ensuring that the derivative expansion is parametrically under control when it is large. The stability of these saddle points, and hence the unitarity of the corresponding defects, depends on certain EFT Wilson coefficients. Nevertheless, we find that stable conformal vortex defects exist in most of the examples we study, including cases in which the defects are not supersymmetric.

Interestingly, the saddle point we consider can be viewed as an analytic continuation of the one describing large-charge operators~\cite{Hellerman:2015nra,Monin:2016jmo,Hellerman:2017veg,Cuomo:2024fuy}. As in that context, the moduli space EFT leads to simple nonperturbative predictions for the defects in terms of the low-energy data on the moduli space. We use this framework to identify universal features of the defect operator spectrum at large $h$ and, in the supersymmetric examples, to obtain new exact results. In particular, we give a simple EFT derivation of the partition function of Gukov-Witten surface operators and compute for the first time the $g$-function for a large class of $1/2$-BPS vortex loops in ABJM theory. 

We also apply the same framework beyond these maximally supersymmetric examples. In the three-dimensional $\mathcal{N}=2$ XYZ model, we define a one-parameter family of $1/2$-BPS vortex loops, labeled by an exactly marginal coupling, and derive exact predictions for their partition function. Remarkably, we find that the latter is \emph{exactly} independent of certain flavor monodromies preserving supersymmetry, despite the absence of any symmetry argument enforcing this independence. We further construct vortex loops in several three-dimensional $\mathcal{N}=1$ SCFTs, which often admit a moduli space thanks to the discrete $\mathbb{Z}_2^R$ symmetry, see e.g.~\cite{Affleck:1982as,Gaiotto:2018yjh,Benini:2018umh,Benini:2018bhk}. We test the EFT predictions using both the $\eps$-expansion and large-$N$ methods, finding agreement in the regimes in which these descriptions are under perturbative control.

The relation between the conformal defects described above and the physics on the moduli space extends beyond the regime of validity of the EFT. To formalize this connection, we introduce an appropriate \emph{macroscopic limit}, inspired by analogous limits studied in CFTs at large scaling dimension and/or at large charge~\cite{Lashkari:2016vgj,Jafferis:2017zna,Caetano:2023zwe,Cuomo:2024fuy}. In the present setting, this limit can be understood as a decompactification limit on the $\mathrm{AdS}_{d-1}\times S^1$ Weyl frame, in which the common radius $R$ is sent to infinity while the dimensionful dilaton vev is held fixed. Under mild assumptions, we use this limit to relate the spectrum of defect operators to the mass spectrum in the corresponding vacuum on the moduli space, thereby providing a direct bridge between DCFT data and observables in the broken phase.

The rest of this paper is organized as follows. In sec.~\ref{sec_moduli_defects} we discuss defect operators in theories with moduli spaces. We first review non-conformal defects and the obstruction to defects of the form \eqref{eq_would_be_intro} in subsec.~\ref{sec_source_defects}, and then present our construction of conformal vortices and their EFT predictions in subsec.~\ref{sec_vortices_general_EFT}. In sec.~\ref{sec_SUSY_vortices} we discuss applications to $1/2$-BPS defects. After reviewing general aspects of supersymmetric monodromy and vortex defects in subsec.~\ref{subsec_BPS}, we study Gukov-Witten defects in $\mathcal{N}=4$ SYM, vortex loops in ABJM theory, and vortex loops in the XYZ model in subsecs.~\ref{sec_GW}, \ref{sec_ABJM_vortex}, and \ref{subsec_XYZ}, respectively. Non-BPS examples are discussed in sec.~\ref{sec_non_susy_vortices}, including the five-field Wess-Zumino (WZ) model in subsec.~\ref{sec_5_field}, its large-$N$ generalization in subsec.~\ref{sec:large_N}, and $\mathcal{N}=1$ SQED in subsec.~\ref{sec_SQED}. The macroscopic limit and its implications are discussed in sec.~\ref{sec_macroscopic_limit}. We finally discuss potential future directions in sec.~\ref{sec_conclusions}. Several appendices complement the text with further technical details.

\section{Moduli spaces of vacua and defects}\label{sec_moduli_defects}

\subsection{Conformal and non-conformal source defects}\label{sec_source_defects}

Some theories admit $p$-dimensional defects whose RG flow never terminates at a conformal fixed point. In such cases, scalar one-point functions can exhibit a slower decay at large distances than allowed by defect conformal invariance,\footnote{The relation between runaway source defects and moduli spaces was discussed previously in app.~B of \cite{Cuomo:2023qvp}.}
\begin{equation}\label{eq_non_conformal_1pt}
\langle\mO_{\Delta,X}(r)\rangle
\sim
\frac{\mu^\delta}{r^{\Delta-\delta}}\,,
\end{equation}
where $\Delta$ and $X$ denote the scaling dimension and the other relevant quantum numbers of the operator, respectively, $r$ is the distance from the defect, $\delta>0$, and $\mu$ is a mass scale. Since the bulk theory is conformal, the scale $\mu$ can be varied arbitrarily. In particular, by scaling $\mu$ appropriately while probing increasingly large distances from the defect, one can obtain arbitrarily large expectation values while remaining arbitrarily far from the defect. This behavior is possible only if the bulk CFT admits a flat direction. 

The simplest example of this sort is provided by a free massless scalar in $d$ dimensions with the source defect
\begin{equation}\label{eq_source}
\mathcal{D}
=\exp\left(\mu^{\delta}\int d^p y\,\phi\right)\,,
\qquad
\delta=p-\frac{d-2}{2}\geq 0\,.
\end{equation}
In this range of $p$, the coupling $\mu^{\delta}$ is relevant and the resulting defect exhibits precisely such a runaway RG flow rather than approaching a conformal defect at large distances~\cite{Cuomo:2021kfm}.

Conversely, given a CFT with a moduli space, one can argue quite generally that non-conformal source defects of the sort described above exist, at least if one allows the defect dimension $p$ to take non-integer values. The argument relies only on the low-energy effective theory on the moduli space. This EFT always contains the dilaton $\Phi$, together with the other massless moduli. We will review the general structure of the moduli-space EFT in detail in the next section. For simplicity, in this section we assume that the moduli space is one-dimensional and that no other light bosonic fields are present; their inclusion does not spoil our argument.\footnote{For instance, an example of such a one-dimensional moduli space occurs in the $3d$ $ABC$ $\mathcal{N}=1$ model~\cite{Gaiotto:2018yjh,Benini:2018bhk}.} To leading order in derivatives, the dilaton action therefore takes the form
\begin{equation}\label{eq_EFT_free_pre}
S=
\frac{1}{2}\int d^d x(\pd\Phi)^2+\ldots\,.
\end{equation}
The EFT is conventionally organized as an expansion around a vacuum with $\langle\Phi\rangle=v$, but it can equally well be used around a slowly varying background, provided that higher-derivative corrections remain small. Schematically, this requires
\begin{equation}
\frac{\pd^2}{|\langle\Phi\rangle|^{\frac{4}{d-2}}}
\ll 1\,.
\end{equation}
Consider now the rotationally invariant solution of the leading-order equations of motion
\begin{equation}
\Phi(r)
=
\frac{\mu^{p-\frac{d-2}{2}}}
{r^{d-p-2}}\,.
\end{equation}
On this background, higher-derivative corrections are suppressed relative to the leading two-derivative term by powers of $(r\mu)^{-4\left(\frac{p}{d-2}-\frac{1}{2}\right)}$. For $p>\frac{d-2}{2}$ these corrections therefore become parametrically small at distances $r\gg 1/\mu$, and the solution is under perturbative control in the infrared. It describes precisely a non-conformal $p$-dimensional source defect whose RG flow runs towards increasingly large values of the dilaton at long distances.

It is more interesting to ask if there exist also \emph{conformal} defects in which the dilaton profile can similarly be made arbitrarily large. As explained in the introduction, such a construction would provide a direct connection between DCFT data and the asymptotic data of the theory on its moduli space. A simple example is the source defect~\eqref{eq_source} with $p=\frac{d-2}{2}$, which induces the following expectation value for the scalar field
\begin{equation}\label{eq_free_pinning}
\langle\phi(r)\rangle=\left(\frac{h}{r}\right)^{\frac{d-2}{2}}\,,
\end{equation}
where the dimensionless parameter $h$ that controls the magnitude of the one-point function is associated with an exactly marginal deformation of the trivial defect.\footnote{Explicitly, the deformation is $\tilde{h}\int d^p y \phi$, with $\tilde{h}=4\pi^{\frac{d+2}{4}}h^{\frac{d-2}{2}}/\Gamma\left(\frac{d-2}{4}\right)$.}

However, this simple construction does not generalize to a generic interacting theory. To see the obstruction, consider expanding the moduli-space EFT around a scale-invariant dilaton profile $\Phi$ coinciding with~\eqref{eq_free_pinning}.
This is a solution of the equations of motion of the free two-derivative action. However, higher-derivative terms in the EFT generically lift the would-be marginal direction parametrized by $h$, generating an effective potential for it \cite{Hinterbichler:2022ids}. This is particularly transparent after a Weyl transformation. Writing the flat-space metric around a planar defect as
\begin{equation}\label{eq_metric_Weyl}
ds^2=ds^2_{\mathds{R}^p}+dr^2+r^2d\Omega_{d-p-1}^2=r^2\left(\frac{ds^2_{\mathds{R}^p}+dr^2}{r^2}+d\Omega_{d-p-1}^2\right)\,,
\end{equation}
where $d\Omega_{d-p-1}^2$ is the sphere metric, the rescaling $ds^2\rightarrow R^2ds^2/r^2$ maps the geometry to $\mathrm{AdS}_{p+1}\times S^{d-p-1}$, with both factors of radius $R$ and the defect at the conformal boundary of $\mathrm{AdS}_{p+1}$. Since $\Phi$ has Weyl weight $(d-2)/2$, in this frame the dilaton profile is constant, $\Phi=(h/R)^{\frac{d-2}{2}}$, while higher-derivative interactions reduce to local curvature invariants multiplied by powers of $\Phi$. Schematically, a term containing $2n$ derivatives gives a contribution of the form
\begin{equation}
\frac{\mathcal{R}^{n}}
{|\Phi|^{\frac{2(2n-d)}{d-2}}}\stackrel{h\gg 1}{\propto} h^{d-2n}
\end{equation}
Evaluated on $\mathrm{AdS}_{p+1}\times S^{d-p-1}$, such terms generate a potential along the putative marginal direction. There is no reason for an extremum of this potential to occur at parametrically large $h$, where the moduli-space EFT is under perturbative control. The existence of a conformal defect with a noncompact marginal parameter therefore requires additional structure beyond the mere existence of a moduli space.

In the next section we will see that we can overcome these obstacles by considering defects which also break rotational symmetry, under the assumption that a continuous internal group is also broken along some direction of the moduli space.

\subsection{Conformal vortices from EFT}\label{sec_vortices_general_EFT}

\subsubsection{General construction}

Let us consider a CFT in $d>2$ with continuous internal symmetry $G$ and a moduli space, such that $G$ is broken to a subgroup $H$ along some direction in the moduli space of vacua. In this setup,~\cite{Cuomo:2024fuy} used the moduli space EFT to relate the properties of operators with large internal charge to the data of the moduli space. Here we use an analogous construction to define a family of conformal defects, whose properties are controlled by the dilaton and the other light moduli.

We briefly review the aspects of the moduli-space EFT that will be useful below. Suppose that in the region of interest, conformal symmetry is spontaneously broken together with an internal symmetry group $G\to H$. We denote by $\Phi$ the dilaton and separate explicitly a set of Goldstone fields $\pi^a$ associated with broken commuting Cartan generators $Q_a$. The remaining scalar fields on the moduli space parametrizing the remaining coordinates on the moduli space, including possible additional Goldstone bosons, will be denoted by $\phi^A$. Additional massless degrees of freedom, such as superpartners or an interacting IR sector, may also be present.

The nonlinear realization of conformal symmetry is most conveniently implemented by introducing the Weyl-invariant metric~\cite{Salam:1969bwb,Isham:1970gz,Ellis:1971sa,Luty:2012ww}
\begin{equation}\label{eq_g_hat}
\hat{g}_{\mu\nu}
=g_{\mu\nu}|\Phi|^{\frac{4}{d-2}}\,.
\end{equation}
We take $\Phi$ to have dimension $(d-2)/2$, while the $\pi^a$ and $\phi^A$ are dimensionless and have vanishing Weyl weight. We choose coordinates on the coset such that the broken Cartan generators act as constant shifts of the corresponding fields, $\pi^a\rightarrow\pi^a+\text{const}$, so that the EFT depends on the $\pi^a$ only through derivatives.

To leading order in derivatives, the scalar sector takes the form
\begin{equation}\label{eq_EFT}
\begin{split}
S_{\rm EFT}
=\int d^d x\sqrt{\hat g}\Bigg[
&
\frac{1}{2}\hat G_{ab}(\phi)\hat g^{\mu\nu}
\pd_\mu\pi^a\pd_\nu\pi^b
+
\hat G_{aB}(\phi)\hat g^{\mu\nu}
\pd_\mu\pi^a\pd_\nu\phi^B
\\
&
+\frac{1}{2}\hat G_{AB}(\phi)\hat g^{\mu\nu}
\pd_\mu\phi^A\pd_\nu\phi^B+
\hat G_{\Phi\Phi}(\phi)
\frac{d-2}{8(d-1)}
\hat{\mathcal R}
\Bigg]+
S_{\rm CFT_{\rm IR}}
+\ldots\,.
\end{split}
\end{equation}
The functions $\hat G_{ab}$, $\hat G_{aB}$, $\hat G_{AB}$ and $\hat G_{\Phi\Phi}$ depend only on the remaining moduli $\phi^A$. In flat space, the last term in~\eqref{eq_EFT} contains the standard kinetic term for the dilaton. Additional symmetries, such as supersymmetry, further constrain the sigma-model metric and the remaining IR sector, which may for instance contain the superpartners of the moduli.

Higher-derivative interactions are constructed analogously from Weyl-invariant geometric quantities and derivatives of the moduli. For instance, at four-derivative order one finds terms of the form
\begin{equation}\label{eq_EFT_4derivatives}
S_4
=\int d^d x\sqrt{\hat{g}}
\left[
c_1(\phi)(\hat{\mR}_{\mu\nu})^2
+
c_2(\phi)(\hat{\mR}_{\mu\nu\rho\sigma})^2
+\ldots
\right]\,.
\end{equation}
In even dimensions, the higher-derivative action also contains Wess-Zumino terms required to match the conformal anomaly, as well as possible 't Hooft anomalies of the internal symmetries. In $d=2n$, such terms arise at order $2n$ in the derivative expansion~\cite{Schwimmer:2010za,Komargodski:2011vj,Baume:2013ika}. We will discuss their role in an explicit example in the next section.

As reviewed around~\eqref{eq_metric_Weyl}, a codimension-two conformal defect can equivalently be viewed as a boundary condition for the theory on $\mathrm{AdS}_{d-1}\times S^1$. We therefore specialize the moduli-space EFT to $\mathrm{AdS}_{d-1}\times S^1$, so that the action reads
\begin{equation}\label{eq_EFT_AdS}
\begin{split}
S_{\rm EFT}
=\int d^d x\sqrt{g}\Bigg[
&
\frac{\Phi^2}{2}\pd_\mu\pi^a\pd^\mu\pi^b \hat{G}_{ab}(\phi)
+
\Phi^2\pd_\mu\pi^a\pd^\mu\phi^B \hat{G}_{aB}(\phi)
\\
&
+\frac{\Phi^2}{2}\pd_\mu\phi^A\pd^\mu\phi^B \hat{G}_{AB}(\phi)+\frac{\hat{G}_{\Phi\Phi}(\phi)}{2}
\left(-\Phi\pd^2\Phi+m_d^2\Phi^2\right)
\Bigg]
+\ldots\,,
\end{split}
\end{equation}
where the conformal mass is negative and saturates the Breitenlohner-Freedman bound for a scalar in $\mathrm{AdS}_{d-1}$:
\begin{equation}
m_d^2=\frac{d-2}{4(d-1)}\mathcal{R}=-\frac{(d-2)^2}{4 R^2} \,.
\end{equation}

We now consider configurations of the form
\begin{equation}\label{eq_AdS_sol}
\Phi=v^{\frac{d-2}{2}}\,,\qquad
\pi^a=\alpha^a \theta+\bar{\pi}^a\,,\qquad
\phi^A=\bar{\phi}^A\,,
\end{equation}
where $\theta\sim\theta+2\pi$ parametrizes the circle and $2\pi\alpha^a$ is the monodromy of the corresponding Goldstone field, $\bar{\pi}^a$ and $\bar{\phi}^A$ are constants. This background preserves the $SO(p+1,1)$ isometry of hyperbolic space, and therefore defines a conformal defect. On the other hand, the winding of the Goldstones breaks the transverse $SO(2)$ rotations generated by $J$, while preserving the diagonal combination
\begin{equation}\label{eq_unbroken_J}
J+\alpha^a Q_a\,.
\end{equation}
Here the $Q_a$ are the broken Cartan generators acting by constant shifts on the Goldstones $\pi^a$.

Because the $\pi^a$ enter the EFT only through derivatives, their equations of motion are automatically satisfied by~\eqref{eq_AdS_sol}. The remaining equations reduce to algebraic conditions analogous to those arising in the large-charge saddle on the cylinder~\cite{Cuomo:2024fuy}, with the timelike Goldstone gradient replaced here by a spacelike winding around $S^1$. At leading order in derivatives, the equation of motion for $\Phi$ gives, for $v\neq0$,
\begin{align}\label{eq_PHI_eom}
\alpha^a \alpha^b \hat{G}_{ab}(\bar{\phi})/R^2-\hat{G}_{\Phi\Phi}(\bar{\phi})|m_d^2| =0\,,
\end{align}
while the remaining moduli obey
\begin{equation}\label{eq_phi_EOM}
\left[|m_d^2|\frac{\pd \hat{G}_{\Phi\Phi}(\phi)}{\pd \phi^A}-\frac{\alpha^a \alpha^b}{R^2}\frac{\pd \hat{G}_{ab}(\phi)}{\pd\phi^A}\right]_{\phi=\bar{\phi}}=0\,.
\end{equation}
Crucially, these equations do not constrain $v$ and simply relate $\alpha^a$ and $\bar{\phi}^A$. For an $n$ dimensional moduli space with (locally) $k$ broken Cartan generators $Q^a$, these provide $n-k$ conditions for the $n-1$ parameters $\{\alpha^a,\bar{\phi}^A\}$, while the constants $\bar{\pi}^a$ are unconstrained. Assuming these conditions admit at least a real stable solution (as we will verify in several examples), $v$ parametrizes a continuous family of conformal defect backgrounds. The EFT description is reliable when the scale set by the dilaton expectation value is much larger than the gradients along the circle,
\begin{equation}
v\gg|\pd\pi|\sim\frac{|\alpha|}{R}\,.
\end{equation}
 
Notice that, in general, the values of $\alpha^a$ solving the equations of motion need not be integers. The corresponding Goldstone fields therefore acquire a nontrivial monodromy upon circling the defect. As a result, the configuration does not generically define a genuine standalone codimension-two defect, but rather it is the boundary of a topological codimension-one defect implementing the corresponding internal symmetry transformation.

Differently from the conformal source defect, higher-derivative interactions, such as those in \eqref{eq_EFT_4derivatives}, simply introduce corrections suppressed by powers of $(Rv)^{-2}$ in the EOMs~\eqref{eq_PHI_eom} and~\eqref{eq_phi_EOM}. Thus, the solution more generally takes the form $\alpha^a=\alpha^a(v)$ and $\bar{\phi}^A=\bar{\phi}^A(v)$, with their large-$v$ limits determined by~\eqref{eq_PHI_eom} and~\eqref{eq_phi_EOM}, while $v$ remains the free parameter labeling the defect family. 

Physically, the family of defects constructed above describes a vortex configuration. As a simple example, a complex free scalar field $\varphi$ on $\mathrm{AdS}_{d-1}\times S^1$ admits the following solution
\begin{equation}
    \varphi=v^{\frac{d-2}{2}}e^{- i\frac{d-2}{2}\theta-i\bar{\pi}}\,.
\end{equation}
Upon a Weyl rescaling to flat space and denoting $z=re^{i\theta}$, this profile describes a conformal one-point function
\begin{equation}\label{eq_free_complex_scalar_profile}
    \varphi=\frac{\lambda}{z^{\frac{d-2}{2}}}\,,\qquad \lambda=(vR)^{\frac{d-2}{2}}e^{-i\bar{\pi}}\,.
\end{equation}
For odd $d$, the exponent is half-integer and the scalar therefore acquires a nontrivial monodromy upon circling the defect. Another notable example of this construction, that we will discuss in detail in sec.~\ref{sec_GW}, is the Gukov-Witten defect in $\mathcal{N}=4$ SYM.

Before discussing the predictions of EFT in greater detail, a few comments are in order.

First, we remark that, although $v$ parametrizes a continuous family of conformal defects, it does not in general correspond to a marginal defect coupling. The reason is that the monodromies $\alpha^a(v)$ generally vary along the family. In the next section, however, we will see that supersymmetry fixes the monodromies to be independent of $v$ for BPS defects. In that case, the dilaton scale $vR$ can be identified with a genuine marginal coupling.

Second, as already anticipated, the vortex configuration constructed above is closely related to the large-charge saddle of~\cite{Hellerman:2017veg,Cuomo:2024fuy}. Notice first that the two configurations preserve the same number of symmetries. The large-charge solution is invariant under $\mathbb{R}_{D-m^aQ_a}\times SO(d)$, while the vortex defect preserves $SO(d-1,1)\times U(1)_{J+\alpha^aQ_a}$. This suggests that these might be related via analytic continuation.

To make this relation explicit, consider the Euclidean $\mathrm{AdS}_{d-1}\times S^1$ metric in global coordinates:
\begin{equation}
ds^2=
R^2\left(
d\rho^2+\sinh^2\rho\,d\Omega_{d-2}^2+d\theta^2
\right)\,.
\end{equation}
We complexify the geometry and perform the analytic continuation
\begin{equation}\label{eq_analytic_cont}
R\rightarrow i\widetilde R\,,
\qquad
\rho\rightarrow i\chi\,,
\qquad
\theta\rightarrow t/
\tilde{R}\,.
\end{equation}
At the same time, we pass to the universal cover of the original $S^1$, so that the periodic coordinate $\theta$ is mapped to the noncompact Lorentzian time $t$. The resulting metric is
\begin{equation}
ds^2
= -dt^2+
\widetilde{R}^2\left(d\chi^2+\sin^2\chi\,d\Omega_{d-2}^2\right)=-dt^2+\widetilde{R}^2 d\Omega_{d-1}^2\,,
\end{equation}
which describes the Lorentzian cylinder $\mathbb{R}\times S^{d-1}$. The vortex background is mapped to
\begin{equation}
\Phi=v^{\frac{d-2}{2}}\,,
\qquad
\pi^a=\alpha^a\theta+\bar{\pi}^a
\quad\longrightarrow\quad
\Phi=v^{\frac{d-2}{2}}\,,
\qquad
\pi^a
=\frac{\alpha^a}{\widetilde R}t+\bar{\pi}^a\,.
\end{equation}
This is precisely the large-charge saddle of~\cite{Cuomo:2024fuy}, with the chemical potentials identified as $m^a=\alpha^a$. Accordingly, the diagonal symmetry $J+\alpha^aQ_a$ preserved by the vortex is mapped to the helical symmetry $D-m^aQ_a$ preserved by the large-charge state. Under the same continuation, the negative conformal mass on $\mathrm{AdS}_{d-1}\times S^1$ becomes the positive conformal mass on the cylinder, and the algebraic equations determining the vortex background map directly to those of the large-charge saddle.

\subsubsection{Observables}

From the EFT we can obtain the defect operator spectrum by expanding the action into fluctuations and analyzing the allowed boundary conditions for the fields. We focus on the universal bosonic sector associated with the moduli. The expansion of their action is obtained by analytic continuation from~\cite{Cuomo:2024fuy}. We report the main step below.

We denote by $\Gb=\hat{G}(\bar{\phi})$ the metric evaluated on the solution of~\eqref{eq_PHI_eom} and \eqref{eq_phi_EOM}, and use additional subscripts to denote derivatives with respect to the $\phi^A$'s, e.g. $\Gb_{ab,C}=\pd \hat{G}_{ab}/\pd\phi^C\vert_{\phi=\bar{\phi}}$, $\Gb_{\Phi\Phi,A}=\pd \hat{G}_{\Phi\Phi}/\pd\phi^A\vert_{\phi=\bar{\phi}}$, and similarly for higher derivatives. We expand the fields as
\begin{align}
\Phi&=v^{\frac{d-2}{2}}+\delta\Phi-\frac12\frac{\Gb_{\Phi\Phi,A}}{\Gb_{\Phi\Phi}}\delta\phi^A\,,\\
\pi^a&=\alpha^a \theta+\bar{\pi}^a+v^{-\frac{d-2}{2}}\left[\alpha^a\left(\delta\pi_0-\frac{\alpha^c\Gb_{cB}}{R^2 |m^2_d|\Gb_{\Phi\Phi}}\delta\phi^B\right)+\delta\pi^a_{\bot}\right]\,,\quad\alpha^a\Gb_{ab}\delta\pi^b_{\bot}=0\,,\\
\phi^A&=\bar{\phi}^A+\delta \phi^A/v^{\frac{d-2}{2}}\,.
\end{align}
Using the equations of motion, we then arrive at
\begin{equation}\label{eq_fluct_action}
\begin{split}
\mL_{\rm fluct}
&=\frac{1}{2}\Gb_{\Phi\Phi}\left[\left(\pd\delta\Phi\right)^2+R^2|m_d^2|(\pd\delta\pi_0)^2+4 |m_d^2|\,\delta\Phi\pd_\theta\delta\pi_0
\right]
\\&
+\frac12\Gb_{ab}\pd^\mu\delta\pi^a_{\bot}\pd_\mu\delta\pi^b_{\bot}
+ \Gb_{aB}\pd_\mu\delta\pi^a_{\bot}\pd^\mu\delta\phi^B
+\frac12 K^2_{AB}\pd^\mu\delta\phi^A\pd_\mu\delta\phi^B
\\&
+\frac{\alpha^a}{R^2}\Gb_{ab,A}\delta\phi^A\pd_\theta\delta\pi^b_{\bot}
+\frac{1}{R}d_{AB}\delta\phi^B\pd_\theta\delta\phi^A
+\frac{1}{2} M^2_{AB}\delta\phi^A\delta\phi^B\,,
\end{split}
\end{equation}
where we defined
\begin{align}
K^2_{AB}&=\Gb_{AB}-\frac{\Gb_{\Phi\Phi,A}\Gb_{\Phi\Phi,B}}{4\Gb_{\Phi\Phi}}
- \frac{\alpha^c\Gb_{cA}\alpha^d\Gb_{dB}}{R^2 |m_d^2|\Gb_{\Phi\Phi}}\,,\\
d_{BC}&=\frac{\alpha^a}{R}\Gb_{aB,C}-\frac{ \alpha^a\Gb_{aB}\Gb_{\Phi\Phi,C}}{R\,\Gb_{\Phi\Phi}}\,,\\
M^2_{AB}&= \frac{\alpha^c \alpha^d}{2R^2}\,\Gb_{cd,AB}-\frac{|m_d^2|}{2}\Gb_{\Phi\Phi,AB}\,.
\end{align}

To obtain the spectrum of defect operators now we simply need to decompose the fields into KK modes over the $S^1$
\begin{equation}
\delta\Phi(x,\theta)=\sum_{\ell\in\mathbb{Z}}\frac{e^{i\ell\theta}}{\sqrt{2\pi}}\delta\Phi^{(\ell)}(x)\,,\quad
\delta\pi_0(x,\theta)=\sum_{\ell\in\mathbb{Z}}\frac{e^{i\ell\theta}}{\sqrt{2\pi}}\delta\pi_0^{(\ell)}(x)\,,\quad
\text{etc}.\,.
\end{equation}
For each $\ell$, we then diagonalize the resulting quadratic operator on $\mathrm{AdS}_{d-1}$. Denoting its mass eigenvalues by $m^2$, the corresponding defect scaling dimensions are obtained from the standard AdS mass-dimension relation
\begin{equation}
    \Delta(\Delta-d+2)=R^2m^2\,.
\end{equation}

From the first line of~\eqref{eq_fluct_action}, we see that $\delta\Phi$ and $\delta\pi_0$ decouple from the other fields to this order, and their quadratic action is identical to that of the fluctuations of a free complex scalar around the profile~\eqref{eq_free_complex_scalar_profile}. The bosonic spectrum therefore always contains two universal KK towers with masses
\begin{equation}\label{eq_KK_masses_general}
R^2m_{\ell,(\pm)}^2=\ell\left[\ell\pm(d-2)\right]
\qquad
\forall\,\ell\in\mathbb{Z}\,,
\end{equation}
up to corrections of order $O\left((Rv)^{-2}\right)$. These masses always satisfy the BF bound $R^2m^2\geq-(d-2)^2/4$. For even $d$, the bound is saturated by the modes $\ell=\mp(d-2)/2$. Their stability, as well as the choice of admissible boundary condition, may then depend on subleading corrections. 
The masses associated with the remaining moduli must similarly be checked against the BF bound. Several examples will be discussed in the next sections. In particular, supersymmetry often protects the spectrum against such instabilities.

In every dimension, one can therefore choose boundary conditions such that the defect operator spectrum contains modes with
\begin{equation}
\Delta^{\ell,(\pm)}=
\frac{d-2}{2}
+
\left|
\ell\pm\frac{d-2}{2}
\right|\,.
\end{equation}
Here the KK momentum $\ell$ measures the eigenvalue under the unbroken combination~\eqref{eq_unbroken_J}. 

Note that at $\ell=0$ we find two neutral marginal deformations: one is associated with the parameter 
\begin{equation}\label{eq_h_par_def}
    h\equiv R v\,,
\end{equation}
while the other shifts the Goldstones as $\pi^a\rightarrow \pi^a+c\,\alpha^a$. More in general, there must be a marginal operator, called \emph{tilt}, for each broken internal generator (see e.g. \cite{Cuomo:2021cnb,Padayasi:2021sik,Drukker:2022pxk}). On the other hand, in the absence of supersymmetry, subleading corrections will make the marginal deformation associated with the dilaton size $h$ either slightly irrelevant or slightly relevant; in the latter case, we expect that integrating this deformation on the defect will trigger an RG flow to a non-conformal defect of the kind discussed around~\eqref{eq_non_conformal_1pt}. In the supersymmetric examples of section~\ref{sec_SUSY_vortices}, this deformation is instead a partner of the tilt, which is exactly marginal (see section~\ref{subsec_BPS}).

Whenever
\begin{equation}
0<\left|\ell\pm\frac{d-2}{2}\right|<1\,,
\end{equation}
the corresponding AdS scalar lies in the alternate-quantization window, and a second admissible conformal boundary condition gives instead
\begin{equation}
\Delta_{\rm alt}^{\ell,(\pm)}=\frac{d-2}{2}-\left|\ell\pm\frac{d-2}{2}\right|\,.
\end{equation}
This happens for odd $d\geq5$, for the modes
\begin{equation}
\ell=\frac{d-3}{2}\,, \frac{d-1}{2}
\quad\text{in the $(-)$ tower}\,,
\qquad
\ell=-\frac{d-3}{2}\,, -\frac{d-1}{2}
\quad\text{in the $(+)$ tower}\,,
\end{equation}
for which
\begin{equation}
\Delta^{\ell,(\pm)}=\frac{d-1}{2}\,,
\qquad\Delta_{\rm alt}^{\ell,(\pm)}=\frac{d-3}{2}\,.
\end{equation}
These two boundary conditions are related by an RG flow triggered by the neutral double-trace operator of dimension $d-3$ at the UV fixed point corresponding to alternate quantization. In $d=3$ the alternate dimension would vanish, which is excluded by unitarity (see app.~\ref{app:one_loop_determinants}).

The masses, and hence the dimensions, of the operators associated with the other moduli depend instead on the metric of the nonlinear sigma model. On general grounds, at large KK momentum their mass squared grows as $\ell^2/R^2$, and therefore the corresponding defect dimensions grow linearly with $|\ell|$. The same conclusion applies to other weakly coupled fields, such as superpartners, when present.\footnote{It is also possible to turn on conical singularities, which modify the spectrum of defect operators associated with spinning fields. We will discuss conical singularity for supersymmetric vortex defects in sec. \ref{sec_SUSY_vortices}.}

The EFT also predicts the scaling of one-point functions with the parameter $h$ defined in~\eqref{eq_h_par_def}. Indeed, barring additional symmetry constraints, an operator of dimension $\Delta$, charges $(q_a)$ under the Cartan fields and spin $J$ can be represented in terms of the light fields on the moduli space as
\begin{equation}
\mO^{(\Delta,q_a)}_{\mu_1\ldots\mu_J}
\simeq
\pd_{\mu_1}\pi^{a_1}\ldots\pd_{\mu_J}\pi^{a_J}
|\Phi|^{\frac{2(\Delta-J)}{d-2}}e^{i q_a\pi^a}
f_{a_1\ldots a_J}(\phi)
+\ldots\,.
\end{equation}
Note that for non-Abelian symmetry groups, $f_{a_1\ldots a_J}(\phi)$ should be chosen such that the operator transforms according to the proper irrep.\footnote{See e.g.~\cite{Monin:2016jmo} for details on this procedure in a similar context.} Evaluating this expression on the saddle point~\eqref{eq_AdS_sol}, we find
\begin{equation}\label{eq_EFT_1pt_general}
\langle \mO^{(\Delta,q_a)}_{\mu_1\ldots\mu_J}\rangle
\propto\delta_{\mu_1}^\theta\ldots\delta_{\mu_J}^\theta
\frac{v^{\Delta-J}e^{i q_a\bar{\pi}^a}}{R^J} e^{iq_a\alpha^a\theta}
\left[
1+O\left(\frac{1}{R^2v^2}\right)
\right]\,.
\end{equation}
This scaling with $v$ holds also for the internal currents associated with the Cartan generators $Q_a$ and the stress tensor, which can be computed directly by Noether procedure from the action~\eqref{eq_EFT}. 
To be explicit, upon Weyl rescaling back to flat space and introducing the transverse coordinate $z=re^{i\theta}$, the one-point function of a scalar operator takes the form
\begin{equation}\label{eq_1pt_fun_EFT}
\left\langle \mO^{(\Delta,q_a)}(z,\bar z)\right\rangle=
C_{\mO}(\bar\phi)
\frac{
h^{\Delta}e^{i q_a\bar{\pi}^a}
}{
z^{\frac{\Delta-q_a\alpha^a}{2}}
\bar z^{\frac{\Delta+q_a\alpha^a}{2}}
}
\left[
1+O\left(h^{-2}\right)
\right]\,,
\end{equation}
where $C_{\mO}(\bar\phi)$ is a $O(1)$ theory-dependent coefficient.

Finally, we consider the partition function of a spherical defect, normalized by the partition function of the bulk CFT. In general, this quantity is ambiguous because of local counterterms supported on the defect worldvolume. For a codimension-two defect, these generate contributions scaling as $(R/\epsilon)^{d-2-2n}$, with integer $n\leq (d-2)/2$, where $\epsilon$ is a UV regulator. In odd $d$, the universal contribution is the finite constant term in $\log\langle\mathcal{D}\rangle$, which is unaffected by such local counterterms. In even $d$, instead, the finite term is scheme dependent, while the universal information is contained in the coefficient of the logarithmic divergence. 
We therefore define $f$ through
\begin{equation}
\log \langle\mathcal{D}\rangle
=
\text{counterterms}
+
\begin{cases}
(-1)^{\frac{d-3}{2}}f
&
d=\text{odd}
\\
(-1)^{\frac{d}{2}}f\log\left(\frac{R}{\epsilon}\right)
&
d=\text{even}\,.
\end{cases}
\end{equation}
For the codimension-two defects considered here, $p=d-2$, this definition reproduces the familiar RG monotones in dimensions $p=1,2,3,4$,
\begin{equation}\label{eq_g_b_def}
f
=
\begin{cases}
\log g & p=1
\\
\frac{b}{3} & p=2
\\
F & p=3
\\
4a & p=4\,,
\end{cases}
\end{equation}
where $g$ is the line-defect $g$-function, $b$ is the Weyl anomaly coefficient of a two-dimensional defect, $F$ is the universal sphere free energy of a three-dimensional defect, and $a$ is the Weyl anomaly coefficient of a four-dimensional defect. The quantity $f$ defined in this way has been proven to be a defect RG monotone for defects of dimension $p\leq4$~\cite{Friedan:2003yc,Jensen:2015swa,Wang:2021mdq,Cuomo:2021rkm,Casini:2023kyj}.

A convenient way to compute $f$ is to evaluate the EFT partition function on $\mathrm{AdS}_{d-1}\times S^1$, and normalize the result by the full theory result in the conformal vacuum. By homogeneity along $\mathrm{AdS}_{d-1}$, the partition function takes the form
\begin{equation}
\log Z=
-\operatorname{Vol}(\mathrm{AdS}_{d-1})\hat{f}\,,
\end{equation}
where $\hat{f}$ denotes the free-energy density along the AdS factor. In the EFT, it admits a semiclassical expansion of the form
\begin{equation}
\hat{f}=2\pi\mL_{\rm classical}+\hat{f}_{\rm 1-loop}+\ldots\,.
\end{equation}
Using the regulated AdS volume reviewed in app.~\ref{App_AdS_volume}, one finds that $f\propto \hat{f}-\hat{f}_{\rm CFT}$ with a positive proportionality coefficient, where $\hat{f}_{\rm CFT}$ is the CFT free energy density in the vacuum, which is independent of $v$. Since in what follows we are only interested in the dependence on the modulus $v$, it is therefore sufficient to study the scaling of $\hat{f}$ itself.

The leading-order classical action~\eqref{eq_EFT_AdS} is quadratic in the dilaton $\Phi$ and vanishes when evaluated on the solution~\eqref{eq_AdS_sol}, as a consequence of the dilaton equation of motion. Barring additional cancellations, the leading classical contribution therefore arises from the first subleading term in the derivative expansion and scales as
\begin{equation}
    \mL_{\rm classical}\propto (Rv)^{d-4}\,.
\end{equation}
The leading quantum correction instead scales as $F_{\rm 1-loop}\propto (Rv)^0$ in odd $d$. In even dimensions it can contain a logarithmic divergence, which is renormalized by a local term in the classical EFT and gives rise to a contribution proportional to $\log(Rv)$. Combining these observations, the leading large-$Rv$ behavior of the universal defect free energy in terms of $h=Rv$ is
\begin{equation}\label{eq_f-EFT_prediction}
    f\propto\begin{cases}
         h^{d-4} &d>4 \\
        \log h & d=4\\
        \text{const.} &d<4\,.
    \end{cases}
\end{equation}
It is important to emphasize that this discussion did not assume any additional symmetry. In particular, when $h$ parametrizes an exactly marginal deformation, as happens for the BPS defects discussed below, $f$ is constant along the corresponding defect conformal manifold. It is nonetheless predicted by the EFT, as we will see in several examples.

\section{Supersymmetric vortices}\label{sec_SUSY_vortices}

\subsection{Generalities on BPS conditions}\label{subsec_BPS}

We start by describing general conditions on a monodromy defect for the R-symmetry to preserve some SUSY. For concreteness we will focus on theories with four supercharges and a $U(1)_R$ symmetry, and discuss 1/2-BPS monodromy defects. For definiteness we state our conventions for the supersymmetry algebra in $4$ dimensions in app. \ref{app:conventions_SUSY}, but our discussion below applies in general dimensions. In particular, we take supercharges to have unit charge under the bulk $U(1)_R$ symmetry.

Consider a monodromy defect for the $R$-symmetry, defined by introducing the twisted boundary conditions around a codimension two locus as
\begin{equation}
\mO(\theta+2\pi)=U_\alpha\mO(\theta)\,,\quad U_\alpha=e^{-2\pi i\alpha R}\,,
\end{equation}
for any operator $\mathcal{O}$ with R-charge $R$. In general, a twist by $g\in G$ breaks the symmetry group to $H\subset G$, where all $h\in H$ are such that $h g=g h$. Since $U_\alpha Q_{\alpha}U^{-1}_{\alpha}=e^{2\pi i \alpha}Q_\alpha$ for any supercharge, it follows that a monodromy defect for the R-symmetry breaks SUSY unless
\begin{equation}
    \alpha=0\mod 1\,.
\end{equation}
Note that integer $\alpha$ does not necessarily correspond to a trivial monodromy, since the R-charges of local operators are not necessarily integerly quantized in our conventions. We will see examples below.

Another way to preserve supersymmetry is to introduce a conical singularity in addition to the twist by the R-symmetry. This is done by replacing the twisted boundary conditions with a modified version:
    \begin{equation}
\mO(\theta+2\pi)=U_{\alpha,\beta}\mO(\theta)\,,\quad
U_{\alpha,\beta}=e^{-2\pi i\alpha R-2\pi i \beta J}\,,
\end{equation}
where $J$ generates rotations around the defect. Choosing 
\begin{equation}
        \beta=\pm 2 \alpha\mod 2
\end{equation} 
may preserve $Q_+$ or $Q_-$, where $\pm$ denotes the charge of $Q$ under the transverse rotation around the defect. Supersymmetric conical defects were recently considered in \cite{Arav:2024exg,Arav:2024wyg}.

In this paper we also impose conformal symmetry, which further constrains the allowed defects. 
Assume the SCFT has a moduli space, which is parametrized by at least a chiral operator $\Phi$ with bottom component $\phi$. Recall that the dimension of a chiral operator in $d$ spacetime dimensions is related to its R-charge by
\begin{equation}\label{eq:BPS_cond}
    \Delta = \frac{d-1}{2}R\;.
\end{equation}
We then consider codimension two defects where the 
bottom component of $\Phi$ acquires a holomorphic expectation value given by
\begin{equation}\label{eq:gen_bg_bps}
    \langle\phi(z,\bar{z})\rangle = \frac{\lambda}{z^\Delta}\;,
\end{equation}
where $z=re^{i\theta}$ is a complex coordinate in the transverse plane to the defect and $\lambda$ is a complex parameter. Its modulus is associated with an unbounded marginal deformation, $|\lambda|\propto h^{\Delta}$ in the notation of the previous section.
In appendix \ref{app:BPS_condition} we show explicitly that the background \eqref{eq:gen_bg_bps} satisfies indeed the 1/2 BPS condition for a chiral field in $d=4$ (and a similar analysis can be done in $3d$). In fact, we show that it preserves a complex superconformal charge $S$ as well. Supersymmetry further implies that the bottom components of the chiral-ring operators $\Phi^n$ generated by $\Phi$ have expectation values fixed by \eqref{eq:gen_bg_bps}:
\begin{equation}
\langle \phi^n(z,\bar{z})\rangle=\langle \phi(z,\bar{z})\rangle^n=\frac{\lambda^n}{z^{n\Delta}}\,.
\end{equation}

When chiral operators have integer dimensions, as in $\mathcal{N}=4$ SYM, the background \eqref{eq:gen_bg_bps} does not require the introduction of a monodromy. In all other cases, the background \eqref{eq:gen_bg_bps} requires an R-monodromy given by
\begin{equation}
    \alpha=\frac{d-1}{2}\,.
\end{equation}
For general spacetime dimension $d$, supersymmetry therefore requires the introduction of a conical defect with parameter
\begin{equation}\label{eq_conical_susy}
    \beta=\pm(d-3)\mod 2\,.
\end{equation}
In practice, this simplifies in physical spacetime dimensions:
\begin{itemize}
    \item In odd dimensions, $\alpha= 0\mod 1$ and no conical defect is needed. Then \eqref{eq:gen_bg_bps} defines a conformal 1/2-BPS monodromy defect.
    \item In even dimensions, in general $\alpha = 1/2\mod 1$ which requires a conical defect with $\beta = 1$. This is a twist by fermion number $(-1)^F$ and is not associated with a geometric singularity of the curvature. In other words, the correct monodromy in this case is
    \begin{equation}
        U_\alpha=e^{-2\pi i \frac{d-1}{2} R}(-1)^F\;.
    \end{equation}
\end{itemize}

As stated above, $\lambda$ parametrizes a conformal manifold of defects, and must therefore be associated with an exactly marginal complex defect operator. Below we show that this follows from supersymmetry and from the breaking of $U(1)_R$ alone; further details on this argument can be found in appendix~\ref{app:susy_tilt}. 

The background \eqref{eq:gen_bg_bps} breaks both $U(1)_R$ and the transverse rotation $J$, and preserves only a linear combination $r$ under which it is invariant; this is the $u(1)$ generator of the preserved $su(1,1|1)$ superconformal algebra (see appendix~\ref{app:BPS_condition}). Since $U(1)_R$ is broken, the defect has a \emph{tilt} operator $\mathcal{T}$ \cite{Cuomo:2021cnb,Padayasi:2021sik,Drukker:2022pxk}, defined by the Ward identity
\begin{equation}\label{eq_tilt_def}
    \partial_\mu j^\mu_R(x)=\delta^{(2)}(z)\,\mathcal{T}(y)\,,
\end{equation}
where $j^\mu_R$ is the $U(1)_R$ current and $y$ denotes the coordinates along the defect. The tilt has protected dimension $p=d-2$, and deforming the defect by it rotates the phase of $\lambda$.

Supersymmetry implies that the tilt is an F-term. Denote the preserved supercharges by $\mathcal Q$ and $\overline{\mathcal Q}$, with $\overline{\mathcal Q}\phi=0$. We then have  
\begin{equation}\label{eq_susy_tilt}
    \mathcal{T}(y)=\left(\overline{\mathcal Q}\hat{\psi}+\mathcal Q\bar{\hat{\psi}}\right)(y)\,,\qquad \mathcal Q\hat{\psi}=0\,,
\end{equation}
where $\hat{\psi}$ is a fermionic defect primary with $\Delta_{\hat{\psi}}=-r_{\hat{\psi}}=\frac12$ for line defects in $d=3$, and with weight $\frac12$ under the $su(1,1|1)$ factor for surface defects in $d=4$. Superconformal representation theory implies $\mathcal Q\hat{\psi}=0$, hence the tilt is part of a short multiplet. The broken $R$-rotation therefore acts on the defect as the F-term of $\hat{\psi}$ with a real coefficient. The same F-term with an imaginary coefficient,
\begin{equation}\label{eq_tilt_partner}
    \mathcal{T}'=i\left(\overline{\mathcal Q}\hat{\psi}-\mathcal Q\bar{\hat{\psi}}\right),
\end{equation}
is a second supersymmetric marginal operator, independent of $\mathcal{T}$ and with the same two-point function. 

Unlike the tilt, $\mathcal{T}'$ is not associated with a symmetry; rather, deforming the defect by it shifts the absolute value $|\lambda|$. It is nevertheless exactly marginal to all orders in conformal perturbation theory. Briefly, this is because for $\mathcal{T}'$ to cease to be marginal the short multiplet $\hat{\psi}$ should recombine with another short multiplet into a long one, but this would also give an anomalous dimension to the tilt operator. The tilt, however, is always exactly marginal due to the broken internal symmetry. We provide more details on this argument in appendix~\ref{app:susy_tilt}.

Note that the above construction assumes that the broken $R$-symmetry maps the preserved supercharges to other preserved ones, $[\mathcal Q,R]\propto\mathcal Q$. In particular, it does not apply to the $1/2$-BPS Maldacena--Wilson line in $\mathcal{N}=4$ SYM, or to similar defects such as the $1/2$-BPS surface operator in the $6d$ $\mathcal{N}=(2,0)$ theory, which also break the R-symmetry and whose tilts were studied in \cite{Drukker:2022pxk}. There every broken $R$-symmetry generator maps the preserved supercharges to broken ones, and hence deforming the defect with the tilt yields a defect preserving a different set of supercharges; as a result the integrated tilt operators are not supersymmetric, and there is no marginal partner of the form~\eqref{eq_tilt_partner}.

\subsection{Gukov-Witten defects in \texorpdfstring{$4d$}{4d}}\label{sec_GW}

A standard example of a conformal defect with noncompact marginal parameters is provided by the half-BPS Gukov--Witten surface operators in four-dimensional $\mathcal{N}=4$ SYM~\cite{Gukov:2006jk,Gukov:2014gja}. We review its definition below following \cite{Choi:2024ktc}. For gauge group $G=U(N)$, one first chooses a partition
\begin{equation}
N=\sum_{I=1}^{M}N_I
\end{equation}
which specifies the Levi subgroup
\begin{equation}
L=\prod_{I=1}^{M}U(N_I)\subset U(N)\,.
\end{equation}
Consider a planar defect supported at $z=0$. The defect is defined by imposing the singular behavior
\begin{equation}
A=
\operatorname{diag}\left(
\alpha_1\mathds{1}_{N_1},
\ldots,
\alpha_M\mathds{1}_{N_M}
\right)d\theta
\end{equation}
for the gauge field and
\begin{equation}
\Phi=
\frac{1}{\sqrt{2}z}
\operatorname{diag}\left(
(\beta_1+i\gamma_1)\mathds{1}_{N_1},
\ldots,
(\beta_M+i\gamma_M)\mathds{1}_{N_M}
\right)
\end{equation}
for the complex scalar $\Phi=(\phi^5+i\phi^6)/\sqrt{2}$. These singularities preserve the Levi subgroup $L$ as the defect is approached. For each $U(N_I)$ factor one can also introduce a two-dimensional theta angle $\eta_I$. The surface operator is therefore characterized by the Levi subgroup $L$ together with $4M$ continuous parameters
\begin{equation}
(\alpha_I,\beta_I,\gamma_I,\eta_I)
\qquad
I=1,\ldots,M\,.
\end{equation}
The parameters $\alpha_I$ and $\eta_I$ are circle valued,
\begin{equation}
\alpha_I\simeq\alpha_I+1\,,
\qquad
\eta_I\simeq\eta_I+1\,,
\end{equation}
whereas $\beta_I$ and $\gamma_I$ are noncompact real parameters. 

The defect preserves superconformal symmetry $PSU(1,1|2)\times PSU(1,1|2)\times U(1)$. All the parameters $(\alpha_I,\beta_I,\gamma_I,\eta_I)$ are couplings for exactly marginal defect operators. Denoting by $F^I$ and $\Phi^I$ the $U(N_I)$ blocks of the bulk fields, these operators are
\begin{equation}
\begin{array}{ccl}
\alpha_I
&:&
\operatorname{Tr}F^I_{z\bar z}
\\[2mm]
\beta_I
&:&
\operatorname{Tr}\left(D_z\bar\Phi^I+D_{\bar z}\Phi^I\right)
\\[2mm]
\gamma_I
&:&
i\operatorname{Tr}\left(D_z\bar\Phi^I-D_{\bar z}\Phi^I\right)
\\[2mm]
\eta_I
&:&
\operatorname{Tr}F^I_{w\bar w}\,,
\end{array}
\end{equation}
where $w$ is a holomorphic coordinate along the defect. These dimension two operators are the top components of short multiplets whose superconformal primaries have weights $(h,\bar h)=(1/2,1/2)$ \cite{Gomis:2016sab,Choi:2024ktc}. Up to normalization, the combinations of these operators that rotate and rescale all the $\beta_I+i\gamma_I$ by a common factor are the tilt of the $U(1)\subset SO(6)_R$ rotating $\Phi$ and its partner \eqref{eq_tilt_partner} (see section~\ref{subsec_BPS}).

When the scalar profile is large and sufficiently generic within the chosen Levi subgroup, the charged $W$ bosons become massive and the defect can be described using the low-energy theory on the Coulomb branch. The Abelian sector consists of $M$ $\N=4$ vector multiplets associated with the $U(1)^M$ center of the Levi subgroup. Each multiplet contains a $U(1)$ gauge field, six real scalars, and their fermionic superpartners. The expectation values of these scalars provide local coordinates on the Abelian part of the Coulomb branch. Its metric is flat and receives no quantum corrections, although its normalization depends on the choice of generators and, in particular, on the multiplicities $N_I$. On the defect background described above, the scalar components of the vector multiplets acquire expectation values $\Phi^I\propto\beta_I+i\gamma_I$ on $\mathrm{AdS}_3\times S^1$.

When $M<N$, the low-energy theory also contains interacting non-Abelian $\mathcal{N}=4$ sectors with gauge group $\prod_{I=1}^M SU(N_I)$. At leading order in the derivative expansion, these sectors decouple from the Abelian multiplets and will not play a role in what follows.

The preserved supersymmetry of the defect allows several defect observables to be computed using localization. Below we check the EFT predictions against these exact results.

An observable of interest is the coefficient of the defect Weyl anomaly $b$. As reviewed in~\eqref{eq_g_b_def}, this can be extracted from the logarithmically divergent contribution to the defect partition function. Since all the defect couplings are exactly marginal, $b$ only depends on the choice of the Levi subgroup $L$. We may therefore perform the computation at generic values of $\{\beta_I,\gamma_I\}$, where the low-energy description is the Coulomb-branch EFT discussed above.

Since the defect is $1/2$ BPS, $b$ can only receive contributions from protected $F$-terms in the EFT. The leading two-derivative action is one such term, but both its classical and quantum contributions vanish. Indeed, in an Abelian gauge theory the Gukov--Witten defect has $b=0$, and therefore the contribution from the $U(1)^M$ vector multiplets vanishes. Additionally, as mentioned above, for a sufficiently generic scalar expectation value the non-Abelian sector completely decouples and is therefore insensitive to the defect background.

Similarly to the large-charge expansion in $\mathcal{N}\geq2$ SCFTs, the remaining contribution arises from the supersymmetric completion of the Wess--Zumino term associated with the four-dimensional Weyl anomaly~\cite{Hellerman:2017sur,Hellerman:2018xpi}. To write this term, we introduce a dimensionless dilaton field $\tau$ and an $R$-symmetry Goldstone $\chi$. Locally on the Coulomb branch, these can be defined in terms of any nonvanishing homogeneous function of the bottom components of the Abelian vector multiplets. For instance, in a patch where $\Phi^1\neq0$, we may write
\begin{equation}
\frac{\Phi^1}{\mu}
f\left(\frac{\Phi^I}{\Phi^1}\right)
=
e^{-\tau-i\chi}\,,
\end{equation}
where $\mu$ is an arbitrary reference scale. On a conformally flat Euclidean background, the bosonic part of the supersymmetric Wess--Zumino action contains the term
\begin{equation}\label{eq_susy_WZ}
S_{\rm WZ}^{E}
=
\frac{\Delta a}{16\pi^2}
\int d^4x\sqrt{g}\,\tau E_4+\ldots\,.
\end{equation}
We provide the complete expression in app.~\ref{app_GW_WZ}. It is important that supersymmetry also fixes the dependence of the Wess-Zumino action on the $R$-symmetry Goldstone $\chi$. In~\eqref{eq_susy_WZ} $\Delta a=a_{\rm UV}-a_{\rm IR}$ is the mismatch of the four-dimensional Euler anomalies between the microscopic theory and the massless theory on the Coulomb branch.  For the breaking $U(N)\rightarrow L=\prod_IU(N_I)$, this is
\begin{equation}\label{eq_Delta_a}
\Delta a=\frac{1}{4}
\left(N^2-\sum_I N_I^2\right)\,.
\end{equation}

In app.~\ref{app_GW_WZ} we compute the $b$ anomaly by evaluating the full Wess-Zumino term on the classical profile. Here we explain the structure of the result. Consider the theory on $S^4$ of radius $R$. The Gukov--Witten scalar profile implies
\begin{equation}
\tau
=
\log(\mu R)+\tau_{\rm prof}\,,
\end{equation}
where $\tau_{\rm prof}$ is independent of both $\mu$ and $R$. The constant part of $\tau$ enters the action only through the Euler term in~\eqref{eq_susy_WZ}, while the nontrivial profile contributes through the terms that we did not write explicitly. The profile-dependent contribution is UV divergent and requires introducing a short-distance cutoff $\epsilon$ around the defect. Using
\begin{equation}
\int_{S^4}d^4x\sqrt{g}\,E_4
=
64\pi^2\,,
\end{equation}
we can therefore write
\begin{equation}
S_{\rm WZ}^{E}
=
4\Delta a\log(\mu R)
+
S_{\rm prof}\left(\frac{R}{\epsilon}\right)\,.
\end{equation}

The explicit evaluation in app.~\ref{app_GW_WZ} gives
\begin{equation}
S_{\rm prof}\left(\frac{R}{\epsilon}\right)
=
-4\Delta a\log\left(\frac{R}{\epsilon}\right)
+\ldots\,,
\end{equation}
so that
\begin{equation}
S_{\rm WZ}^{E}
=
4\Delta a\log(\mu\epsilon)
+\ldots\,,
\end{equation}
where the dots stand for finite contributions which are independent of the defect cutoff $\epsilon$. Thus, the coefficient of the UV divergent logarithm is fixed by the bulk anomaly mismatch between the UV theory and the Coulomb branch.\footnote{The dependence on the macroscopic scale $R$ of the normalized defect partition function arises instead from the Weyl anomaly of the massless IR theory relative to that of the UV theory. In the present case $\log Z_L-\log Z_{U(N)}=4\Delta a\log(\mu R)+\ldots$. Combining this with $\log Z_{\rm WZ}=-S_{\rm WZ}^{E}=-4\Delta a\log(\mu\epsilon)+\ldots$ gives $\log\langle\mathcal{D}\rangle=4\Delta a\log(R/\epsilon)+\ldots$.} We conclude that the universal contribution to the normalized defect expectation value is
\begin{equation}\label{eq_app_b_sphere}
\log\langle\mathcal{D}\rangle_{\rm WZ}
=
4\Delta a
\log\left(\frac{R}{\epsilon}\right)
+
O(1)\,,
\end{equation}
from which we obtain
\begin{equation}\label{eq_GW_b_result}
b
=
12\Delta a
=
3\left(
N^2-\sum_{I=1}^{M}N_I^2
\right)\,.
\end{equation}
This agrees with the known exact result \cite{Jensen:2018rxu,Chalabi:2020iie,Wang:2020xkc,Choi:2024ktc}.

Another interesting class of observables consists of the one-point functions of protected operators, which were studied in~\cite{Choi:2024ktc}. These are the chiral primaries
\begin{equation}
\mathcal{O}_{\Delta,k}=\frac{1}{\sqrt{\Delta}}\left(\frac{8\pi^2}{g^2N}\right)^{\frac{\Delta}{2}}C^{\Delta,k}_{i_1\ldots i_\Delta}\operatorname{Tr}\left(\phi^{i_1}\cdots\phi^{i_\Delta}\right),
\end{equation}
where $g$ is the gauge coupling, normalized so that the scalar kinetic term is $\frac{1}{g^2}\operatorname{Tr}(\partial\phi^i)^2$, and $C^{\Delta,k}$ is a unit-normalized symmetric traceless tensor that is invariant under the $SO(4)$ rotating $\phi^1,\ldots,\phi^4$ and has charge $k$ under the $U(1)$ rotating $\Phi$. Conformal symmetry fixes $\langle\mathcal{O}_{\Delta,k}\rangle=C_{\Delta,k}\,z^{-(\Delta+k)/2}\bar z^{-(\Delta-k)/2}$. 

Since the K\"ahler potential is exact at tree-level in $\mathcal{N}=4$ SYM, at leading order in the EFT the one-point function is the value of the operator on the classical profile, which gives
\begin{equation}\label{eq_GW_1pt_tree}
C_{\Delta,k}=\left(\frac{4\pi^2}{g^2N}\right)^{\frac{\Delta}{2}}\sqrt{\frac{(\Delta+k+2)(\Delta-k+2)}{4\Delta(\Delta+1)}}\,\sum_{I=1}^M N_I\left(\beta_I+i\gamma_I\right)^{\frac{\Delta+k}{2}}\left(\beta_I-i\gamma_I\right)^{\frac{\Delta-k}{2}}+\ldots\,.
\end{equation}
This reproduces the leading term of the exact results of~\cite{Drukker:2008wr,Choi:2024ktc} for all $\Delta$, $k$ and Levi group $L$. For $k=\pm\Delta$ it is exact, while for $|k|<\Delta$ the exact result is a polynomial in $g^2$ of degree $(\Delta-|k|)/2$, after factoring out the prefactor $g^{-\Delta}$~\cite{Choi:2024ktc}. The corrections are suppressed by $g^2N/(\beta_I^2+\gamma_I^2)$ and arise from loops of the massive $W$ bosons in the full theory. In the EFT they are therefore associated with higher derivative corrections, as we show explicitly below in a concrete example.

Let us consider the first correction to $C_{2,0}$. The operator $\mathcal{O}_{2,0}$ is the superconformal primary of the stress tensor multiplet, and supersymmetry relates its one-point function to that of the stress tensor~\cite{Drukker:2008wr,Choi:2024ktc}:
\begin{equation}\label{eq_GW_T}
\langle T_{\mu\nu}\rangle=-\frac{d_2}{36\pi^2}\frac{\delta_{\mu\nu}}{|z|^4}\,,\qquad
\langle T_{ij}\rangle=-\frac{d_2}{36\pi^2}\frac{4n_in_j-3\delta_{ij}}{|z|^4}\,,\qquad
d_2=3\sqrt{6}\,N\,C_{2,0}\,,
\end{equation}
where $\mu,\nu$ are directions along the defect, $i,j$ are transverse directions and $n_i$ is the unit normal. Here $d_2$ is a defect Weyl anomaly coefficient in the normalization of~\cite{Jensen:2018rxu}, where it was shown that $d_2\geq0$ if the averaged null energy condition holds. The two-derivative action of the Abelian multiplets gives the classical contribution $d_2=\frac{24\pi^2}{g^2}\sum_IN_I(\beta_I^2+\gamma_I^2)$. The leading correction arises from the Wess--Zumino term~\eqref{eq_susy_WZ}. In app.~\ref{app_GW_WZ} we evaluate its stress tensor on the Gukov--Witten profile and find a contribution $12\Delta a$ to $d_2$, independent of $\beta_I$ and $\gamma_I$. Using~\eqref{eq_GW_b_result} we obtain
\begin{equation}\label{eq_GW_d2}
\begin{split}
d_2&=\frac{24\pi^2}{g^2}\sum_{I=1}^MN_I\left(\beta_I^2+\gamma_I^2\right)+b\,,\\
C_{2,0}&=\frac{8\pi^2}{\sqrt{6}\,g^2N}\sum_{I=1}^MN_I\left(\beta_I^2+\gamma_I^2+\frac{g^2}{8\pi^2}\left(N-N_I\right)\right).
\end{split}
\end{equation}
Since the perturbative expansion for $C_{2,0}$ terminates after the first correction relative to the leading term~\cite{Choi:2024ktc}, these two contributions give the complete result. It agrees with the supergravity calculation of~\cite{Drukker:2008wr} and with the value of $d_2$ quoted in~\cite{Jensen:2018rxu}.\footnote{The correction to $C_{2,0}$ appears in~\cite{Choi:2024ktc} with the opposite sign. That sign would give $d_2=\frac{24\pi^2}{g^2}\sum_IN_I(\beta_I^2+\gamma_I^2)-b$, which is negative for small $\beta_I$ and $\gamma_I$, in contradiction with $d_2\geq0$.} Thus the Wess--Zumino term determines both $b$ and the quantum part of $d_2$, and $d_2-b$ is classical.

\subsection{Vortex loops in ABJM theory}\label{sec_ABJM_vortex}

Another important example of the vortex defects discussed above is provided by the BPS vortex loops of ABJM theory~\cite{Drukker:2008jm}. Consider the three-dimensional $\N=6$ theory with gauge group
\begin{equation}
U(N)_k\times U(N)_{-k}\,,
\end{equation}
and denote the four complex bifundamental scalars by $C^I$, $I=1,\ldots,4$. The moduli space is~\cite{Aharony:2008ug}
\begin{equation}
\mathcal M_N
=
\operatorname{Sym}^N\left(\mathbb C^4/\mathbb Z_k\right)\,.
\end{equation}
Vortex loops are defined by prescribing singular boundary conditions for the scalar and gauge fields along a line. We focus below on the $1/2$ BPS family. The discussion differs for even and odd $k$.

\paragraph{Vortex defects for even $k$} Let us begin by discussing the defect for even $k$. Let
\begin{equation}
A^\pm=A\pm\widehat A
\end{equation}
denote the sum and difference of the two gauge fields. For even $k$, the general $1/2$ BPS vortex is specified by a partition
\begin{equation}\label{eq_ABJM_partition}
N=N_0+\sum_{l=1}^{M}N_l\,.
\end{equation}
We consider the scalar singularity
\begin{equation}\label{eq_ABJM_vortex_profile}
C^1
=
\frac{1}{\sqrt z}
\operatorname{diag}\left(
0_{N_0},
\beta^{(1)}\mathds{1}_{N_1},
\ldots,
\beta^{(M)}\mathds{1}_{N_M}
\right)\,,
\qquad
C^{2,3,4}=0\,,
\end{equation}
where the $M$ parameters $\beta^{(l)}$ can be taken to be real and positive using the $U(1)^M$ gauge transformations acting on the nontrivial blocks. The gauge field satisfies $A^-=0$, while $A^+$ carries an independent flat holonomy around the defect,
\begin{equation}\label{eq_Aplus_ABJM}
A^+_z
=
-\frac{i}{2kz}
\operatorname{diag}\left(
0_{N_0},
\alpha^{(1)}\mathds{1}_{N_1},
\ldots,
\alpha^{(M)}\mathds{1}_{N_M}
\right)\,,
\end{equation}
parametrized by periodic variables $\alpha^{(l)}\sim\alpha^{(l)}+1$. The defect preserves twelve supercharges and the bosonic symmetry
\begin{equation}
SU(1,1)\times SU(3)\times U(1)_d\,,
\end{equation}
where $U(1)_d$ is a diagonal combination of transverse rotations and an $R$-symmetry transformation. Since the ABJM scalars have dimension $1/2$, the scalar singularity in \eqref{eq_ABJM_vortex_profile} is precisely of the form \eqref{eq:gen_bg_bps}.

In the gauge \eqref{eq_ABJM_vortex_profile}, the matter fields associated with each nontrivial block obey antiperiodic boundary conditions around the defect, while those in the trivial $N_0$ block remain periodic. For even $k$, the sign change belongs to the gauged $\mathbb Z_k$ and can be traded, by a singular block-diagonal gauge transformation acting nontrivially only on the nonzero blocks, for a nontrivial flat $A^-$ holonomy. Thus, although the elementary matter fields are not single-valued in this gauge, gauge-invariant local operators retain trivial monodromy around the defect \cite{Drukker:2008jm,Aharony:2008ug}. The vortex therefore defines a genuine defect. We will review below how this construction must be modified for odd $k$.

As an application, we discuss the calculation of the partition function $g$ of the above defect when all the nontrivial blocks in~\eqref{eq_ABJM_vortex_profile} and~\eqref{eq_Aplus_ABJM} are one-dimensional, i.e. $N_{i\geq 1}=1$. The parameters $\{\alpha^{(i)},\beta^{(i)}\}$ parametrize a continuous family of $1/2$ BPS conformal defects \cite{Drukker:2008jm}, and are therefore expected to be exactly marginal defect parameters. We consequently expect the $g$-function to be independent of them. Similarly to the Gukov--Witten example, we may therefore compute it in the regime where all the $\beta^{(i)}$ are large and mutually distinct, where the scalar profile allows studying the defects via the moduli space EFT. 

At a generic point of the moduli space, the off-diagonal fields connecting different eigenvalues acquire masses set by the separation of the corresponding scalar expectation values, $m_{ij}^2\propto (|\beta^{(i)}|^2-|\beta^{(j)}|^2)^2$ \cite{Berenstein:2008dc}, and can therefore be integrated out. The unbroken gauge group is
\begin{equation}
U(N-M)_k\times U(N-M)_{-k}\times U(1)^M\,,
\end{equation}
where each $U(1)$ factor is the diagonal subgroup associated with one of the nonzero one-dimensional blocks. Indeed, in each such block $A^-$ becomes massive, and integrating it out generates a Maxwell kinetic term for $A^+$. Thus the low-energy theory associated with every nonzero eigenvalue is an Abelian $\mathcal N=8$ vector multiplet \cite{Mukhi:2008ux,Pang:2008hw}. After dualizing the Abelian gauge field, this is equivalently described by a free $\mathcal N=8$ multiplet containing eight real scalars and eight real fermions. The $U(N-M)_k\times U(N-M)_{-k}$ sector associated with the vanishing block remains at the origin of the moduli space and decouples from the Abelian sectors when the scalar expectation values $\beta^{(i)}$ are parametrically separated from each other.

We may therefore describe the defect in terms of the low-energy EFT, which at leading order in derivatives takes the form
\begin{equation}
\mL_{\rm EFT}
=
\mL_{\mathrm{ABJM}(N-M,k)}
+
\sum_{l=1}^M\sum_{I=1}^4
\left|\pd C^I_{(l)}\right|^2
+
\text{fermions}
+\ldots\,,
\end{equation}
where the ellipsis denotes higher-derivative interactions suppressed at large $\beta^{(l)}$. The fields $C^I_{(l)}$ parametrize the local flat cover of the corresponding $\mathbb C^4/\mathbb Z_k$ factor of the moduli space; the $\mathbb Z_k$ quotient is therefore implicit in this description. The vortex selects the sector in which the moduli space fields obey boundary conditions
\begin{equation}
C^I_{(l)}(\theta+2\pi)
=
-C^I_{(l)}(\theta)\,.
\end{equation}
The parameters $\beta^{(l)}$ determine the classical scalar profiles,
\begin{equation}
C^1_{(l)}
=
\frac{\beta^{(l)}}{\sqrt z}\,,
\qquad
C^{2,3,4}_{(l)}=0\,,
\end{equation} 
and are associated with exactly marginal deformations \cite{Giombi:2021uae}. The holonomies $\alpha^{(l)}$ are instead not visible in the two-derivative local scalar Lagrangian obtained after dualizing the photon. 

It is now simple to compute the $g$-function. Recall that we assume that the result is independent of the continuous parameters $\{\alpha^{(i)},\beta^{(i)}\}$. The classical EFT action yields no contribution of order $(\beta^{(i)})^0$ (see the discussion around~\eqref{eq_f-EFT_prediction}); therefore the constant term is entirely determined by the one-loop EFT partition function, normalized by the full theory partition function in the conformal vacuum. Using that $AdS_2\times S^1$ is Weyl equivalent to $S^3$, we arrive at the following result
\begin{equation}
\log g
=
F_N(k)-F_{N-M}(k)
-M\left(F_{\rm free}-\log g_{\rm free}\right)\,.
\end{equation}
Here
\begin{equation}
F_N(k)
=
-\log Z_{S^3}^{\operatorname{ABJM}(N,k)}
\end{equation}
denotes the ABJM sphere free energy, which is known exactly from localization \cite{Drukker:2010nc,Fuji:2011km,Marino:2011eh}. The sphere free energy of a single free $\mathcal N=8$ multiplet is
\begin{equation}\label{eq_ABJM_Ffree}
F_{\rm free}
=
4\left(
F_{\rm scalar}+F_{\rm Dirac}
\right)
=
\log 4\,,
\end{equation}
where \cite{Klebanov:2011gs}
\begin{equation}
F_{\rm scalar}
=
\frac{\log 2}{4}
-
\frac{3\zeta(3)}{8\pi^2}\,,
\qquad
F_{\rm Dirac}
=
\frac{\log 2}{4}
+
\frac{3\zeta(3)}{8\pi^2}\,.
\end{equation}
The required defect partition functions are those of the standard $\mathbb Z_2$ monodromy defects of a free complex scalar and a free Dirac fermion \cite{Giombi:2021uae,Barkeshli:2025cjs}. One finds
\begin{equation}\label{eq_free_g}
\log g_{\rm{scalar}}
=
\frac{\log 2}{4}
-
\frac{7\zeta(3)}{8\pi^2}\,,
\qquad
\log g_{\rm Dirac}
=
\frac{\log 2}{4}
+
\frac{7\zeta(3)}{8\pi^2}\,.
\end{equation}
Consequently, for the full free $\mathcal N=8$ multiplet,
\begin{equation}\label{eq_ABJM_gfree}
\log g_{\rm free}
=
4\left(
\log g_{\rm scalar}
+
\log g_{\rm Dirac}
\right)
=
\log 4
=
F_{\rm free}\,.
\end{equation}
The contribution of each free Abelian sector therefore cancels between the sphere free energy and its defect $g$-function, and the final result is simply
\begin{equation}\label{eq_ABJM_g_even}
\log g
=
F_N(k)-F_{N-M}(k)\,.
\end{equation}

It is instructive to compare this result with holography. At large $N$ and fixed $k$, the ABJM sphere free energy behaves as \cite{Drukker:2010nc,Marino:2011eh}
\begin{equation}\label{eq_large_N_F_ABJM}
F_N(k)
=
\frac{\pi\sqrt{2k}}{3}N^{3/2}
+
O(N^{1/2})\,.
\end{equation}
For fixed $M$, our result therefore gives
\begin{equation}
\log g
=
M\pi\sqrt{\frac{kN}{2}}
+
O(N^{-1/2})\,.
\end{equation}
This agrees precisely with the semiclassical holographic result for even $k$ and $M=1$ obtained in \cite{Drukker:2008jm,Drukker:2023bip}. In particular, the absence of an $O(N^0)$ term in our result reproduces the vanishing one-loop correction found holographically.

Finally, note that when the blocks in the partition \eqref{eq_ABJM_partition} have dimension larger than one, the low-energy theory has gauge group
\begin{equation}
U(N_0)_k\times U(N_0)_{-k}\times \prod_{l=1}^M U(N_l)\,.
\end{equation}
Each $U(N_l)$ factor describes maximally supersymmetric Yang--Mills theory, with gauge coupling scaling as $
g_{{\rm YM},l}^2\sim \frac{|\beta^{(l)}|^2}{k^2}$ \cite{Mukhi:2008ux,Pang:2008hw}. The $SU(N_l)$ sector becomes strongly coupled at energies well below $g_{{\rm YM},l}^2$ and flows to an interacting three-dimensional $\mathcal N=8$ superconformal field theory, while the center-of-mass $U(1)$ sector remains free. It would be interesting to study the $g$-function and other defect observables using this interacting low-energy theory, possibly via supersymmetric localization.

\paragraph{Vortex defects for odd $k$.}
For odd $k$, the gauged $\mathbb Z_k$ does not contain the $-1$ element. Consequently, the construction described above induces a nontrivial monodromy on gauge-invariant local operators. One can define a genuine defect operator instead by requiring all the nontrivial blocks in the partition \eqref{eq_ABJM_partition} to have even dimension,
\begin{equation}
N_l\in 2\mathbb Z\,,
\qquad
l=1,\ldots,M\,,
\end{equation}
and supplementing the vortex with a non-Abelian gauge twist \cite{Drukker:2008jm}. The scalar singularity is taken to be
\begin{equation}\label{eq_ABJM_odd_k_scalar_profile}
C^1
=
\frac{1}{\sqrt z}
\operatorname{diag}\left(
0_{N_0},
\beta^{(1)}\sigma_3\otimes\mathds{1}_{N_1/2},
\ldots,
\beta^{(M)}\sigma_3\otimes\mathds{1}_{N_M/2}
\right)\,,
\qquad
C^{2,3,4}=0\,.
\end{equation}
In addition, the holonomy of $A^+$ around the defect contains a non-Abelian component,
\begin{equation}
\operatorname{Hol}(A^+)
=
\operatorname{diag}\left(
\mathds{1}_{N_0},
e^{i\pi\alpha^{(1)}/k}\sigma_1\otimes\mathds{1}_{N_1/2},
\ldots,
e^{i\pi\alpha^{(M)}/k}\sigma_1\otimes\mathds{1}_{N_M/2}
\right)\,.
\end{equation}
The holonomy specifies how the fields are glued after one circuit around the defect. The scalar must therefore obey
\begin{equation}\label{eq_ABJM_odd_k_holonomy_on_C}
C^1(\theta+2\pi)
=
g\,C^1(\theta)\,g^{-1}\,,
\qquad
g=\sigma_1\,,
\end{equation}
where the overall phase $e^{i\pi\alpha^{(l)}/k}$ in each nontrivial block in the holonomy acts trivially by conjugation. This condition is easily checked to be satisfied by the scalar profile~\eqref{eq_ABJM_odd_k_scalar_profile} using $\sigma_1\sigma_3\sigma_1=-\sigma_3$. This configuration is globally well defined as a gauge-theory background, and gauge-invariant local operators are single-valued \cite{Drukker:2008jm}.

Similarly to even $k$, we can compute the $g$-function of this defect when all the nontrivial blocks are two-dimensional using the moduli-space EFT. We assume again that the continuous parameters $\{\alpha^{(l)},\beta^{(l)}\}$ are exactly marginal and take the different blocks to be widely separated on the moduli space. The off-diagonal modes connecting distinct blocks are then gapped as before, and the light modes associated with each block are semiclassically described by a $U(2)$ gauge theory with coupling $g_{\mathrm{YM},l}^2\sim |\beta^{(l)}|^2/k^2$.

The semiclassical analysis of the perturbative fields, however, does not completely specify the IR configuration within each $U(2)$ block. To see this, recall that on the Coulomb branch of maximally supersymmetric $SU(2)$ SYM the gauge group is broken to $U(1)$, and the remaining photon can be dualized to a compact scalar $\sigma\sim\sigma+2\pi$. Together with the seven adjoint scalars $\vec X$, this gives the moduli space \cite{Seiberg:1997ax}
\begin{equation}
    \frac{\mathbb R^7\times S^1_{\sigma}}{\mathbb Z_2}\,,
\end{equation}
where the Weyl group identifies $(\vec X,\sigma)\sim(-\vec X,-\sigma)$. Nonperturbatively, $\sigma$ is measured by the phase of monopole operators and is meaningful also at $\vec X=0$. The point $(\vec X=0,\sigma=0)$ is the strongly coupled origin at which SYM flows to the interacting $\mathcal N=8$ SCFT, while away from this point the low-energy dynamics is described by a free $\mathcal N=8$ superconformal multiplet. This remains true also at the orbifold singular point $(\vec X=0,\sigma=\pi)$.

The vortex profile~\eqref{eq_ABJM_odd_k_scalar_profile} places the seven scalars charged under $SU(2)$ at the origin $\vec{X}=0$. To determine the EFT we therefore need to compute the value of the modulus $\sigma_{\mathrm{SU}(2)}$ associated with the vortex-loop configuration. We presently do not have a direct field-theoretical argument fixing this datum. It is, however, particularly transparent in the holographic description. The dual photon $\sigma_{\mathrm{SU}(2)}$ has a geometric interpretation as the relative position of the two constituent branes along the M-theory circle \cite{Seiberg:1997ax}. For the vortex loop, each two-dimensional block is described holographically by a doubly wrapped M2-brane \cite{Drukker:2008jm}. Denoting by $w^I$, $I=1,\ldots,4$, the complex transverse coordinates on $\mathbb C^4/\mathbb Z_k$, the M2 embedding is mapped after one turn around the vortex as $w^I\to-w^I$. For odd $k$ this transformation is not contained in the $\mathbb Z_k$ orbifold, so the brane closes only after a second turn. The two sheets at fixed position along the vortex are therefore related by the antipodal map on the covering M-theory circle and are separated by half of its period. In the normalization $\sigma\sim\sigma+2\pi$, this implies
\begin{equation}
    \sigma_{\mathrm{SU}(2)}=\pi\,.
\end{equation}
Thus each two-dimensional block sits at the second Weyl-invariant point $(\vec X=0,\sigma=\pi)$ of the semiclassical $SU(2)$ moduli space, rather than at the interacting point $(\vec X=0,\sigma=0)$.

Similarly to even $k$, we may therefore compute the $g$-function using a fully Abelian EFT consisting of $2M$ free $\mathcal N=8$ vector multiplets and a decoupled ABJM sector with gauge group $U(N-2M)_k\times U(N-2M)_{-k}$. The crucial difference is that the two multiplets descending from each $U(2)$ block are not separately periodic around the defect, but are exchanged by the Weyl monodromy.\footnote{This follows from~\eqref{eq_ABJM_odd_k_holonomy_on_C} and identifying the $U(2)$ $\mathcal{N}=8$ SYM moduli space coordinates with the diagonal entries of $C$ in each block.} Denoting them by $C_1^{(l)}$ and $C_2^{(l)}$, and suppressing the internal $SU(4)_R$ index, we have
\begin{equation}
    C_1^{(l)}(\theta+2\pi)
    =
    C_2^{(l)}(\theta)\,,
    \qquad
    C_2^{(l)}(\theta+2\pi)
    =
    C_1^{(l)}(\theta)\,.
\end{equation}
It is convenient to diagonalize this action by introducing
\begin{equation}
    C_{\pm}^{(l)}
    =
    \frac{C_1^{(l)}\pm C_2^{(l)}}{\sqrt{2}}\,.
\end{equation}
These combinations obey
\begin{equation}
    C_+^{(l)}(\theta+2\pi)
    =
    C_+^{(l)}(\theta)\,,
    \qquad
    C_-^{(l)}(\theta+2\pi)
    =
    -C_-^{(l)}(\theta)\,.
\end{equation}
Thus every two-dimensional block gives one untwisted free multiplet and one free multiplet with monodromy. The former only contributes with its untwisted sphere partition function, whereas the latter also gives the contribution in \eqref{eq_ABJM_gfree}. Proceeding as before, we then obtain
\begin{equation}
\begin{aligned}
    \log g
    &=
    F_N(k)-F_{N-2M}(k)-2M F_{\rm free}
    +M\log g_{\rm free}
    \\
    &=
    F_N(k)-F_{N-2M}(k)-M\log 4\,.
\end{aligned}
\end{equation}

We may compare again with the holographic one-loop computation in \cite{Drukker:2023bip}. Using \eqref{eq_large_N_F_ABJM}, we find for fixed $k$ and $M$
\begin{equation}\label{eq_ABJM_g_largeN}
\log g
=
2M\pi\sqrt{\dfrac{kN}{2}}-M\log4+O(N^{-1/2})\,,
\end{equation}
which agrees with \cite{Drukker:2023bip} for $M=1$. In particular, for odd $k$ the doubly wrapped M2-brane contributes the universal one-loop term $-\log 4$, differently than for even $k$.

\subsection{\texorpdfstring{$1/2$}{1/2} BPS vortex loop in the \texorpdfstring{$3d$}{3d} \texorpdfstring{$\mathcal{N}=2$}{N=2} XYZ model}\label{subsec_XYZ}

\subsubsection{Non-perturbative results in \texorpdfstring{$d=3$}{d=3}}

As a final example, we consider a vortex defect in the $3d$ $\mathcal{N}=2$ SCFT arising as the IR limit of a theory of three chiral fields with superpotential
\begin{equation}
W=\frac{g}{2}XYZ\,.
\end{equation}
Each field has R-symmetry charge $2/3$. In addition, the model enjoys an internal symmetry group $U(1)^2\rtimes S_3$, where the $U(1)^2$ factors act by rotating the chiral fields with phases whose product is unity, while $S_3$ permutes $X$, $Y$, and $Z$. The BPS bound implies that the chiral fields above have scaling dimension equal to their R-charge
\begin{equation}
    \Delta=R=\frac{2}{3}\,.
\end{equation}
The model also admits an infinite chiral ring  \cite{Luty:1995sd}, which is spanned by the identity together with the operators $\{X^n,Y^n,Z^n\}_{n\geq 1}$, with scaling dimension $\Delta_n=\frac23 n$.

The $F$-term equations admit solutions in which one chiral field acquires a vev while the other two vanish. The theory therefore admits a moduli space given by the union of three complex branches meeting at the origin,
\begin{equation}
\mathcal{M}=\mathbb{C}_X\cup \mathbb{C}_Y\cup \mathbb{C}_Z\,.
\end{equation}
Supersymmetry and the internal symmetries ensure that this moduli space is not lifted at the quantum level \cite{Aharony:1997bx}. The theory also admits an exactly marginal deformation which, in the UV description, is associated with the superpotential coupling $W=h( X^3+Y^3+Z^3)$. The latter partially breaks the internal symmetry and lifts the moduli space \cite{Strassler:1998iz,Baggio:2017mas}. In what follows, we set $h=0$ and focus on the theory with the moduli space $\mathcal M$ above.

We now consider a defect obtained by imposing a unit R-symmetry monodromy, which is nontrivial since chiral ring operators have charges $\in \frac23\mathbb{Z}$, and expanding around the background
\begin{equation}\label{eq_X_defect}
    \langle X\rangle=\frac{\lambda}{z^{2/3}}\,.
\end{equation}
We also allow for an arbitrary monodromy $\nu\in (-1/2,1/2]$ for the $U(1)$ flavor symmetry rotating the chiral fields $Y$ and $Z$ as $(Y,Z)\rightarrow (e^{i\alpha} Y,e^{-i\alpha}Z)$.\footnote{One may also consider a twist under the $\mathbb{Z}_2$ exchanging $Y\leftrightarrow Z$.} We shall argue below that for arbitrary $\nu$ and for sufficiently large $\lambda$ the defect preserves an $su(1,1|1)$ superconformal subalgebra, where the vev of the chiral field is controlled by an exactly marginal complex defect deformation. The defect R-symmetry in particular is a linear combination of the bulk R-symmetry and transverse rotations,
\begin{equation}
    r=R-J\,.
\end{equation}
Below we describe the definition and properties of this defect using the moduli space EFT. We will then check our claims using a weakly coupled analytic continuation of the $XYZ$ model near four spacetime dimensions.

To argue that \eqref{eq_X_defect} indeed defines a $1/2$-BPS defect with a complex marginal parameter, we consider large $|\lambda|\gg 1$. Working as usual in the AdS$_2\times S^1$ Weyl frame, the large chiral field expectation value effectively moves the theory to the $X$-branch. The flavor charged fields $Y$ and $Z$ then acquire a large gap and can be integrated out. The low energy modes are described by the axio-dilaton $\Phi$ and its Dirac superpartner $\Psi$, whose action at leading order in derivatives reads: 
\begin{equation}\label{eq_EFT_XYZ}
    \mL=|\pd \Phi|^2+m_d^2|\Phi|^2+\bar{\Psi}\slashed{\nabla}\Psi+\ldots\,,
\end{equation}
where $R^2m_d^2=-\frac{(d-2)^2}{4}$
is the conformal mass on AdS$_2\times S^1$. The defect is obtained by expanding around the solution
\begin{equation}\label{eq_XYZ_dilaton}
    \Phi=v^{1/2}e^{-i\theta/2}
    \qquad\Longleftrightarrow\qquad
    \Phi\vert_{\rm flat}=\frac{(Rv)^{1/2}}{\sqrt{z}}\,,
\end{equation}
where the local coordinate $\Phi$ on the moduli space is related to the chiral operator by
\begin{equation}
    X\propto\Phi^{4/3}\,.
\end{equation}
It follows that $\lambda\propto(Rv)^{2/3}$.

Supersymmetry ensures that, to all orders in $1/v$, the EFT does not introduce a potential for the modulus $v$, which is hence associated with a protected complex marginal operator. This is the top component of a chiral primary $\hat{\psi}$ with
\begin{equation}
    \Delta_{\hat{\psi}}=-r_{\hat{\psi}}=\frac{1}{2}\,,
\end{equation}
which is the lightest operator appearing in the bulk-to-defect OPE of the Dirac field $\Psi$ at half-integer $U(1)$ twist \cite{Barkeshli:2025cjs}. This $\hat{\psi}$ is the defect fermion in the supersymmetric tilt equation \eqref{eq_susy_tilt}: the tilt rotates the phase of $\lambda$, while its partner \eqref{eq_tilt_partner} changes $v$, and is exactly marginal independently of the $1/v$ expansion.

As a check, we can consider the leading higher-derivative correction to the action \eqref{eq_EFT_XYZ}. In flat space, it takes the form \cite{Hellerman:2017veg}
\begin{equation}\label{eq_FTPR}
    \mL_{\pd^4}\vert_{\mathrm{flat}}
    =
    \frac{1}{\bar{\Phi}}\pd^2\pd^2\frac{1}{\Phi}\,.
\end{equation}
We refer to \cite{Hellerman:2017veg} for its curved-space completion. Since $\Phi$ is holomorphic in flat space, \eqref{eq_FTPR} vanishes when evaluated on the profile \eqref{eq_XYZ_dilaton}.\footnote{Alternatively, the vanishing of \eqref{eq_FTPR} follows from the fact that this term vanishes on the large-charge helical profile \cite{Hellerman:2017veg}, which is related to \eqref{eq_XYZ_dilaton} by analytic continuation, as explained around \eqref{eq_analytic_cont}.} The term \eqref{eq_FTPR} therefore does not generate a potential for $v$, which remains exactly marginal at this order in the derivative expansion.

We similarly expect nonperturbative corrections in $1/v$ not to generate a potential for $v$. Such effects arise from massive degrees of freedom propagating around the compact circle \cite{Grassi:2019txd,Hellerman:2021duh}. For supersymmetric boundary conditions, bosons and fermions contribute with opposite signs to the effective potential, and hence their contribution cancels when summed over.

Note that the flavor monodromy only affects the massive degrees of freedom and does not feature in the low energy description. This implies that the low lying defect spectrum and several correlation functions only depend on $\nu$ at nonperturbative order in $1/v$ (when they are not protected). 

We therefore found a three-parameter family of $1/2$ BPS defects labeled by
\begin{equation}
    (|\lambda|,\arg\lambda,\nu)\,,
\end{equation}
where the first two span a defect conformal manifold as described above. Strictly speaking, our arguments establish the existence of these defects only for sufficiently large $\lambda$, but in practice we expect this family to be smoothly connected to a family of monodromy defects with $\langle X\rangle=0$. We will provide evidence for this conjecture below.

A remarkable prediction of the EFT is that the defect free energy, and hence its $g$-function, is not only independent of $\lambda$, as expected from the exact marginality of the coupling, but it is also independent of the flavor monodromy! Indeed marginality of $\lambda$ implies that this is independent of $v$, and hence only receives contributions at one-loop within the EFT. To compute it, we normalize the free chiral field partition with a $1/2$ $U(1)$ twist by the full CFT result as in the previous section. Using the known value of $g$ for a free scalar and a free fermion at half integer $U(1)$ monodromy \cite{Giombi:2021uae,Barkeshli:2025cjs}, we find
\begin{equation}\label{eq_g_XYZ_3d}
    \log g=F_{XYZ}-F_{\rm scalar}-F_{\rm fermion}+\frac{\log 2}{2}=F_{XYZ}\,,
\end{equation}
where we used $F_{\rm scalar}+F_{\rm fermion}=\frac12\log 2$ \cite{Klebanov:2011gs}, and $F_{XYZ}$ is the $XYZ$-CFT sphere free energy. This is known from localization~\cite{Jafferis:2010un},
\begin{equation}
    F_{XYZ}
    = \frac12\log 3+\frac{3}{2\pi}\operatorname{Cl}_2\left(\frac{2\pi}{3}\right)
    \simeq 0.872372\,.
\end{equation}
The result \eqref{eq_g_XYZ_3d} holds for arbitrary $\lambda \in \mathbb{C}$ and $\nu\in(-1/2,1/2]$.  
We will check the independence of the defect free energy from the flavor monodromy in the $\eps$-expansion in the next section.

Evidence for the connection between our defect and known supersymmetric monodromy defects at $\lambda=0$ comes from localization. Supersymmetric monodromy defects were referred to as ``vortex loops'' in \cite{Drukker:2012sr}, and are defined by prescribing a background flavor holonomy around the line, but no nonzero scalar profile is imposed.
For each chiral multiplet there are two useful supersymmetric boundary prescriptions \cite{Hosomichi:2021gxe}: BC1 keeps the scalar regular while allowing a fermionic component to be mildly singular, whereas BC2 allows a mildly singular scalar and imposes regularity on the complementary fermionic component.\footnote{BC1 should not be identified with standard quantization for every field: the prescription for the fermionic fields depends on the holonomy. We will see explicitly in the $\varepsilon$-expansion in the next subsection that supersymmetry sometimes requires us to pick the alternate conditions for some fermions even when all the scalars obey standard boundary conditions.}

For the $XYZ$ model, we consider the flavor monodromies
\begin{gather}
    X(\theta+2\pi)=e^{2\pi i h_X}X(\theta)\,,\quad
    Y(\theta+2\pi)=e^{2\pi i h_Y}Y(\theta)\,,\quad
    Z(\theta+2\pi)=e^{2\pi i h_Z}Z(\theta)\,,
\end{gather}
where 
\begin{equation}
     h_X+h_Y+h_Z=0\,.
\end{equation}
Specializing the localization formulas of
\cite{Drukker:2012sr,Hosomichi:2021gxe} to the round sphere gives
\begin{equation}
    g_{\rm loc}
    =e^{F_{XYZ}}
    \left|
    \prod_{a=X,Y,Z}
    s_1\!\left(i\left[\frac13-h_a+k_a\right]\right)
    \right|\,,
    \qquad
    k_a=
    \begin{cases}
        \lfloor h_a\rfloor&\text{for }\mathrm{BC1}\,,\\
        \lceil h_a\rceil&\text{for }\mathrm{BC2}\,,
    \end{cases}
    \label{eq_XYZ_g_loc_general}
\end{equation}
where $s_1$ is the double-sine function and the normalization is relative to the sphere partition function without a defect.

We can reproduce the monodromies of our defect by choosing
\begin{equation}
    (h_X,h_Y,h_Z)
    =\left(\frac13,\frac13-\nu,\nu-\frac23\right)\,.
\end{equation}
We restrict to $0\leq\nu\leq 1/2$; negative $\nu$ follows by exchanging $Y$ and $Z$. Then a candidate for the $\lambda=0$ limit of our defect is obtained by imposing BC1 on $X$ and $Z$, and choosing instead 
\begin{equation}
    Y:\qquad
    \begin{cases}
        \mathrm{BC1}&\text{for }0\leq\nu\leq\frac13\,,\\
        \mathrm{BC2}&\text{for }\frac13<\nu\leq\frac12\,.
    \end{cases}
\end{equation}
For this prescription $(k_X,k_Y,k_Z)=(0,0,-1)$ throughout the interval, and the localization expression \eqref{eq_XYZ_g_loc_general} is continuous across the point where its
holonomy becomes trivial. Consequently,
\begin{equation}
    g_{\rm loc}
    =e^{F_{XYZ}}\left|s_1(0)s_1(i\nu)s_1(-i\nu)\right|
    =e^{F_{XYZ}}\,,
    \label{eq_XYZ_g_loc_matching}
\end{equation}
where we used $s_1(0)=1$ and $s_1(x)s_1(-x)=1$. The agreement of \eqref{eq_XYZ_g_loc_matching} with \eqref{eq_g_XYZ_3d}, including its independence
of $\nu$, supports the conjecture that these monodromy defects are connected to the large-$\lambda$ family described above.

\subsubsection{Perturbative analysis in \texorpdfstring{$4-\eps$}{4-epsilon} dimensions}

We now study this defect in the weakly coupled continuation of the XYZ model to $d=4-\eps$. Following~\cite{Sharon:2020mjs}, we continue each $3d$ chiral multiplet to a $4d$ chiral multiplet. The number of spinor components is therefore kept fixed in the
continuation; in three dimensions a four-dimensional Weyl spinor becomes a
complex two-component spinor.\footnote{Note that this analytic continuation breaks parity away from $d=3$. An alternative continuation preserving parity would consist in introducing $N_D$ $4d$ Dirac fields for each $3d$ fermion, and then continue $N_D\rightarrow 1/2$.} Writing the three chiral multiplets as $(x,\psi_x)$, $(y,\psi_y)$, and
$(z,\psi_z)$, the Euclidean flat space Lagrangian is
\begin{align}
\mL=&\sum_{\varphi=x,y,z}
\left(|\pd \varphi|^2+i\bar\psi_\varphi\bar\sigma^\mu
\pd_\mu\psi_\varphi\right)
+\frac{|g|^2\mu^\eps}{4}
\left(|xy|^2+|yz|^2+|zx|^2\right)
\nonumber\\
&-\frac{g\mu^{\eps/2}}{2}
\left(x\psi_y\psi_z+y\psi_z\psi_x+z\psi_x\psi_y\right)
-\frac{g^*\mu^{\eps/2}}{2}
\left(\bar x\bar\psi_y\bar\psi_z
+\bar y\bar\psi_z\bar\psi_x
+\bar z\bar\psi_x\bar\psi_y\right)\,,
\label{eq:XYZ_eps_action}
\end{align}
where $\mu$ is the sliding scale. The above theory admits a fixed point at~\cite{Ferreira:1996az}
\begin{equation}
 \frac{|g_*|^2}{(4\pi)^2}=\frac{2\eps}{3}+O(\eps^2)\,.
 \label{eq:XYZ_eps_fixed_point}
\end{equation}
By field redefinitions we can always make the cubic coupling real and positive. In what follows we set $\mu=1$ and take $g\in \mathbb{R}^+$ with no loss of generality.

We work in the Weyl frame $\mathrm{AdS}_{d-1}\times S^1$, where the conformal coupling gives every scalar a negative mass $m^2_d=-(d-2)^2/(4R^2)$. We allow an R-symmetry monodromy $\alpha$ and a flavor monodromy $\nu$ so that the scalar boundary conditions are
\begin{align}
 x(\theta+2\pi)&=e^{-\frac{4\pi i}{3}\alpha}x(\theta)\,,
 &y(\theta+2\pi)&=e^{-\frac{4\pi i}{3}\alpha-2\pi i\nu}y(\theta)\,,
 &z(\theta+2\pi)&=e^{-\frac{4\pi i}{3}\alpha+2\pi i\nu}z(\theta)\,.
 \label{eq:XYZ_eps_scalar_monodromies}
\end{align}
We additionally allow for a conical defect with parameter $\beta$, so that the Weyl fermions obey
\begin{align}
 \psi_x(\theta+2\pi)
 &=-e^{\frac{2\pi i}{3}\alpha}e^{-2\pi i\beta J}
 \psi_x(\theta)\,,\nonumber\\\nonumber
 \psi_y(\theta+2\pi)
 &=-e^{\frac{2\pi i}{3}\alpha-2\pi i\nu}e^{-2\pi i\beta J}
 \psi_y(\theta)\,,\\
 \psi_z(\theta+2\pi)
 &=-e^{\frac{2\pi i}{3}\alpha+2\pi i\nu}e^{-2\pi i\beta J}
 \psi_z(\theta)\,,
 \label{eq:XYZ_eps_fermion_monodromies}
\end{align}
where $J$ is the spin generator. As discussed in sec. \ref{subsec_BPS}, supersymmetry fixes the conical parameter in terms of the R-monodromy as in \eqref{eq_conical_susy}. We will soon specialize to this value.

To study the vortex defect, we expand around the configuration
\begin{equation}
    \langle x\rangle=v\,e^{-\frac{2i}{3}\alpha\theta}\,,
    \qquad
    \langle y\rangle=\langle z\rangle=0\,,
    \label{eq:XYZ_eps_profile}
\end{equation}
where, with no loss of generality, we take $v$ real and positive to avoid cluttering of notation. We treat \eqref{eq:XYZ_eps_profile} as an off-shell background and compute the defect free energy at one-loop. It is natural to work in the following scaling limit 
\begin{equation}\label{eq_XYZ_scaling_limit}
g^2\ll 1\,,\quad   v^2 R^{d-2}\gg 1\quad\text{with}\quad
g^2v^2 R^{d-2}=\text{fixed}\,,
\end{equation}
which is analogous to the large charge double-scaling limit \cite{Badel:2019oxl}, that was studied in~\cite{Sharon:2020mjs} for this model. In this limit the defect free energy takes the form
\begin{equation}\label{eq_free_energy_XYZ_expansion}
    F(v,\alpha,\beta,\nu)=R\operatorname{Vol}(\mathrm{AdS}_{d-1})\left[\frac{f_{-1}(m,\alpha,\beta,\nu)}{g^2}+f_0(m,\alpha,\beta,\nu)+\ldots\right]\,\,
\end{equation}
where we defined
\begin{equation}
    m=\frac{g v R}{2}\,.
\end{equation}
Here the leading term simply follows from the classical action and is independent of $\beta$ and $\nu$
\begin{equation}
  f_{-1}(m,\alpha,\beta,\nu)=2\pi g^2\mL_{\rm classical}=8\pi m^2\left[\frac49\alpha^2-\left(\frac{d-2}{2}\right)^2\right]\,.
\end{equation}
Specializing to the fixed point \eqref{eq:XYZ_eps_fixed_point} near four dimensions and extremizing the defect free energy we obtain $\alpha$ as a function $\beta,\nu$ and $v$. At leading order this is equivalent to solving the classical equation of motion for $x$:
\begin{equation}
    \left[
    \frac{4}{9}\alpha^2-\frac{(d-2)^2}{4}
    \right]v=0\,.
\end{equation}
For $v\neq0$, we expand around the solution
\begin{equation}\label{eq:XYZ_alpha_tree}
    \alpha_{\rm tree}
    =\frac{3(d-2)}{4}
    =\frac32-\frac{3}{4}\eps\,.
\end{equation}
This agrees with the supersymmetric value
$ \alpha_{\rm BPS} =\frac{d-1}{2} =\frac32-\frac12\eps$ to leading order $\eps$.

We now discuss the mass spectrum around the profile \eqref{eq:XYZ_eps_profile}. Expanding the fields in $KK$ modes along the circle with momentum $n\in\mathbb{Z}$ (on top of the monodromy), from the bosonic quadratic action we find the following masses:
\begin{align}
 R^2M_{x,n}^2
 &=\left(n-\frac{2\alpha}{3}\right)^2-\left(\frac{d-2}{2}\right)^2\,,
 \nonumber\\
 R^2M_{y,n}^2
 &=\left(n-\frac{2\alpha}{3}-\nu\right)^2+m^2-\left(\frac{d-2}{2}\right)^2,
 \nonumber\\
 R^2M_{z,n}^2
 &=\left(n-\frac{2\alpha}{3}+\nu\right)^2+m^2-\left(\frac{d-2}{2}\right)^2,
 \qquad n\in\mathbb Z.
 \label{eq:XYZ_eps_bosonic_masses}
\end{align} 

For general conical parameter $\beta$ we need to treat the different spin components of the fermions separately. Things simplify for integer $\beta$, for which \eqref{eq:XYZ_eps_fermion_monodromies} reads
\begin{align}
 \psi_x(\theta+2\pi)
 &=(-1)^{\beta+1}e^{\frac{2\pi i}{3}\alpha}
 \psi_x(\theta)\,,\nonumber\\ \nonumber
 \psi_y(\theta+2\pi)
 &=(-1)^{\beta+1}e^{\frac{2\pi i}{3}\alpha-2\pi i\nu}
 \psi_y(\theta)\,,\\ 
 \psi_z(\theta+2\pi)
 &=(-1)^{\beta+1}e^{\frac{2\pi i}{3}\alpha+2\pi i\nu}
 \psi_z(\theta)\,.
 \label{eq:XYZ_eps_fermion_monodromies_susy}
\end{align}
Luckily for us, \eqref{eq_conical_susy} implies that the BPS point is at 
\begin{equation}
    \beta=1+O(\eps)\,,
\end{equation}
so that to the order of interest we can take $\beta \in \mathbb{Z}$. 

In this case we can straightforwardly decompose into KK modes and obtain the fermions' mass spectrum. To this order $\psi_x$ behaves as a free field, hence the associated KK tower has mass
\begin{equation}
    R M_{\psi_x,n}=-\frac{\alpha}{3}-\left(n-\frac{1}{2}-\frac{\beta}{2}\right)\,,\quad n,\beta\in\mathbb{Z}\,.
\end{equation}
To derive the masses of $\psi_y$ and $\psi_z$, we note that their quadratic Lagrangian reads
\begin{equation}
 \mL^{(2)}_{\psi_y,\psi_z}
 =i\bar\psi_y\bar\sigma^\mu\nabla_\mu\psi_y
  +i\bar\psi_z\bar\sigma^\mu\nabla_\mu\psi_z
  -\frac{gv}{2}e^{-\frac{2i}{3}\alpha\theta}\psi_y\psi_z
  -\frac{gv}{2}e^{\frac{2i}{3}\alpha\theta}
    \bar\psi_y\bar\psi_z\,.
 \label{eq:XYZ_eps_quadratic_fermions}
\end{equation}
The $\theta$ dependence in the quadratic term cancels on the chosen boundary conditions, and we obtain the following two component mass matrix
\begin{equation}
 R\mathcal M_n
 =-\frac{\alpha}{3}\,\mathds 1
 +\begin{pmatrix}
 n-\frac12-\frac\beta2 -\nu&m\\
 m&-n+\frac12+\frac\beta2 +\nu
 \end{pmatrix}\,,\quad
 n,\,\beta\in\mathbb{Z}\,.
 \label{eq:XYZ_eps_Weyl_mass_matrix}
\end{equation}
Its two eigenvalues are
\begin{equation}
 RM_{n}^{\pm}
 =-\frac{\alpha}{3}
 \pm\sqrt{\left(n-\frac12-\frac\beta2-\nu\right)^2+m^2}\,,\quad n,\,\beta\in\mathbb{Z}\,.
 \label{eq:XYZ_eps_Weyl_masses}
\end{equation}

We can now check explicitly that the spectrum organizes into supersymmetric multiplets for the BPS values of the monodromies. To leading nontrivial order we can set $d=4$, $\alpha=3/2$ and $\beta=1$. We see that the \(y_n\) and \(z_{2-n}\) modes are degenerate in mass:
\begin{equation}
 R^2M_{y,n}^2
 =
 R^2M_{z,2-n}^2
 =
 (n-1-\nu)^2+m^2-1\,,
\end{equation}
while the two fermionic eigenvalues become
\begin{equation}
    RM_n^\pm=-\frac12\pm\sqrt{(n-1-\nu)^2+m^2}\,.
\end{equation}
They therefore obey
\begin{equation}
 R^2M_{y,n}^2
 =
 R^2M_{z,2-n}^2
 =
 (RM_n^\pm)^2+RM_n^\pm-\frac34\,,
 \label{eq:XYZ_eps_susy_mass_relation}
\end{equation}
which is the supersymmetric relation between scalar and fermion masses on \(\mathrm{AdS}_3\). Equivalently, their conformal dimensions satisfy
\begin{equation}\label{eq_XYZ_Delta_susy_eps}
    \Delta_n^\pm=\Delta_{B,n}\mp\frac12\,,
    \qquad
    \Delta_{B,n}=1+\sqrt{(n-1-\nu)^2+m^2}\,.
\end{equation}
Note that the supersymmetric pairing holds for arbitrary flavor monodromy \(\nu\) as expected.
A similar check applies to the massless multiplet parametrizing the \(X\)-branch:
\begin{equation}
 R^2M_{x,n}^2=(n-1)^2-1\,,
 \qquad
 RM_{\psi_x,n}=-(n-1)-\frac12\,,
\end{equation}
which again obey \eqref{eq:XYZ_eps_susy_mass_relation}.

In the previous discussion we neglected an important subtlety concerning boundary conditions. The relations \eqref{eq_XYZ_Delta_susy_eps} always hold for sufficiently large $m$, where both fermions and bosons admit a unique boundary condition.
More generally, we can always keep standard boundary conditions for the bosons, but slightly more care is needed with the fermionic partners. Recall that a Dirac fermion of mass $M$ admits a standard boundary condition associated with the dimension $\Delta_{\rm std}=1+|M|$, and an alternate boundary condition associated with the dimension $\Delta_{\rm alt}=1-|M|$. The fermions with mass $M^{-}_n$ have always negative mass and we thus can consistently choose standard for all values of $m$, ensuring that $\Delta^{-}_n=\Delta_{B,n}+\frac12$. Whenever $0\leq \sqrt{(n-1-\nu)^2+m^2}< 1/2$, the other fermionic mass $M^+_n$ changes sign instead, and the supersymmetric choice
$\Delta_n^+=\Delta_{B,n}-\frac12$ corresponds instead to alternate quantization for the fermion with mass $M_n^+$. Standard quantization would instead give
\begin{equation}\label{eq_XYZ_std}
    \Delta_{n,\mathrm{std}}^+
    =\frac32-\sqrt{(n-1-\nu)^2+m^2}
    =\frac52-\Delta_{B,n}\,,
\end{equation}
which does not obey the supersymmetric pairing above.
In the range $0\leq\nu\leq1/2$, only the $n=1$ mode can enter this regime. Thus, whenever
\begin{equation}
    0\leq\sqrt{\nu^2+m^2}<\frac12\,,
\end{equation}
preserving supersymmetry with standard scalar boundary conditions requires alternate quantization for the fermion with mass $M_1^+$.

The subleading correction in \eqref{eq_free_energy_XYZ_expansion} is the sum of the one-loop determinants and the wave-function counterterm:
\begin{equation}
    f_0(m,\alpha,\beta,\nu)=f_0^{(\mathrm{det})}(m,\alpha,\beta,\nu)+f_0^{(\mathrm{ct})}(m,\alpha,\beta,\nu)\,.
\end{equation}
The counterterm arises from the free field kinetic term and in the minimal subtraction scheme is given by
\begin{equation}
    f_0^{(\mathrm{ct})}(m,\alpha,\beta,\nu)=2\pi\,\delta Z_x\mL_{\rm classical}\,,\qquad \delta Z_x=-\frac{g^2}{2(4\pi)^2\eps}\,,
\end{equation}
where we used the known value of the chiral field wavefunction \cite{Ferreira:1996az}. Denoting with $f_S(M)$ and $f_D(M)$ the free energy contributions normalized by the AdS volume from, respectively, a real scalar and an AdS$_3$ Dirac fermion of mass $M$, the determinants contribution can be formally written as an infinite sum
\begin{equation}
\begin{split}
    f_0^{(\mathrm{det})}(m,\alpha,1,\nu)=\sum_{n\in \mathbb{Z}}&\left[2f_S(M_{x,n})+2f_S(M_{y,n})+2f_S(M_{z,n})\right.\\
    &\left.+f_D(M_{\psi_x,n}) +f_D(M_{n}^+)   +f_D(M_{n}^-)      \right]\,,
    \end{split}
\end{equation}
where we specialized to $\beta=1$ and, for $\sqrt{\nu^2+m^2}<1/2$, the $M_1^+$ fermion should be considered at the alternate fixed point for the boundary conditions to preserve supersymmetry. Here and in what follows we normalize the free energy so that it vanishes for the trivial defect, i.e.\ we subtract the same sum evaluated at $m=\nu=0$, $\alpha=3/2$, with standard boundary conditions for all fields.
We discuss in app. \ref{app:one_loop_determinants} how to regulate and compute this sum. We find that this has a divergent part given by
\begin{equation}
    f_0^{(\mathrm{det})}(m,\alpha,1,\nu)\Big\vert_{\rm div}=\frac{\left(\frac{4\alpha ^2}{9}-1\right) m^2}{4 \pi  \eps}\,,
\end{equation}
which precisely cancels the $1/\eps$ pole from the counterterm. Overall, we find remarkably that the supersymmetric free energy is $\nu$-independent and takes the following simple form
\begin{equation}
    f_0\left(m,\frac32,1,\nu\right)\Bigg\vert_{d=4}=-\frac{m^2}{4 \pi }+
\frac{1}{12 \pi }\,.
\end{equation}

To obtain the defect free energy at the fixed point, we need to add the classical term and specialize to the fixed point. Expanding the R-monodromy as
\begin{equation}
    \alpha=\frac{3}{2}-\delta\alpha \,\eps\,,
\end{equation}
the leading classical term contributes as 
\begin{equation}
    \frac{f_{-1}\left(m,\frac{3}{2}-\delta\alpha \,\eps,1,\nu\right)}{g_*^2}=0\times \frac{1}{\eps}+\frac{(3-4 \delta\alpha) m^2}{4 \pi }+O(\eps)\,.
\end{equation}
Overall the defect free energy therefore reads
\begin{equation}\label{eq:XYZ_defect_F_1loop}
    \frac{F\left(v,\frac32-\delta\alpha\,\eps,1,\nu\right)}{\operatorname{Vol}(\mathrm{AdS}_{d-1})}=\frac{1-2\delta\alpha}{2\pi}m^2
    +\frac{1}{12 \pi } +O(\eps)\,.
\end{equation}
It is now immediate to see that extremizing \eqref{eq:XYZ_defect_F_1loop} over $m$ we find
\begin{equation}
    \delta\alpha=\frac12\quad\implies \quad\alpha=\alpha_{\rm BPS}\,.
\end{equation}
The conformal defect is $1/2$ BPS for this choice of R-monodromy. Additionally, for $\alpha=\alpha_{\rm BPS}$ the defect free energy is independent of both $v$ and the flavor monodromy as expected
\begin{equation}\label{eq_XYZ_F_eps_exp}
    \frac{F\left(v,\alpha_{\rm BPS},\beta_{\rm BPS},\nu\right)}{\operatorname{Vol}(\mathrm{AdS}_{d-1})}=\frac{1}{12\pi}+O(\eps)\,.
\end{equation}
This establishes in particular the existence of a conformal manifold associated with $v$.

We can interpret the result \eqref{eq_XYZ_F_eps_exp} using the $X$-branch EFT. To the leading nontrivial order the defect is two-dimensional and its partition function admits a universal logarithmically divergent contribution with coefficient $b/3$, as reviewed around \eqref{eq_g_b_def}. Since the regulated AdS volume is logarithmically divergent for $d=4$ (see app. \ref{App_AdS_volume}), $b$ can be extracted from the free-energy
\begin{equation}
    b=6\pi \frac{F\left(v,\alpha_{\rm BPS},\beta_{\rm BPS},\nu\right)}{\operatorname{Vol}(\mathrm{AdS}_{d-1})}=\frac12\,.
\end{equation}
In $d=4$ the EFT fields are free and their monodromy is trivial. Therefore the only contribution to the defect free energy arises from the supersymmetric Wess-Zumino term for the Weyl anomaly. Proceeding as in sec. \ref{sec_GW} around \eqref{eq_GW_b_result}, we obtain:
\begin{equation}
    \Delta b=12 \Delta a\,,
\end{equation}
where the $a$-anomaly difference from integrating out the two chiral multiplets $Y$ and $Z$ is
\begin{equation}
    \Delta a=2 a_{\rm chiral}=\frac{1}{24}\quad\implies \quad b=\frac12\,.
\end{equation}
This is in perfect agreement with the explicit one-loop result.

It is instructive to compare the free energy for the BPS defect with the analogous result for non-supersymmetric boundary conditions. The simplest way to break supersymmetry is to change the boundary conditions for one of the KK modes. For instance, insisting on standard boundary conditions for the $M_1^+$ fermion for all values of $m$, we find
\begin{equation}\label{eq:XYZ_defect_F_1loop_std}
    \frac{F_{\rm std}\left(v,\frac32-\delta\alpha\,\eps,1,\nu\right)}{\operatorname{Vol}(\mathrm{AdS}_{d-1})}=\frac{1-2\delta\alpha}{2\pi}m^2
    +\begin{cases} \displaystyle
\frac{m^2+\nu ^2}{2 \pi }-\frac{\left(m^2+\nu ^2\right)^{3/2}}{3 \pi } &\text{if }\sqrt{m^2+\nu ^2}<\frac12 \\[0.5em]\displaystyle
\frac{1}{12 \pi } &\text{if }\sqrt{m^2+\nu ^2}\geq\frac12\,,
\end{cases}
\end{equation}
up to corrections of order $O(\eps)$. As expected, the result only differs from \eqref{eq:XYZ_defect_F_1loop} for $\sqrt{m^2+\nu ^2}<\frac12$, where our choice of boundary conditions yields the fermionic dimension \eqref{eq_XYZ_std}, breaking the supersymmetric pairing \eqref{eq_XYZ_Delta_susy_eps}.

We also consider the free energy at $\beta=0$, hence without the conical defect required by supersymmetry near four dimensions. We find that for 
\begin{equation}
    \sqrt{m^2+\left(\frac12-\nu\right)^2}\geq \frac12\,,
\end{equation}
the free energy can be written in terms of an infinite sum as
\begin{equation}\label{eq_XYZ_F_non_susy}
\frac{F\left(v,\frac32-\delta\alpha\,\eps,0,\nu\right)}{\operatorname{Vol}(\mathrm{AdS}_{d-1})}=
\frac{1-2\delta\alpha}{2\pi}m^2
+
\frac{23}{192\pi}
-
\frac{2m^2}{\pi^3}
\sum_{k=0}^{\infty}
\frac{\cos\left(2\pi(2k+1)\nu\right)}{(2k+1)^2}
K_2\left(2\pi(2k+1)m\right)\,.
\end{equation}
The sum over the Bessel function represents the contribution from the massive multiplets winding $2k+1$ times around the compact direction analogously to the discussion in~\cite{Grassi:2019txd,Hellerman:2021duh}.\footnote{Note that the infinite sum in \eqref{eq_XYZ_F_non_susy} vanishes at $\nu=\frac14$, due to the propagation phases of the free particles. We do not expect similar cancellations to occur at subleading order.} The bosonic and fermionic contributions canceled out for supersymmetric boundary conditions, while here they provide exponentially small corrections to the free energy. Indeed at large \(m\), the result can be expanded as\footnote{The constant term here can be understood from EFT as the sum of the Wess-Zumino contribution as before and the partition function of a free fermion with half-integer $U(1)$ monodromy.}
\begin{equation}
\begin{split}
    \frac{F\left(v,\frac32-\delta\alpha\,\eps,0,\nu\right)}{\operatorname{Vol}(\mathrm{AdS}_{d-1})}&=
    \frac{1-2\delta\alpha}{2\pi}m^2+\frac{23}{192\pi}\\ &-\frac{\cos(2\pi\nu)}{\pi^3}m^{3/2}e^{-2\pi m}\left[
1+\frac{15}{16\pi m}
+O\left(m^{-2}\right)\right]+O\left(m^{3/2}e^{-6\pi m}\right)\,.
\end{split}
\end{equation}
Extremizing we find
\begin{equation}
    \delta\alpha
    =
    \frac12
    +
    \frac{\cos(2\pi\nu)}{\pi}\sqrt{m}\,e^{-2\pi m}
    \left[
        1+\frac{3}{16\pi m}
        +O\left(m^{-2}\right)
    \right]+O\left(m^{3/2}e^{-6\pi m}\right)\,,
\end{equation}
so that $\alpha=\alpha_{\rm BPS}$ only up to exponentially small corrections. This non-BPS defect nonetheless exists and may be studied reliably within EFT. We expect that two-loop contributions will introduce also power corrections in $1/m$ to the value of $\alpha$. Indeed the boundary conditions break supersymmetry, and thus quantum corrections will in general spoil $\alpha=\alpha_{\rm BPS}$ and the exact marginality of $v$ already within the EFT.

\section{Non-BPS examples}\label{sec_non_susy_vortices}

In this section we study non-BPS vortex defects in $3d$ $\mathcal{N}=1$ SCFTs with moduli spaces. We will find that these always exist in the $5$-field Wess-Zumino (WZ) model, and for a range of monodromies in its $2N+1$-field generalization, while the stability of vortices in SQED with $N_f$ flavors depends on the value of a certain Wilson coefficient. We will also benchmark the EFT predictions in the WZ model against controlled perturbative calculations in the $\eps$-expansion and at large $N$. Throughout this section we set the common radius of $\mathrm{AdS}_{d-1}$ and $S^1$ to $R=1$.

\subsection{5-field WZ model}\label{sec_5_field}

\subsubsection{EFT analysis in \texorpdfstring{$3d$}{3d}}

We now turn to a non-supersymmetric vortex defect in the $3d$ $\mathcal{N}=1$ SCFT describing the infrared limit of the following five-field Wess-Zumino model
\begin{equation}
    W=\frac{g}{2}A\left(|X|^2-|Y|^2\right)\,.
\end{equation}
Here $A$ is a real superfield, while $X$ and $Y$ are complex. Their scalar components are denoted by $a,x,y$, respectively. The continuous global symmetry is $U(1)_x\times U(1)_y$, acting independently on $X$ and $Y$. Including discrete symmetries, the internal symmetry group can be written as
\begin{equation}
    G_{\rm int}
    =
    \left[O(2)_x\times O(2)_y\right]\rtimes\mathbb Z_2^{\rm exch}\,,
\end{equation}
where the reflections inside the two $O(2)$ factors act by complex conjugation, while $\mathbb Z_2^{\rm exch}$ exchanges $X$ and $Y$ together with $A\rightarrow-A$. Finally, the theory has a $\mathbb Z_2^R$ symmetry under which $A\rightarrow-A$. These symmetries protect the classical moduli space, where either $\langle a\rangle \neq 0$ or $\langle |x|^2\rangle=\langle |y|^2\rangle\neq 0$, from quantum corrections.

To obtain the vortex defect, we consider the moduli space EFT in the branch $\langle |x|^2\rangle=\langle |y|^2\rangle\neq0$. The low-energy bosonic degrees of freedom are therefore the dilaton $\Phi$ and two $2\pi$-periodic Goldstone fields $\pi_x$ and $\pi_y$. The discrete symmetries discussed above constrain the leading two-derivative action to take the form~\cite{Cuomo:2024fuy}
\begin{equation}\label{eq:fiveWZ_EFT}
\mL_{\rm EFT}=
    \frac12\left[(\pd\Phi)^2-\frac{1}{4}\Phi^2\right]
    +\frac{c}{2}\Phi^2
    \left[(\pd\pi_x)^2+(\pd\pi_y)^2\right]
    +\ldots\,,
\end{equation}
where $c$ is a positive Wilson coefficient, and the dots denote higher-derivative terms as well as the fermionic superpartners. Notice that, to this order, the EFT depends on a single nontrivial Wilson coefficient. 

We can use \eqref{eq:fiveWZ_EFT} to look for vortex defects directly in the strongly coupled $3d$ theory. On $\mathrm{AdS}_2\times S^1$ we consider the configuration
\begin{equation}
    \Phi=v^{1/2}\,,
    \qquad
    \pi_x=-\alpha_x\theta\,,
    \qquad
    \pi_y=-\alpha_y\theta\,.
    \label{eq:fiveWZ_EFT_vortex}
\end{equation}
The Goldstone equations are automatically satisfied, while the dilaton equation for $v\neq0$  implies
\begin{equation}
c\left(\alpha_x^2+\alpha_y^2\right)=\frac14\,.
    \label{eq:fiveWZ_EFT_saddle}
\end{equation}
Thus there is a continuous family of defects parametrized by the direction of $(\alpha_x,\alpha_y)$ and $v$.

It is interesting to study the bosonic fluctuations around \eqref{eq:fiveWZ_EFT_vortex}. Writing
\begin{equation}
    \Phi=v^{1/2}+\delta\Phi\,,
    \qquad
    \pi_i=-\alpha_i\theta+\frac{\delta\pi_i}{\sqrt{c\,v}}\,,
\end{equation}
we introduce the combinations parallel and orthogonal to the winding,
\begin{equation}
    \delta\pi_\parallel
    =
    \frac{\alpha_x\delta\pi_x+\alpha_y\delta\pi_y}
    {\sqrt{\alpha_x^2+\alpha_y^2}}\,,
    \qquad
    \delta\pi_\perp
    =
    \frac{-\alpha_y\delta\pi_x+\alpha_x\delta\pi_y}
    {\sqrt{\alpha_x^2+\alpha_y^2}}\,.
\end{equation}
Expanding \eqref{eq:fiveWZ_EFT} to quadratic order and using the saddle-point condition \eqref{eq:fiveWZ_EFT_saddle}, we obtain
\begin{equation}
    \mL_{\rm EFT}^{(2)}
    =
    \frac12(\pd\delta\Phi)^2
    +\frac12(\pd\delta\pi_\parallel)^2
    +\frac12(\pd\delta\pi_\perp)^2
    -\delta\Phi\,\pd_\theta\delta\pi_\parallel\,.
    \label{eq:fiveWZ_EFT_quadratic}
\end{equation}
Thus the orthogonal Goldstone $\delta\pi_\perp$ decouples, while $\delta\Phi$ mixes with $\delta\pi_\parallel$. Decomposing in $KK$ modes along the $S^1$, we obtain the following masses 
\begin{equation}
    M_{1,n}^2=n(n-1)\,,
    \qquad
    M_{2,n}^2=n(n+1)\,,
    \qquad
    M_{3,n}^2=n^2\,,
    \label{eq_5field_KK_masses_3d}
\end{equation}
where $n\in\mathbb{Z}$ is the KK momentum. Remarkably, once the saddle-point equation is imposed, the spectrum is independent both of the Wilson coefficient $c$ and of the ratio $\alpha_x/\alpha_y$. The first two entries in \eqref{eq_5field_KK_masses_3d} agree with the general prediction \eqref{eq_KK_masses_general}. Note also that all the masses are non-negative, and hence lie above the BF bound.

\subsubsection{Perturbative analysis in \texorpdfstring{$4-\eps$}{4-eps} dimensions}\label{sec:5_field_eps}

We now construct the vortex defect studied above using a weakly coupled continuation to $d=4-\eps$ dimensions of the 5-field WZ model. To this aim, following~\cite{ThomasSeminar,Cuomo:2024vfk}, we first consider a non-supersymmetric generalization with $N_f$ copies of each Majorana fermion. We then analytically continue the fermionic sector to $4-\eps$ dimensions by introducing $N_D=N_f/4$ Dirac fermions. The continuation corresponding to the original $3d$ $\mathcal N=1$ model is finally obtained by taking the formal limit $    N_D\rightarrow\frac14\,.$

The bosonic action on $\mathrm{AdS}_{d-1}\times S^1$ is
\begin{equation}
    \mL_{\rm bos}
    =
    |\pd x|^2+|\pd y|^2+\frac12(\pd a)^2
    +m_d^2\left(|x|^2+|y|^2+\frac{a^2}{2}\right)
    +\frac{g^2}{8}\left(|x|^2-|y|^2\right)^2
    +\frac{g^2}{4}a^2\left(|x|^2+|y|^2\right)\,.
\end{equation}
The model was studied perturbatively in~\cite{Cuomo:2024fuy}. In the supersymmetric limit $N_D\rightarrow1/4$, it admits a perturbative fixed point at
\begin{equation}
    g_*^2=\frac{16\pi^2}{3}\eps+O(\eps^2)\,.
    \label{eq:fiveWZ_fixed_point}
\end{equation}
At this fixed point the scalar components $x,y$ have scaling dimensions $\Delta_X,\Delta_Y$ given by
\begin{equation}
    \Delta_X=\Delta_Y
    =1-\frac{5}{12}\eps+O(\eps^2)\,.
    \label{eq:fiveWZ_DeltaX}
\end{equation}

We now introduce independent monodromies for the $U(1)_x$ and $U(1)_y$ symmetries and consider the classical configuration
\begin{equation}
     x=v_x e^{-i\alpha_x\theta}\,,
    \qquad
     y=v_y e^{-i\alpha_y\theta}\,,
    \qquad
     a=0\,,
    \label{eq:fiveWZ_vortex_profile}
\end{equation}
where $v_x$ and $v_y$ can be chosen real and non-negative. We work in a double scaling limit analogous to \eqref{eq_XYZ_scaling_limit}, where $v_x\sim v_y\sim 1/g$.
The equation of motion for $a$ is automatically satisfied, while those for $x$ and $y$ reduce to
\begin{align}
    \left[
        \alpha_x^2+m_d^2
        +\frac{g^2}{4}\left(v_x^2-v_y^2\right)
    \right]v_x&=0\,,
    \nonumber\\
    \left[
        \alpha_y^2+m_d^2
        -\frac{g^2}{4}\left(v_x^2-v_y^2\right)
    \right]v_y&=0\,.
    \label{eq:fiveWZ_saddle_equations}
\end{align}
For a saddle with $v_x,v_y\neq0$, these equations imply
\begin{equation}
    \alpha_x^2+\alpha_y^2=-2m_d^2
    =\frac{(d-2)^2}{2}\,,
    \label{eq:fiveWZ_alpha_constraint}
\end{equation}
together with
\begin{equation}
    g^2\left(v_x^2-v_y^2\right)
    =
    2\left(\alpha_y^2-\alpha_x^2\right)\,.
    \label{eq:fiveWZ_v_constraint}
\end{equation}
The saddle-point equations therefore fix the difference $v_x^2-v_y^2$, while leaving the sum $v_x^2+v_y^2$ as an undetermined free parameter.

Let us briefly discuss the stability of these solutions. We expand
\begin{equation}
    x=e^{-i\alpha_x\theta}
    \left(v_x+\frac{\delta x_1+i\delta x_2}{\sqrt2}\right)\,,
    \qquad
    y=e^{-i\alpha_y\theta}
    \left(v_y+\frac{\delta y_1+i\delta y_2}{\sqrt2}\right)\,.
\end{equation}
Using the saddle-point equations \eqref{eq:fiveWZ_saddle_equations}, the mass matrix for a KK mode with momentum $n\in\mathbb Z$, in the basis $(\delta x_1,\delta x_2,\delta y_1,\delta y_2)$, is
\begin{equation}
\mathcal M_{B,n}^2=
\begin{pmatrix}
n^2+\frac{g^2v_x^2}{2}
&
-2in\alpha_x
&
-\frac{g^2v_xv_y}{2}
&
0
\\
2in\alpha_x
&
n^2
&
0
&
0
\\
-\frac{g^2v_xv_y}{2}
&
0
&
n^2+\frac{g^2v_y^2}{2}
&
-2in\alpha_y
\\
0
&
0
&
2in\alpha_y
&
n^2
\end{pmatrix}.
\label{eq:fiveWZ_general_bosonic_matrix}
\end{equation}
The fluctuation of $a$ decouples from this sector.

Generic solutions with $\alpha_x^2\neq\alpha_y^2$ are unstable near four dimensions. Indeed, it is sufficient to consider the $n=1$ sector. In $d=4$, using \eqref{eq:fiveWZ_alpha_constraint} and \eqref{eq:fiveWZ_v_constraint}, we find
\begin{equation}
    \det\left(\mathcal M_{B,1}^2+\mathds 1\right)
    =
    -8\left(\alpha_x^2-\alpha_y^2\right)^2\,,
    \label{eq:fiveWZ_asymmetric_tachyon}
\end{equation}
which implies that at least one eigenvalue of $\mathcal M_{B,1}^2$ violates the BF bound $M^2\geq -1$. By continuity, this instability persists near four dimensions for generic fixed values of the parameters. Based on the EFT analysis in $d=3$, we expect no instability for $d$ sufficiently far from $d=4$.

In what follows therefore we focus on the stable symmetric saddle
\begin{equation}
    \alpha_x=\alpha_y=\alpha\,,
    \qquad
    v_x=v_y=v\,.
    \label{eq:fiveWZ_symmetric_saddle}
\end{equation}
Choosing $\alpha>0$, the equations of motion imply
\begin{equation}
    \alpha_{\rm tree}
    =
    \frac{d-2}{2}
    =
    1-\frac{\eps}{2}\,.
    \label{eq:fiveWZ_alpha_tree}
\end{equation}
Below we determine the one-loop corrected value of $\alpha$.

Defining
\begin{equation}
    m=gv\,,
\end{equation}
the defect free energy admits an expansion analogous to \eqref{eq_free_energy_XYZ_expansion},
\begin{equation}
    F(v,\alpha)
    =
    \operatorname{Vol}(\mathrm{AdS}_{d-1})
    \left[
        \frac{f_{-1}(m,\alpha)}{g^2}
        +f_0(m,\alpha)
        +O(g^2)
    \right]\,.
    \label{eq:fiveWZ_free_energy_expansion}
\end{equation}
The leading term follows from evaluating the classical action on \eqref{eq:fiveWZ_symmetric_saddle},
\begin{equation}
    f_{-1}(m,\alpha)
    =
    4\pi m^2
    \left[
        \alpha^2-\left(\frac{d-2}{2}\right)^2
    \right]\,.
    \label{eq:fiveWZ_classical_free_energy}
\end{equation}
The subleading term $f_0$ receives contributions from the one-loop fluctuation determinants and from the wave-function counterterm. We relegate the details of this calculation to app. \ref{App_5_field_eps}. Below we report the main results.

To determine the corrected saddle near four dimensions, we write
\begin{equation}
    \alpha=1-\delta\alpha\,\eps\,.
    \label{eq:fiveWZ_alpha_expansion}
\end{equation}
Using \eqref{eq:fiveWZ_fixed_point}, the classical contribution becomes
\begin{equation}
    \frac{f_{-1}(m,\alpha)}{g_*^2}
    =
    \frac{3(1-2\delta\alpha)}{4\pi}m^2
    +O(\eps)\,.
\end{equation}
For $m\ll1$, the renormalized one-loop contribution is
\begin{equation}
    f_0(m,1)
    =
    \frac{m^2}{8\pi}
    -\frac{1+1/\sqrt2}{12\pi}m^3
    +O(m^4,\eps)\,.
    \label{eq:fiveWZ_small_m_f0}
\end{equation}
Here we have taken the limit $\alpha=1+O(\eps)$. The nonanalytic cubic term originates from bosonic modes which become massless as $\alpha\rightarrow1$.

Combining the classical and one-loop contributions, we obtain
\begin{equation}
    \frac{F(v,\alpha)}{\operatorname{Vol}(\mathrm{AdS}_{d-1})}
    =
    \frac{7-12\delta\alpha}{8\pi}m^2
    -\frac{1+1/\sqrt2}{12\pi}m^3
    +O(m^4,\eps)\,.
    \label{eq:fiveWZ_small_m_free_energy}
\end{equation}
Extremizing with respect to $m$ gives
\begin{equation} \label{eq:fiveWZ_dalpha_small_m}
    \delta\alpha
    =
    \frac{7}{12}
    -\frac{1+1/\sqrt2}{12}m
    +O(m^2)\,.
\end{equation}

The $m\rightarrow0$ limit admits a simple interpretation. For $\alpha=1$ the monodromy is trivial, and the expectation value in \eqref{eq:fiveWZ_symmetric_saddle} is triggered by deforming the defect by the operators $\pd_z x$ and $\pd_z y$. At the trivial defect these operators have dimension $\Delta_X+1$. Turning on a small monodromy $\alpha-1$ continuously shifts the transverse spin of these defect modes and, to the order of interest, their dimensions become
\begin{equation}
    \Delta^{(\alpha)}_{\pd_zx}
    =
    \Delta_X+1-(1-\alpha)+O(\eps^2)
    =
    1+\alpha-\frac{5}{12}\eps+O(\eps^2)\,,
\end{equation}
where we used \eqref{eq:fiveWZ_DeltaX}.
The onset of the nontrivial saddle occurs when this deformation becomes marginal, namely when its dimension coincides with the defect dimension,
\begin{equation}
   \Delta^{(\alpha)}_{\pd_zx}=d-2\,,
\end{equation}
which gives
\begin{equation}
    \alpha
    =
    1-\frac{7}{12}\eps+O\left(\eps^2\right)\quad\implies\quad\delta\alpha=\frac{7}{12}+O\left(\eps\right)\,,
\end{equation}
in precise agreement with \eqref{eq:fiveWZ_dalpha_small_m} for $m\rightarrow 0$.

Let us finally focus on the main regime of interest, $m\gg1$. For $\alpha=1+O(\eps)$ we find
\begin{equation}
    f_0(m,\alpha)
    =
    -\frac{m^2}{16\pi}
    -\frac{1}{16\pi}\log m
    +O\left(m^0,\eps\right)\,.
    \label{eq:fiveWZ_large_m_f0}
\end{equation}
Combining this result with the classical contribution gives
\begin{equation}
    \frac{F(v,\alpha)}{\operatorname{Vol}(\mathrm{AdS}_{d-1})}
    =
    \frac{11-24\delta\alpha}{16\pi}m^2
    -\frac{1}{16\pi}\log m
    +O\left(m^0,\eps\right)\,.
    \label{eq:fiveWZ_large_m_free_energy}
\end{equation}
Extremizing with respect to $m$ therefore yields
\begin{equation}
    \delta\alpha
    =
    \frac{11}{24}
    -\frac{1}{48m^2}
    +O\left(\frac{\log m}{m^4}\right)\,,
\end{equation}
and hence
\begin{equation}
    \alpha
    =
    1-\frac{11}{24}\eps
    +\frac{\eps}{48m^2}
    +O\left(\frac{\eps\log m}{m^4},\eps^2\right)\,.
    \label{eq:fiveWZ_alpha_large_m}
\end{equation}

For this value of $\alpha$ the free energy decreases logarithmically with $m$,
\begin{equation}
    \frac{F(v,\alpha(v))}{\operatorname{Vol}(\mathrm{AdS}_{d-1})}
    =
    -\frac{1}{16\pi}\log m
    +O\left(m^0,\eps\right)\,.
    \label{eq:fiveWZ_large_m_free_energy_EFT}
\end{equation}
To leading order in the $\eps$-expansion, the defect Weyl anomaly coefficient $b$, as defined in \eqref{eq_g_b_def}, therefore behaves as
\begin{equation}
    b=-\frac{3}{8}\log m+O(m^0,\eps)\,,
    \label{eq:fiveWZ_large_m_b}
\end{equation}
where we used the regulated volume of $\mathrm{AdS}_3$ in \eqref{eq_app_Vol_AdS3}. Thus $b$ becomes arbitrarily negative in the limit $m\rightarrow\infty$, and we expect that the defect is RG stable and is the IR fixed point obtained from a relevant deformation of the standard monodromy defect at large $m$.

The asymptotic value in \eqref{eq:fiveWZ_alpha_large_m} has a simple interpretation in the moduli-space EFT. For the symmetric vortex, the EFT saddle-point condition~\eqref{eq:fiveWZ_EFT_saddle} in $d$-dimensions reads 
\begin{equation}
    2c\,\alpha^2
    =
    \left(\frac{d-2}{2}\right)^2\,.
    \label{eq:fiveWZ_EFT_saddle_eps}
\end{equation}
Near four dimensions the Wilson coefficient $c$ was computed perturbatively in~\cite{Cuomo:2024fuy},
\begin{equation}
    c=\frac12-\frac{\eps}{24}+O(\eps^2)\,.
\end{equation}
It follows that
\begin{equation}
    \alpha_{\rm EFT}
    =
    \frac{d-2}{2\sqrt{2c}}
    =
    1-\frac{11}{24}\eps+O(\eps^2)\,,
    \label{eq:fiveWZ_alpha_EFT}
\end{equation}
in precise agreement with the $m\rightarrow\infty$ limit of the microscopic calculation.

The logarithmic term in \eqref{eq:fiveWZ_large_m_f0} also has a simple origin within the moduli-space EFT, analogous to the discussion of subleading corrections to the dimension of large charge operators in this model discussed in \cite{Cuomo:2024fuy}. In general dimension, imposing the EFT saddle-point equation, the three light bosonic towers~\eqref{eq_5field_KK_masses_3d} have masses
\begin{equation}
    M_{1,n}^2
    =
    n^2+(d-2)n\,,
    \qquad
    M_{2,n}^2
    =
    n^2-(d-2)n\,,
    \qquad
    M_{3,n}^2=n^2\,,
    \qquad n\in\mathbb Z\,.
    \label{eq:fiveWZ_EFT_spectrum_d}
\end{equation}
The logarithm originates entirely from the last tower. Specializing to $d=4$ and using the techniques in app. \ref{app:one_loop_determinants}, its one-loop contribution to the defect free energy is formally written as an infinite sum
\begin{equation}
    \frac{F(v,\alpha(v))}{\operatorname{Vol}(\mathrm{AdS}_3)}
    \supset
    -\frac{1}{12\pi}
    \sum_{n\in\mathbb Z}
    \left(n^2+1\right)^{3/2}\,.
    \label{eq:fiveWZ_EFT_mode_partition}
\end{equation}
At large KK momentum, we can expand as
\begin{equation}
    \left(n^2+1\right)^{3/2}
    =
    |n|^3+\frac32|n|
    +\frac{3}{8|n|}
    +O\left(|n|^{-3}\right)\,.
\end{equation}
The first two terms lead to power divergences in the KK sum, and can be removed using local counterterms. The $1/|n|$ term gives rise instead to a logarithmic divergence. Since the heavy fields that have been integrated out have masses of order $m$, the effective description applies up to $|n|\lesssim m$. Imposing this cutoff, we find
\begin{equation}
    \frac{F(v,\alpha(v))}{\operatorname{Vol}(\mathrm{AdS}_3)}
    \supset
    -\frac{1}{16\pi}\log m\,,
\end{equation}
in precise agreement with \eqref{eq:fiveWZ_large_m_f0}. The other two light bosonic towers as well as the fermions (whose masses are in app. \ref{App_5_field_eps}) give only polynomial functions of $|n|$ at large momentum and therefore do not generate a logarithm. Note that the logarithmic dependence of the free energy on $m$ is in agreement with the general EFT expectation \eqref{eq_f-EFT_prediction} in $d=4$.

\subsection{\texorpdfstring{$2N+1$}{2N+1}-field WZ model at large \texorpdfstring{$N$}{N}}\label{sec:large_N}

We now consider the $3d$ $\mathcal{N}=1$ theory with one real superfield $A$ and real superfields $X_i,Y_i$, $i=1,\ldots,N$, in the fundamental of two $SO(N)$ factors, with superpotential
\begin{equation}
    W=A(X_i^2-Y_i^2)\;.
\end{equation}
Note that for $N=2$ this reduces to the theory considered in section~\ref{sec_5_field}, up to the normalization of the superpotential coupling, which does not affect the infrared fixed point. In general, this model admits a moduli space of vacua where $\langle A\rangle\neq 0$ or $\langle X_i^2\rangle =\langle Y_i^2\rangle\neq 0$. We analyze this model at large $N$, using methods from \cite{Alvarez-Gaume:2019biu}.

We derive the large-$N$ effective action of this theory in appendix \ref{app:large_N_EFT}. The result is
\begin{equation}
\begin{split}
    S_{\rm eff} &= \frac{N}{2} \operatorname{Tr}\log \left[-\Delta + F_A + a^2 + m_d^2\right] + \frac{N}{2} \operatorname{Tr}\log \left[-\Delta - F_A + a^2 + m_d^2 \right]  \\
    &- \frac{N}{2} \operatorname{Tr}\log\left[i \gamma^\mu \partial_\mu + a \right]  - \frac{N}{2} \operatorname{Tr}\log\left[i \gamma^\mu \partial_\mu - a \right] \\
    &+\int d^3x\sqrt{g}\left[\left(F_A+a^2+m_d^2\right)|x|^2+\left(-F_A+a^2+m_d^2\right)|y|^2\right]\;.
\end{split}
\end{equation}
Here $a$ is the bottom component of the superfield $A$ and $F_A$ is its top component. We denote the scalar components of $X_i,Y_i$ by $x_i,y_i$, and $m_d^2=-1/4$ is the conformal mass term on $\mathrm{AdS}_2\times S^1$ with unit radii. We have also separated a complex field $x=(x_1+ix_2)/\sqrt2$ and similarly for $y$. The first line is the result of integrating out the fluctuations of the bottom components of $X_i,Y_i$ around this background, and the second line is the result of integrating out their fermionic components. At leading order in $1/N$ we do not distinguish the $N-2$ components without a classical profile from all $N$ components, since the difference is suppressed by $1/N$. In the fermionic determinants, $\operatorname{Tr}$ denotes the trace for a two-component Dirac operator. Since the microscopic fermions are Majorana, the contribution of $N$ Majorana fermions is $N/2$ times the Dirac determinant.

Next we introduce monodromies $2\pi\alpha_j$ for the various Cartans of $SO(N)\times SO(N)$, where we take $N$ even. Pairing the fields $x_i$ into complex pairs $\varphi_j = (x_{2j-1}+ix_{2j})/\sqrt2$ ($j=1,\ldots,N/2$), this amounts to a change of the kinetic terms:
\begin{equation}
    D_{\theta}^{j}\varphi_{j}=(\partial_{\theta}+i\alpha_j)\varphi_j\;,
    \qquad D_{m}^{j}\varphi_{j}=\partial_m\varphi_j\quad (m\in \mathrm{AdS}_2)\;.
\end{equation}
For simplicity we will assume that for each $SO(N)$, all Cartans have the same monodromy, so that $\alpha^x_j\equiv \alpha_x$ and $\alpha^y_j\equiv \alpha_y$, although the more general case can also be discussed. We take $0\leq\alpha_x,\alpha_y<1$; charge conjugation relates $\alpha$ and $1-\alpha$, but it will be useful to keep the full range.
The large $N$ effective action now is
\begin{equation}
\begin{split}
 S_{\rm eff} &= \frac{N}{2} \operatorname{Tr}\log \left[-\Delta_x + F_A + a^2 + m_d^2\right] + \frac{N}{2} \operatorname{Tr}\log \left[-\Delta_y - F_A + a^2 + m_d^2 \right] \\
&- \frac{N}{2} \operatorname{Tr}\log\left[i \gamma^\mu D^x_\mu + a \right]  - \frac{N}{2} \operatorname{Tr}\log\left[i \gamma^\mu D^y_\mu - a \right] \\
&+\int d^3x\sqrt{g}\left[\left(F_A+\alpha^2_x+a^2+m_d^2\right)|x|^2+\left(-F_A+\alpha^2_y+a^2+m_d^2\right)|y|^2\right]\;.
\end{split}
\end{equation}

We can now look for conformal defects. Defining 
\begin{equation}\label{eq_largeN_I}
\begin{split}
    \mathcal{I}&=\frac{N}{2} \operatorname{Tr}\log \left[-\Delta_x + F_A + a^2 + m_d^2\right] + \frac{N}{2} \operatorname{Tr}\log \left[-\Delta_y - F_A + a^2 + m_d^2 \right]  \\
    &- \frac{N}{2} \operatorname{Tr}\log\left[i \gamma^\mu D^x_\mu + a \right]  - \frac{N}{2} \operatorname{Tr}\log\left[i \gamma^\mu D^y_\mu - a \right]\;,
\end{split}
\end{equation}
and denoting its bosonic part (first line) by $\mathcal{I}_{\rm B}$, the equations of motion are
\begin{equation}\label{eq:largeN_eqs} 
    \begin{split}
    \partial_{a}\mathcal{I} + 4\pi\operatorname{Vol}(\mathrm{AdS}_2)\,a(|x|^2+|y|^2)=0\;,\\
        \partial_{F_A} \mathcal{I}_{\rm B} + 2\pi\operatorname{Vol}(\mathrm{AdS}_2)(|x|^2-|y|^2)=0\;,\\
        (F_A+\alpha_x^2+a^2+m_d^2)x=0\;,\\
        (-F_A+\alpha_y^2+a^2+m_d^2)y=0 \;.  \end{split}
\end{equation}
Since the defect preserves the $\mathbb Z_2^R$ symmetry $A\rightarrow-A$ of section \ref{sec_5_field}, $\partial_a \mathcal{I}|_{a=0}=0$ and we can set $a=0$, which solves the first equation. Since $\mathcal{I}$ is of order $N$, the solutions below have $x,y$ of order $\sqrt N$.

For $x,y\neq0$ the last two equations fix the monodromies in terms of $F_A$,
\begin{equation}\label{eq:largeN_alphas}
    \alpha_x(F_A)=\sqrt{\frac14-F_A}\,,
    \qquad
    \alpha_y(F_A)=\sqrt{\frac14+F_A}\,,
\end{equation}
so that $\alpha_x^2+\alpha_y^2=-2m_d^2=1/2$. Along this locus the determinants in $\mathcal{I}_{\rm B}$ are real only for $|F_A|\leq 9/64$, i.e.\ for $\sqrt7/8\leq\alpha_{x,y}\leq 5/8$ (see appendix \ref{app:large_N_I}). At $F_A=9/64$ ($F_A=-9/64$) the $n=-1$ KK mode of the components of $y_i$ ($x_i$) orthogonal to the condensate saturates the BF bound on AdS$_2$, and beyond this point the defect is unstable.

\begin{figure}
    \centering
    \includegraphics[width=0.5\linewidth]{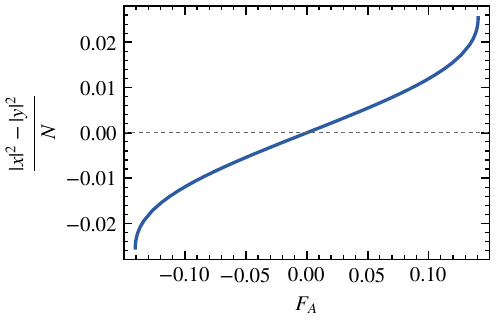}
    \caption{$(|x|^2-|y|^2)/N$ on the large-$N$ saddle as a function of $F_A$, for $|F_A|\leq 9/64$.}
    \label{fig:I}
\end{figure}

The second equation in \eqref{eq:largeN_eqs} fixes $|x|^2-|y|^2$ as a function of $F_A$. We bring $\pd_{F_A}\mathcal{I}_B$ in a convergent integral form in appendix \ref{app:large_N_I}, see \eqref{eq:largeN_diff}, and present the solution for $|x|^2-|y|^2$ as a function of $F_A$ in figure \ref{fig:I}. We see that $|x|^2-|y|^2$ is odd and monotonic in $F_A$, and it remains finite at the onset of the tachyonic instability, $F_A=\pm 9/64$, where $|x|^2-|y|^2=\pm 0.0254\,N$. Therefore $|x|^2-|y|^2$ is bounded. On the other hand the sum $|x|^2+|y|^2\geq\left||x|^2-|y|^2\right|$ is not fixed and provides an exact modulus. Note in particular that the monodromies are independent of the value of $|x|^2+|y|^2$, which is hence associated with an exactly marginal deformation at leading order in $N$.

Thus for every $|F_A|\leq 9/64$ there is a solution to the equations of motion with non-vanishing $x,y$. Mapping back to flat space, this leads to a family of conformal defects of the schematic form
\begin{equation}
    x=\frac{\lambda_x}{r^{1/2}}e^{i\alpha_x\theta}\;,\quad y=\frac{\lambda_y}{r^{1/2}}e^{i\alpha_y\theta}\;,
\end{equation}
where $\alpha_{x,y}$ and $|\lambda_x|^2-|\lambda_y|^2$ are fixed in terms of $F_A$, while $|\lambda_x|^2+|\lambda_y|^2$ is arbitrary and is associated with an exactly marginal deformation.

The above solution admits a simple interpretation within the moduli space EFT, which holds at large $|x|^2+|y|^2$. Generalizing~\eqref{eq:fiveWZ_EFT}, the low energy action consists of a dilaton and two $SO(N)/SO(N-1)$ nonlinear sigma models in the form
\begin{equation}
\mL_{\rm EFT}=\frac12\left[(\pd\Phi)^2-\frac{1}{4}\Phi^2\right]
+\frac{c}{2}\Phi^2
\left[\left(\pd\hat{n}_x\right)^2+\left(\pd\hat{n}_y\right)^2\right]    +\ldots\,,
\end{equation}
where $\hat{n}_x^2=\hat{n}_y^2=1$ are $N$-dimensional unit vectors. Expanding around the solution
\begin{equation}\label{eq:largeN_WZ_EFT_vortex}
    \Phi=v^{1/2}\,,
    \qquad
    \hat{n}_x=\left(\cos\alpha_x\theta,\sin\alpha_x \theta,\vec{0}_{N-2}\right)\,,
    \qquad
   \hat{n}_y=\left(\cos\alpha_y\theta,\sin\alpha_y \theta,\vec{0}_{N-2}\right)\,,
\end{equation}
the dilaton's equation of motion gives (for $v\neq 0$) the same condition as in~\eqref{eq:fiveWZ_EFT_saddle}. This coincides with the above condition $\alpha_x^2+\alpha_y^2=1/2$ for 
\begin{equation}
    c=\frac12\,.
\end{equation}
Additionally, expanding into KK modes around the $S^1$, we find that the fluctuations of the nonlinear sigma model lead to modes whose masses on AdS$_2$ read $M_n^2=n(n+2\alpha_{x,y})$. In particular, the $n=-1$ mode violates the BF bound $M^2\geq-1/4$ for $\alpha_{x,y}>5/8$. Using the condition $\alpha_x^2+\alpha_y^2=1/2$, this gives precisely the window $\sqrt7/8\leq\alpha_{x,y}\leq 5/8$.

\begin{figure}[t]
    \centering  
    \subfloat[The bosonic contribution to $\log(g)/N$.\label{fig:Logg}]{
        \includegraphics[width=0.48\textwidth]{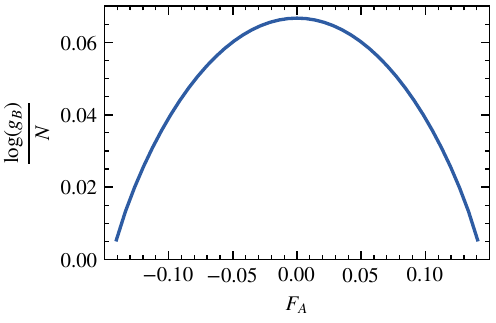}
    }
    \hfill
    \subfloat[Comparing with the standard monodromy defect.\label{fig:Loggcompare}]{
        \includegraphics[width=0.48\textwidth]{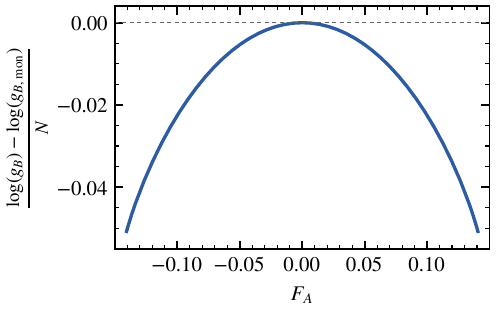}
    }
    
    \caption{Results for the bosonic contribution to $\log g/N$.}
    \label{fig:main}
\end{figure}

We can also use the large $N$ expansion to compute the $g$-function of the defect. Note that this is independent of the exactly marginal parameter associated with $|x|^2+|y|^2$ at leading order in $N$. The $g$-function is the sum of the bosonic and the fermionic fluctuation determinants. The bosonic contribution is
\begin{equation}\label{eq:largeN_gB}
    \log g_{\rm B}(F_A)
    =-\left[\mathcal{I}_{\rm B}(F_A,\alpha_x(F_A),\alpha_y(F_A))-\mathcal{I}_{\rm B}(0,0,0)\right]\,,
\end{equation}
where $\mathcal{I}_{\rm B}$ denotes the first line of~\eqref{eq_largeN_I} as before, and the monodromies $\alpha_x$ and $\alpha_y$ are specified by the condition \eqref{eq:largeN_alphas}. We plot the bosonic contribution to $\log g$ (normalized by $N$) as a function of $F_A$ in figure \ref{fig:Logg}.
Since the fermionic determinants are independent of $F_A$ except through the monodromies, the fermionic contribution is just the sum of that of free Dirac fields with monodromies specified by $\alpha_x(F_A)$ and $\alpha_y(F_A)$.\footnote{The free Dirac field $g$-function for $0\leq \alpha\leq 1/2$ can be obtained by integrating the $U(1)$ current one-point function on $\mathrm{AdS}_2\times S^1$,
\begin{equation}
\langle J_\theta\rangle
=
\frac{1}{8}\left(1-4\alpha^2\right)\tan(\pi \alpha)
=
\frac{1}{2\pi}\pd_\alpha\log g_{\rm D}(\alpha)\,,
\end{equation}
computed in~\cite{Barkeshli:2025cjs}. Note that, for standard boundary conditions as we assume above, $\log g_{\rm D}(\alpha)=\log g_{\rm D}(1-\alpha)$ by charge conjugation invariance.} Denoting the latter as $\log g_{\rm D}(\alpha)$, the full answer for the $g$-function is
\begin{equation}\label{eq:largeN_g_full}
    \log g(F_A)
    =\log g_{\rm B}(F_A)
    +\frac{N}{2}
    \left[
    \log g_{\rm D}(\alpha_x(F_A))
    +\log g_{\rm D}(\alpha_y(F_A))
    \right]
    \;.
\end{equation}

In figure \ref{fig:Loggcompare} we compare~\eqref{eq:largeN_g_full} with the result~$\log g_{\rm mon}$ for the standard monodromy defect, namely the saddle with $x=y=0$, the same monodromies, and $F_A$ fixed by $\partial_{F_A}\mathcal{I}_{\rm B}=0$~\cite{Giombi:2021uae}. Note that, since the fermionic contribution is identical for our defect and for the standard monodromy defect, it cancels from the ratio of their $g$-functions:
\begin{equation}\label{eq:largeN_g_comparison}
    \log g-\log g_{\rm mon}
    =
    \log g_{{\rm B}}-\log g_{{\rm B},\rm mon}\,,
\end{equation}
where the subscript $\rm B$ denotes the bosonic contribution as above.
Note that the difference in~\eqref{eq:largeN_g_comparison} is a strictly concave function of $F_A$. It reaches a maximum at $F_A=0$, where the difference vanishes, while $\log g<\log g_{\rm mon}$ for $F_A\neq0$, as shown in figure \ref{fig:Loggcompare}. This suggests that for $F_A=0$ our defect is smoothly connected to the standard monodromy defect, while for $F_A\neq 0$ the defect we constructed may be an IR fixed point of an RG flow starting from the standard monodromy defect.

\subsection{SQED}\label{sec_SQED}

We now consider the $3d$ $\mathcal N=1$ SCFT arising as the infrared limit of $U(1)$ SQED with $N_f$ charge-one matter multiplets and a vanishing Chern-Simons level. Time-reversal and flavor symmetries protect its moduli space from being lifted~\cite{Gaiotto:2018yjh,Benini:2018bhk}. At a generic point, the light bosonic fields are the dilaton $\Phi$ and $2N_f-2$ Goldstone bosons parametrizing $\mathbb{CP}^{N_f-1}$.

To study conformal vortex defects we put the theory on $\mathrm{AdS}_2\times S^1$ and use its moduli-space EFT. The leading two-derivative action takes the form
\cite{Cuomo:2024fuy}
\begin{equation}
\label{eq_SQED_EFT}
\mL_{\rm EFT}=
\frac12\left[(\pd\Phi)^2+m_d^2\Phi^2\right]
+2c\Phi^2\gamma_{i\bar j}\pd_\mu w_i\pd^\mu\bar w_j+\ldots\,,
 \qquad
\gamma_{i\bar j}=\frac{(1+|w|^2)\delta_{ij}-\bar w_iw_j}{(1+|w|^2)^2}\,,
\end{equation}
where $i=1,\ldots,N_f-1$, $c$ is a positive Wilson
coefficient, and $m_d^2=-1/4$. The dots denote
higher-derivative terms and fermionic superpartners.

We consider the configuration
\begin{equation}
\label{eq_SQED_vortex}
    \Phi=v^{1/2}\,,
    \qquad
    w_1=e^{-i\alpha\theta}\,,
    \qquad
    w_a=0\,,
    \quad a=2,\ldots,N_f-1\,.
\end{equation}
The monodromy acts through the $U(1)$ isometry rotating
$w_1$, with the other coordinates invariant.
The equations for the $w_i$ are automatically satisfied,
while the dilaton equation requires
\begin{equation}
\label{eq_SQED_saddle}
    c\,\alpha^2=\frac14\,.
\end{equation}
We take $\alpha=1/(2\sqrt c)$ below.

To study the spectrum of operators on the defect, we consider fluctuations by writing
\begin{equation}
\label{eq_SQED_fluctuations}
    \Phi=v^{1/2}+\delta\Phi\,,
    \qquad
    w_1=
    \left(1+\frac{\eta}{\sqrt{ c\,v}}\right)
    e^{-i\alpha\theta+i\delta\pi/(\sqrt{ c\,v})}\,,
    \qquad
    w_a=\frac{\zeta_a}{\sqrt {c\,v}}\,.
\end{equation}
Using~\eqref{eq_SQED_saddle}, the quadratic action becomes
\begin{equation}
\label{eq_SQED_quadratic}
\begin{split}
    \mL_{\rm EFT}^{(2)}
    ={}&
    \frac12(\pd\delta\Phi)^2
    +\frac12(\pd\delta\pi)^2
    -\delta\Phi\,\pd_\theta\delta\pi
    +\frac12(\pd\eta)^2-\frac{\alpha^2}{2}\eta^2
    \\
    &+
    \sum_{a=2}^{N_f-1}
    \left[
        |\pd\zeta_a|^2
        +\frac{i\alpha}{2}
        \left(
            \zeta_a\pd_\theta\bar\zeta_a
            -\bar\zeta_a\pd_\theta\zeta_a
        \right)
    \right]\,.
\end{split}
\end{equation}
Decomposing in KK modes along $S^1$, we obtain
\begin{equation}
\label{eq_SQED_bosonic_masses}
\begin{aligned}
    M_{1,n}^2&=n(n-1)\,,
    &\qquad M_{2,n}^2&=n(n+1)\,,
    \\
    M_{\eta,n}^2&=n^2-\frac{1}{4c}\,,
    & M_{a,\pm,n}^2&=n^2\pm\frac{n}{2\sqrt c}\,,
\end{aligned}
\end{equation}
where $n\in\mathbb Z$. The first two towers agree with
the universal prediction~\eqref{eq_KK_masses_general}.
The last pair occurs for each complex field $\zeta_a$
and is absent for $N_f=2$.

The mass spectrum of the KK modes of $\eta$ supply a nontrivial stability constraint. Its zero mode satisfies the
$\mathrm{AdS}_2$ BF bound only if
\begin{equation}
\label{eq_SQED_BF}
    M_{\eta,0}^2=-\frac{1}{4c}\geq-\frac14
    \qquad\Longleftrightarrow\qquad c\geq1\,.
\end{equation}
All other modes satisfy the bound in this range.
For fixed $c>1$, the saddle is therefore perturbatively
stable at sufficiently large $v$, while at $c=1$
subleading corrections must be examined.

The available controlled results, however, lie below
this threshold. For $N_f=2$, computations in the $\epsilon$-expansion of the dual WZ model near four dimensions give $
 c=1-\frac{\epsilon}{3}+O(\epsilon^2)\,$ \cite{Cuomo:2024fuy} (a direct $\epsilon$-expansion of SQED was attempted in~\cite{Breitstein:2024pwi}).\footnote{In $d$ dimensions, the mass spectrum reads
 \begin{equation}
\begin{aligned}
M_{1,n}^2&=n\bigl(n-d+2\bigr)\,,
&M_{2,n}^2&=n\bigl(n+d-2\bigr)\,,&
M_{\eta,n}^2&=n^2-\frac{(d-2)^2}{4c}\,,
&M_{a,\pm,n}^2&=n^2\pm\frac{d-2}{2\sqrt c}\,n\,.
\end{aligned}
\end{equation}
In particular, the $\eta$ zero mode satisfies the BF bound only for $c\geq1$ for any value of $d$.} Moreover, the large-$N_f$ analysis of \cite{Popov:2026ofo} gives $c=1/4+O(1/N_f)$. This vortex is thus unstable at large $N_f$ as well as in the $\epsilon$ expansion. This is the unique vortex saddle for $N_f=2$. For $N_f\geq 3$ different vortex configurations exist, but we have not found one which is stable for $c<1$.

\section{Macroscopic limit}\label{sec_macroscopic_limit}

In sec.~\ref{sec_vortices_general_EFT} we showed that in theories with moduli spaces of vacua there exists a class of vortex defects for which the operators whose vevs provide local coordinates along the moduli space acquire expectation values of the form
\begin{equation}
\langle \mO_{\Delta,q_a}(z,\bar z)\rangle
=
\frac{c_{\mO}(\bar{\phi})h^{\Delta}e^{i q_a\bar{\pi}^a}}
{z^{\frac{\Delta-q_a\alpha^a(h)}{2}}
\bar z^{\frac{\Delta+q_a\alpha^a(h)}{2}}}\,,
\end{equation}
in the notation of~\eqref{eq_1pt_fun_EFT}. At large $h$, the local physics is expected to approach that of the flat-space theory in a nontrivial vacuum on the moduli space. Indeed, the dilaton expectation value $v\propto h/R$ sets a dimensionful scale on $\mathrm{AdS}_{d-1}\times S^1$ much larger than the geometric one $\sim 1/R$. The EFT makes this statement manifest in the regime $1/v\ll |x|\ll R$, where both the massive degrees of freedom and the curvature of the background can be neglected, but we expect this statement to hold more generally.

These ideas can be formalized through the existence of a suitable \emph{macroscopic limit}~\cite{Lashkari:2016vgj,Jafferis:2017zna}, analogous to that explored for large-charge operators in theories with a moduli space of vacua in~\cite{Caetano:2023zwe,Cuomo:2024fuy,Ivanovskiy:2024vel}. Namely, we consider
\begin{equation}\label{eq_vortex_macroscopic_limit}
R\rightarrow\infty\,,
\qquad
h\rightarrow\infty\,,
\qquad
v=\frac{h}{R}
\quad\text{fixed}\,.
\end{equation}
In this limit correlation functions of local operators at fixed physical separations approach those evaluated in a nontrivial vacuum of the flat-space theory, at the point on the moduli space selected by the vortex saddle-point.

To see how the macroscopic limit constrains the large-$h$ behavior of DCFT correlators, consider first the one-point function of a scalar primary $\mO_{\Delta,q_a}$ of dimension $\Delta$ and charges $q_a$. For the moment we do not assume the validity EFT and simply analyze the consequences of the limit~\eqref{eq_vortex_macroscopic_limit}.
Defect conformal symmetry and the unbroken combination $J+\alpha^a(h)Q_a$ fix its one-point function to take the form 
\begin{equation}\label{eq_macro_scalar_1pt}
\langle \mO_{\Delta,q_a}(z,\bar z)\rangle_{h,\mathrm{flat}}
=
\frac{a_{\mO}(h)}
{z^{\frac{\Delta-q_a\alpha^a(h)}{2}}
\bar z^{\frac{\Delta+q_a\alpha^a(h)}{2}}}
\quad\implies\quad
\langle \mO_{\Delta,q_a}\rangle_{h,\mathrm{AdS}_{d-1}\times S^1}
=
\frac{a_{\mO}(h)}{R^\Delta}
e^{i q_a\alpha^a(h)\theta}\,,
\end{equation}
where here and in what follows we neglect the dependence on the dimensionless moduli $\bar{\phi}^A$ and $\bar{\pi}^a$.
In the macroscopic limit physical distances are kept fixed. In particular, introducing the dimensionful coordinate $y=R\theta$ along the circle, we take $R\rightarrow\infty$ at fixed $y$, so that
$\theta\rightarrow 0$ and the phase disappears.
Denoting by $c_{\mO}$ the dimensionless expectation value on the moduli space,
\begin{equation}
\langle \mO_{\Delta,q_a}\rangle_{\rm vac}
=
c_{\mO}v^\Delta\,,
\end{equation}
the macroscopic limit imposes that the DCFT one-point coefficient satisfies
\begin{equation}\label{eq_macro_scalar_prediction}
\lim_{h\rightarrow\infty}
h^{-\Delta}a_{\mO}(h)
=
c_{\mO}\,.
\end{equation}
Thus the leading large-$h$ one-point coefficients of scalar operators are fixed by their expectation values in the corresponding vacuum on the moduli space. 
Similarly, since Poincar\'e invariance of the vacuum implies that operators transforming nontrivially under the Lorentz group have vanishing expectation value in the moduli space, the macroscopic limit implies that
\begin{equation}\label{eq_macro_spin_prediction}
\lim_{h\rightarrow\infty}
h^{-\Delta}a_{\mO_{\Delta,J}}(h)=0\,,
\qquad J>0\,.
\end{equation}
These considerations naturally agree with the EFT result~\eqref{eq_EFT_1pt_general}, which provides a more precise prediction. 

The existence of a macroscopic limit is more constraining for higher-point correlators. As an illustration, following the treatment of heavy-light-light-heavy correlators in~\cite{Komargodski:2021zzy,Cuomo:2024fuy}, we consider the two-point function of scalar bulk operators and use it to relate the spectrum of massive particles in the moduli-space vacuum to the dimensions of defect operators.

Consider two scalar primaries $\mO_1$ and $\mO_2$ inserted at $(x_1,\theta_1)$ and $(x_2,\theta_2)$ on $\mathrm{AdS}_{d-1}\times S^1$. As above, the macroscopic limit is defined by~\eqref{eq_vortex_macroscopic_limit} while keeping fixed the dimensionful distances. Explicitly, we consider the limit
\begin{equation}\label{eq_macro_2pt_limit}
R,h\rightarrow\infty\,,
\qquad
v=\frac{h}{R}
\quad\text{fixed}\,,
\qquad
s_{\rm AdS}(x_1,x_2),\,
y\equiv R(\theta_1-\theta_2)
\quad\text{fixed}\,,
\end{equation}
where $s_{\rm AdS}(x_1,x_2)$ denotes the dimensionful geodesic distance between the two points along $\mathrm{AdS}_{d-1}$. In this limit the geometry becomes locally flat, and the separation between the insertions is
\begin{equation}\label{eq_flat_separation_macro}
X^2=s_{\rm AdS}^2(x_1,x_2)+y^2\,.
\end{equation}
The existence of the macroscopic limit then amounts to
\begin{equation}\label{eq_macro_2pt}
\left\langle
\mO_1(x_1,\theta_1)
\mO_2(x_2,\theta_2)
\right\rangle_{h,\mathrm{AdS}_{d-1}\times S^1}
\xrightarrow{\rm macro.}
\left\langle
\mO_1(X)\mO_2(0)
\right\rangle_{v,\mathbb{R}^d}\,.
\end{equation}
It is instructive to formulate this statement in terms of cross ratios. Denote the dimensions and charges of the operators as, respectively, $\Delta_i$ and $q_{ia}$. In flat space their two-point function can be written as
\begin{equation}\label{eq_vortex_2pt_crossratios}
\langle \mO_1(x_1,\theta_1)\mO_2(x_2,\theta_2)\rangle_{h,\mathrm{flat}}=\frac{e^{i\alpha^a(h)(q_{1a}\theta_1+q_{2a}\theta_2)}}{
r_1^{\Delta_1}r_2^{\Delta_2}}
\mathcal G_{12,h}(\chi,\theta_{12})\,,
\qquad
\chi=
\frac{x_{\parallel,12}^2+r_1^2+r_2^2}{r_1r_2}\,,
\end{equation}
where $\theta_{12}=\theta_1-\theta_2$ and $x_{||,12}^2$ denotes the distance in the coordinates parallel to the defect. On $\mathrm{AdS}_{d-1}\times S^1$ we have
$\chi=2\cosh(s_{\rm AdS}/R)$, and~\eqref{eq_macro_2pt} is thus equivalent to
\begin{equation}\label{eq_macro_2pt_crossratios}
\lim_{h\rightarrow\infty}
h^{-(\Delta_1+\Delta_2)}\,
\mathcal G_{12,h}\left(
2\cosh\left(\frac{v s}{h}\right),
\frac{v y}{h}
\right)
=
v^{-(\Delta_1+\Delta_2)}
\left\langle \mO_1(X)\mO_2(0)\right\rangle_{v,\mathbb R^d}\,,
\end{equation}
where $X^2=s^2+y^2$ as above.

The relation~\eqref{eq_macro_2pt_crossratios} connects DCFT data and the spectrum in the broken vacuum. To make this connection explicit, we use the K\"all\'en--Lehmann decomposition to express the moduli space correlator as
\begin{equation}\label{eq_KL_dec}
    \left\langle
\mO_1(X)\mO_2(0)
\right\rangle_{v,\mathbb{R}^d}=c_{\mO_1}c_{\mO_2}v^{\Delta_1+\Delta_2}+\int_0^\infty dm^2\rho_{\mO_1\mO_2}(m^2) G^{(d)}_{m}(|X|)\,,
\end{equation}
where $\rho_{\mO_1\mO_2}(m^2)$ is the spectral density and 
\begin{equation}
G^{(d)}_\mu(s)
=
\frac{1}{(2\pi)^{\frac{d}{2}}}
\left(\frac{\mu}{s}\right)^{\frac{d-2}{2}}
K_{\frac{d-2}{2}}(\mu s)\,,
\end{equation}
is the Euclidean massive propagator.

We now study~\eqref{eq_macro_2pt_crossratios} using the conformal block decomposition in the bulk-to-defect channel~\cite{Billo:2016cpy}:\footnote{Note that the breaking of transverse rotations does not modify the form of the conformal blocks, since the preserved diagonal generator $J+\alpha^a Q_a$ plays the role of the transverse $SO(2)$ generator.}
\begin{equation}\label{eq_defect_channel_macro}
\mathcal G_{12,h}(\chi,\theta)
=
a_{\mO_1}(h)a_{\mO_2}(h)
+
\sum_{\widehat{\mO}\neq \mathds{1}}
B_{12,\widehat{\mO}}(h)\,
\widehat f_{\widehat\Delta,\ell}(\chi,\theta)\,,
\end{equation}
where we separated the identity contribution, $B_{12,\widehat{\mO}}$ denotes the appropriate product of bulk-to-defect OPE coefficients, $\widehat\Delta$ is the exchanged defect operator dimension, and $\ell$ the quantum number under the modified transverse rotation generator. Since the external operators are scalars, only defect primaries which are scalars under the rotation group parallel to the defect contribute. Their conformal blocks factorize as
\begin{equation}\label{eq_defect_block_AdS}
\widehat f_{\widehat\Delta,\ell}(\chi,\theta)
=
e^{i\ell\theta}\,
\chi^{-\widehat\Delta}
{}_2F_1\left(
\frac{\widehat\Delta}{2},
\frac{\widehat\Delta+1}{2};
\widehat\Delta+2-\frac d2;
\frac{4}{\chi^2}
\right)\,.
\end{equation}
Up to a $\widehat\Delta$-dependent normalization, the second factor in~\eqref{eq_defect_block_AdS} is simply the scalar bulk-to-bulk propagator on $\mathrm{AdS}_{d-1}$, with
\begin{equation}
m_{\rm AdS}^2R^2
=
\widehat\Delta(\widehat\Delta-d+2)\,.
\end{equation}

The above considerations make it clear what is the macroscopic limit of the conformal block. To obtain finite dimensionful quantum numbers, we consider families of defect operators such that
\begin{equation}\label{eq_macro_defect_quantum_numbers}
\mu^2
\equiv
\frac{\widehat\Delta(\widehat\Delta-d+2)}{R^2}\,,
\qquad
k\equiv\frac{\ell}{R}
\end{equation}
remain fixed as $R\rightarrow\infty$. This implies $ e^{i\ell\theta}\rightarrow e^{iky}$, while the propagator in hyperbolic space approaches the massive flat-space propagator (see e.g. \cite{Komatsu:2020sag}). Overall, we find
\begin{equation}\label{eq_macro_defect_block}
\widehat f_{\widehat\Delta,\ell}
\left(
2\cosh\frac{s}{R},\frac{y}{R}
\right)
\xrightarrow{\rm macro.}
2\pi^{\frac{d-2}{2}}
R^{\frac{d-2}{2}}
\mu^{-\frac{d-4}{2}}\,
e^{iky}G^{(d-1)}_\mu(s)\,.
\end{equation}

It is then convenient to introduce the following distribution
\begin{equation}\label{eq_macro_defect_density}
\mathcal K_{12}^{(h)}(\mu,k)
=
2\pi^{\frac{d-2}{2}}
R^{\frac{d-2}{2}-\Delta_1-\Delta_2}
\sum_{\widehat{\mO}\neq\mathds{1}}
B_{12,\widehat{\mO}}(h)\,
\mu_{\widehat{\mO}}^{-\frac{d-4}{2}}\,
\delta(\mu-\mu_{\widehat{\mO}})
\delta\left(k-\frac{\ell_{\widehat{\mO}}}{R}\right)\,.
\end{equation}
Using $h=vR$, together with the macroscopic limit of the one-point functions,
\begin{equation}
\lim_{h\rightarrow\infty}
h^{-(\Delta_1+\Delta_2)}
a_{\mO_1}(h)a_{\mO_2}(h)
=
c_{\mO_1}c_{\mO_2}\,,
\end{equation}
the defect-channel decomposition gives
\begin{equation}\label{eq_macro_defect_spectral}
v^{-\Delta_1-\Delta_2}
\left\langle\mO_1(X)\mO_2(0)\right\rangle_{v,\mathbb R^d}
=
c_{\mO_1}c_{\mO_2}
+
v^{-\Delta_1-\Delta_2}
\int_0^\infty d\mu\int_{-\infty}^{+\infty}dk\,
\mathcal K_{12}(\mu,k)\,
e^{iky}G^{(d-1)}_\mu(s)\,,
\end{equation}
where
\begin{equation}
\mathcal K_{12}(\mu,k)
=
\lim_{h\rightarrow\infty}\mathcal K_{12}^{(h)}(\mu,k)\,.
\end{equation}

Finally, using
\begin{equation}\label{eq_d_prop_decomposition}
G^{(d)}_m(X)=
\int_{-\infty}^{+\infty}\frac{dk}{2\pi}
e^{iky}
G^{(d-1)}_{\sqrt{m^2+k^2}}(s)\,.
\end{equation}
in the K\"all\'en--Lehmann representation~\eqref{eq_KL_dec} and changing variables from $m^2$ to $\mu=\sqrt{m^2+k^2}$, we find the following relation
\begin{equation}\label{eq_K_macro}
\mathcal K_{12}(\mu,k)
=
\frac{\mu}{\pi}\,
\Theta(\mu^2-k^2)\,
\rho_{\mO_1\mO_2}(\mu^2-k^2)\,.
\end{equation}
Note that the emergence of $d$-dimensional rotational invariance in the macroscopic limit imposes a nontrivial constraint on the large-$h$ defect spectrum and bulk-to-defect OPE coefficients: their limiting spectral density can depend on $\mu$ and $k$ only through the combination $\mu^2-k^2$, up to the prefactor in~\eqref{eq_K_macro}.

As a simple example, suppose that the spectral density in the broken vacuum contains a stable scalar particle of mass $M$,
\begin{equation}
\rho_{\mO_1\mO_2}(m^2)
\supset
Z_{12}\,\delta(m^2-M^2)\,.
\end{equation}
Its contribution to the defect spectral density is then
\begin{equation}\label{eq_macro_single_particle_density}
\mathcal K_{12}(\mu,k)
\supset Z_{12}\frac{\mu}{\pi}\Theta(\mu^2-k^2)\delta\left(\mu^2-M^2-k^2\right)=
\frac{Z_{12}}{2\pi}\,
\delta\left(\mu-\sqrt{M^2+k^2}\right)\,.
\end{equation}
Therefore the massive particle in the moduli spaces implies that there exists a Regge trajectory of defect operators with quantum numbers scaling as $\widehat\Delta,\ell\sim h$. More precisely, the relativistic dispersion relation implies that, as $h\rightarrow\infty$, the dimensions of the operators in this Regge trajectory satisfy
\begin{equation}\label{eq_macro_dispersion_relation}
\lim_{\substack{h\rightarrow\infty\\
\ell/h=\kappa}}
\frac{\widehat\Delta(h,\ell)}{h}
=
\sqrt{\kappa^2+\frac{M^2}{v^2}}\,,
\end{equation}
where $\kappa$ physically represents the momentum measured in units of the vev $v$.
Additionally, assuming the massive particle is encoded in a single defect operator for each $\ell$, comparison with~\eqref{eq_macro_defect_density} also fixes the asymptotic bulk-to-defect OPE coefficients. Defining the dimensionless residue
\begin{equation}
z_{12}\equiv
Z_{12}v^{d-2-\Delta_1-\Delta_2}\,,
\end{equation}
one finds that for $h\rightarrow\infty$
\begin{equation}
B_{12,\widehat{\mO}_\ell}(h)
\sim
\frac{z_{12}}{4\pi^{d/2}}\,
h^{(\Delta_1+\Delta_2)-\frac{d}{2}}
\left(
\kappa^2+\frac{M^2}{v^2}
\right)^{\frac{d-4}{4}}\,.
\end{equation}
Hence both the mass and the residue of a particle in the moduli-space vacuum are encoded in the DCFT data at large $h$.

In summary, the spectrum of particles in the moduli space is related to large-$h$ limit of DCFT data. In particular, the bulk-to-defect conformal block expansion of the correlator~\eqref{eq_defect_channel_macro} reduces to the K\"all\'en-Lehmann decomposition~\eqref{eq_KL_dec} in the macroscopic limit. Note that both the defect and the moduli space correlator can be also expressed as infinite sum using the OPE between the two bulk fields. This implies that the moduli space bootstrap equation studied in~\cite{Karananas:2017zrg,Cuomo:2024vfk}, relating OPE and  K\"all\'en-Lehmann decomposition in the moduli space, is the macroscopic limit of the bootstrap equation for bulk two-point functions in the presence of a defect~\cite{Liendo:2012hy,Billo:2016cpy}.

\section{Outlook}\label{sec_conclusions}

In this paper we constructed a general class of vortex defects in CFTs with moduli spaces, whose DCFT data encode the physics of the broken vacuum in the macroscopic limit described in sec.~\ref{sec_macroscopic_limit}. As a byproduct, the moduli space EFT provides a convenient tool to study the corresponding DCFT data, see sec.~\ref{sec_moduli_defects}. Together with previous works~\cite{Hellerman:2017veg,Karananas:2017zrg,Cuomo:2024vfk,Cuomo:2024fuy,Ivanovskiy:2024vel}, our construction provides another illustration of the fact that the existence of a moduli space is reflected nontrivially in the spectrum of both local and nonlocal conformal operators, directly in the conformal vacuum. Our main results were summarized in greater detail in sec.~\ref{Sec_Intro}. Below we discuss a few directions for future work.

As explained in the main text, the EFT methodology developed in this work closely parallels the large-charge expansion in SCFTs, which has found several applications in recent years, see e.g.~\cite{Hellerman:2017veg,Hellerman:2017sur,Hellerman:2018xpi,Grassi:2019txd,Sharon:2020mjs}. It is therefore natural to expect that our methods will provide new insights into strongly coupled conformal defects, particularly in supersymmetric theories. In particular, we expect that the moduli space EFT can be used to extract further nontrivial information about Gukov-Witten surface operators, including non-protected observables, as well as vortex loops in ABJM theory, beyond the results obtained in secs.~\ref{sec_GW} and~\ref{sec_ABJM_vortex}. It might be particularly interesting to compare the EFT predictions with the chiral-algebra results for Gukov-Witten operators~\cite{Beem:2013sza,Cordova:2017mhb,Bianchi:2019sxz,Bianchi:2026ukq}. We also expect that the construction of the half-BPS vortex loop in the $3d$ $\mathcal{N}=2$ XYZ model of sec.~\ref{subsec_XYZ} can be generalized to other theories with at least four supercharges. 

The same construction applies to the $6d$ $\mathcal{N}=(2,0)$ theories, whose tensor branch plays the role of the Coulomb branch of $\mathcal{N}=4$ SYM and has been studied with the moduli-space EFT at large charge~\cite{Heckman:2024erd}. We expect to find BPS codimension-two vortex defects in these theories, in direct analogy with the Gukov-Witten operators of $4d$ $\mathcal{N}=4$ SYM discussed in sec.~\ref{sec_GW}. Half-BPS codimension-two defects of the $A_{N-1}$ theory labelled by a Levi subgroup $S(U(N_1)\times\cdots\times U(N_n))$ have been studied extensively, both through class-$\mathcal{S}$ punctures~\cite{Chacaltana:2012zy} and holographically~\cite{Capuozzo:2023fll}, and reduce to these Gukov-Witten operators for the compactified theory on $T^2$. It would be interesting to identify the vortex defects predicted by the EFT among these objects, and to compute their defect Weyl anomalies from the EFT as we did in $4d$ in sec.~\ref{sec_GW}.

More generally, it is natural to ask whether the analytic continuation described in sec.~\ref{sec_vortices_general_EFT} can be promoted to an exact dictionary relating observables of vortex defects to those of large-charge operators, allowing results in either setting to be translated into the other.

The simplest bootstrap equation for CFTs with moduli spaces, studied in~\cite{Karananas:2017zrg,Cuomo:2024vfk}, relates the OPE expansion of two-point functions of local CFT operators in the broken vacuum to their K\"all\'en-Lehmann spectral decomposition. As discussed in sec.~\ref{sec_macroscopic_limit}, this equation can be viewed as the macroscopic limit of the defect crossing equation equating the bulk OPE and bulk-to-defect OPE expansions of two-point functions~\cite{Liendo:2012hy,Billo:2016cpy}. On the other hand, scattering amplitudes and form factors in the broken vacuum are subject to additional constraints from crossing symmetry, unitarity, and analyticity, as explored in the S-matrix and form factor bootstrap~\cite{Karateev:2019ymz,Kruczenski:2022lot}. These constraints are expectedly related to those obeyed by four-point functions of defect operators and correlators involving both bulk and defect operators. It would be interesting to understand this map in detail.

Our results also motivate a defect analogue of the CFT distance conjecture~\cite{Perlmutter:2020buo}. For the supersymmetric vortices of sec.~\ref{sec_SUSY_vortices}, where $h$ is exactly marginal, the moduli-space EFT implies that the limit $h\to\infty$ lies at infinite distance on the defect conformal manifold, and that in this limit the defect operators associated with the light fields on the moduli space become generalized free fields: their correlation functions approach those of free fields on $\mathrm{AdS}_{d-1}\times S^1$, with corrections suppressed by inverse powers of $h$. We therefore conjecture that every infinite-distance limit on a conformal manifold of defects is accompanied by a sector of defect operators whose correlation functions become Gaussian. Note that the Gaussian sector may also consist of defect degrees of freedom which are decoupled from the bulk.
The moduli space EFT establishes this conjecture
for the limit $h\to\infty$ of all the supersymmetric defects constructed in this work, and it would be interesting to establish it 
for general extended operators
(a defect version of the CFT distance conjecture was also mentioned in~\cite{Diatlyk:2026gab}, 
motivated by interfaces
in the Gross-Neveu-Yukawa model).

Finally, perhaps the most important open question is: can the consistency conditions obeyed by the conformal defects studied in this work be used to constrain the dynamics on the moduli space? Answering this question, even in specific strongly coupled examples such as the $XYZ$ model studied in sec.~\ref{subsec_XYZ}, would be very valuable. Recent progress on the bootstrap problem for defects seems promising in this direction~\cite{Gabai:2023lax,Lanzetta:2025xfw}. A similar question arises for the large-charge operators studied in~\cite{Cuomo:2024fuy}, although in that context the controlling parameter $Q$ is discrete rather than associated with a continuous (quasi-)marginal deformation.

\section*{Acknowledgments}
We thank F.~Apruzzi, F.~Benini, M.~Bertolini, L.~Bianchi, T.~Dumitrescu, F.~Galvagno, S.~Harder, Z.~Komargodski, D.~Orlando, A.~Sever, and Y.~Wang for useful discussions. We are especially grateful to L.~Rastelli for discussions on related topics that inspired this work. We acknowledge the use of ChatGPT (OpenAI) and Claude (Anthropic) to assist with technical calculations and proofreading of this work. A.S. is supported by the Mani L. Bhaumik Institute for Theoretical Physics.

\appendix

\section{Details on supersymmetry}

\subsection{Conventions for the \texorpdfstring{$4d$}{4d} \texorpdfstring{$\mathcal{N}=1$}{N=1} superalgebra}\label{app:conventions_SUSY}

The 4d $\mathcal{N}=1$ superconformal algebra in Euclidean signature is
\begin{equation}
\begin{split}
    \{Q_\alpha,\bar Q_{\dot\beta}\}
&=2(\sigma^\mu)_{\alpha\dot\beta}P_\mu,
\\
\{\bar S^{\dot\alpha},S^\beta\}
&=2(\bar\sigma^\mu)^{\dot\alpha\beta}K_\mu,
\\
\{Q_\alpha,S^\beta\}
&=2\delta_\alpha{}^\beta D
+2(\sigma^{\mu\nu})_\alpha{}^\beta M_{\mu\nu}
+3\,\delta_\alpha{}^\beta R,
\\
\{\bar Q_{\dot\alpha},\bar S^{\dot\beta}\}
&=2\delta_{\dot\alpha}{}^{\dot\beta} D
-2(\bar\sigma^{\mu\nu})^{\dot\beta}{}_{\dot\alpha} M_{\mu\nu}
-3\,\delta_{\dot\alpha}{}^{\dot\beta} R .
\end{split}
\end{equation}
We use the basis
\begin{equation}
    \sigma^\mu = (i\sigma^1,i\sigma^2,i\sigma^3,I)\;,\quad 
    \bar\sigma^\mu = (-i\sigma^1,-i\sigma^2,-i\sigma^3,I)
\end{equation}
with $\sigma^{\mu\nu}=\frac14\left(\sigma^\mu\bar\sigma^\nu-\sigma^\nu\bar\sigma^\mu\right)$ and $\bar\sigma^{\mu\nu}=\frac14\left(\bar\sigma^\mu\sigma^\nu-\bar\sigma^\nu\sigma^\mu\right)$.

The components $\phi,\psi_\alpha,F$ of a chiral multiplet transform as
\begin{align}\label{eq:Q_transfs}
Q_\alpha \phi &= \sqrt{2}\,\psi_\alpha,
&
\bar Q_{\dot\alpha}\phi &=0,
\notag\\
Q_\alpha \psi_\beta &= \sqrt{2}\,\epsilon_{\alpha\beta}F,
&
\bar Q_{\dot\alpha}\psi_\beta
&=
\sqrt{2}\,(\sigma^\mu)_{\beta\dot\alpha}\partial_\mu\phi,
\\
Q_\alpha F&=0,
&
\bar Q_{\dot\alpha}F
&=
\sqrt{2}\,(\bar\sigma^\mu)_{\dot\alpha}{}^{\beta}\partial_\mu\psi_\beta.\notag
\end{align}

A chiral superconformal primary also has the transformation rules
\begin{align}\label{S_transfs}
S^\alpha \phi &= 0,
&
\bar S^{\dot\alpha}\phi
&=
\sqrt{2}\,x_\mu(\bar\sigma^\mu)^{\dot\alpha\beta}\psi_\beta,\notag
\\
S^\alpha \psi_\beta
&=
3\sqrt{2}\,R\,\delta^\alpha{}_\beta\,\phi
+
\sqrt{2}\,
\bigl(x_\mu \bar\sigma^\mu \sigma^\nu\bigr)^\alpha{}_\beta\,\partial_\nu\phi,
&
\bar S^{\dot\alpha}\psi_\beta
&=
-\sqrt{2}\,x_\mu(\bar\sigma^\mu)^{\dot\alpha}{}_\beta\,F,
\\
S^\alpha F
&=
-\sqrt{2}\,(3R-2)\,\psi^\alpha
+
\sqrt{2}\,x_\mu(\bar\sigma^\mu)^{\dot\beta\alpha}
(\bar\sigma^\nu)_{\dot\beta}{}^{\gamma}\,\partial_\nu\psi_\gamma,
&
\bar S^{\dot\alpha}F &= 0.\notag
\end{align}
and obeys
\begin{equation}\label{eq:BPS_chiral}
\Delta_\Phi=\frac{3}{2}R_\Phi\,.
\end{equation}

\subsection{BPS condition on vortex defects}\label{app:BPS_condition}

We perform the computation for $4d$ $\mathcal{N}=1$ chiral superfields. The calculation for $3d$ $\mathcal{N}=2$ case is similar and we discuss some aspects of it at the end.

Take a $4d$ $\mathcal{N}=1$ superconformal chiral multiplet $\Phi$ with dimension $\Delta = \frac{3}{2}R$ and components $\phi,\psi,F$. Set the background values
\begin{equation}\label{eq:bg}
    \phi = \frac{\lambda}{z^\Delta}\;,\quad \psi = F =0\;.
\end{equation}
Here we have introduced a codimension 2 defect along the directions $x_3,x_4$ and $z=x_1+ix_2$ is the complex coordinate in the transverse plane. We also define $w=x_3+ix_4$ to be the complex coordinate along the defect. 
We now argue that this configuration preserves a complex supercharge $Q$ (so that it is 1/2 BPS) and also classically preserves a complex superconformal charge $S$. This is done by showing that the transformations \eqref{eq:Q_transfs},\eqref{S_transfs} vanish for a set of supercharges.

We start with the preserved $Q$. On the background \eqref{eq:bg}, the only nontrivial equation is
\begin{equation}
    \bar Q_{\dot\alpha}\psi_\beta
=
\sqrt{2}(\sigma^\mu)_{\beta\dot\alpha}\partial_\mu\phi=0\;.
\end{equation}
and its conjugate. On this background this becomes
\begin{equation}
\sqrt{2}(\sigma^z)_{\beta\dot\alpha}\partial_z\phi=0\;,
\end{equation}
where $\sigma^z=\sigma^1+i\sigma^2=\begin{pmatrix}
    0 & 2i \\ 0 &0
\end{pmatrix}$. Then the supercharge $\bar{Q}_{\dot{1} }$ is preserved. Similarly the conjugate equation
\begin{equation}
    Q_\alpha\bar{\psi}_{\dot\beta}=\sqrt 2(\sigma^\mu)_{\alpha\dot\beta}\partial_\mu\bar\phi=0
\end{equation}
tells us that $Q_1$ is preserved as well. 

Similarly for the superconformal charges, the only nontrivial equation to check is 
\begin{equation}
    S^\alpha \psi_\beta
=
3\sqrt{2}\,R\,\delta^\alpha{}_\beta\,\phi+\sqrt{2}\,
\bigl(x_\mu \bar\sigma^\mu \sigma^\nu\bigr)^\alpha{}_\beta\,\partial_\nu\phi=0\;,
\end{equation}
plugging in the background this becomes
\begin{equation}
3\,R\,\delta^\alpha{}_\beta\,\phi+\,
\bigl(x_\mu \bar\sigma^\mu \sigma^z\bigr)^\alpha{}_\beta\,\partial_z\phi=0\;.
\end{equation}
An explicit computation gives
\begin{equation}
    (x_\mu \bar\sigma^\mu \sigma^z\bigr)^\alpha{}_\beta=2\begin{pmatrix}
        z & 0\\ -\bar{w} & 0
    \end{pmatrix}
\end{equation}
We cannot preserve $S^2$, but using 
\begin{equation}
    z\partial_z\phi =-\Delta \phi
\end{equation}
on the background, the equation for $S^1$ becomes
\begin{equation}
    (3R-2\Delta)\phi=0
\end{equation}
which is obeyed due to \eqref{eq:BPS_chiral}. We thus preserve $S^1$, and a similar calculation for the conjugate equation
\begin{equation}
\bar S^{\dot\alpha}\,\bar\psi_{\dot\beta}
=
3\sqrt{2}\,R\,\delta^{\dot\alpha}{}_{\dot\beta}\,\bar\phi
+
\sqrt{2}\,\bigl(x^\mu \sigma_\mu \bar\sigma^\nu\bigr)^{\dot\alpha}{}_{\dot\beta}\,\partial_\nu \bar\phi
\end{equation}
tells us that $\bar{S}^{\dot{1}}$ is also preserved. 

To summarize, in our conventions the background \eqref{eq:bg} preserves two real supercharges $Q_1,\bar{Q}_{\dot 1}$ and two superconformal partners $S^1,\bar{S}^{\dot 1}$. The algebra of preserved supercharges is
\begin{equation}
\begin{split}
    \{Q_1,\bar Q_{\dot 1}\}&=2P_w,
\\
\{S^{1},\bar S^{\dot 1}\}&=2K_w, \\
\{Q_1,S^{1}\}&=2D+2i(M_{12}+M_{34})+3R,\\
\{\bar Q_{\dot 1},\bar S^{\dot 1}\}&=2D-2i(M_{12}-M_{34})-3R,
\end{split}
\end{equation}
Thus the defect preserves a global $su(1,1|1)$ superconformal algebra (there is also a right-moving sector which preserves the bosonic algebra $su(1,1)$). In particular, we have $\mathcal{N}=(2,0)$ SUSY on the defect, and we can read off the R-symmetry on the defect:
\begin{equation}
    r=\frac32R-J\;,
\end{equation}
where $J=-iM_{12}$ is the Hermitian generator of rotations around the defect, as in the main text.

The discussion for the $3d$ $\mathcal{N}=2$ case is similar. The algebra is also $su(1,1|1)$, and in this case the preserved defect R-symmetry is 
\begin{equation}
    r=R-J
\end{equation}
where $J$ generates rotations around the defect.

The background \eqref{eq:bg} also fixes the expectation values of all the chiral-ring operators $\Phi^n$ generated by $\Phi$. To show this, work again in $4d$ and consider $n$ insertions of the bottom component $\phi$ at the same transverse position $z$ and at separated positions $w_i$ along the defect,
\begin{equation}
    G_n(w_1,\ldots,w_n;z)
    =
    \left\langle
    \phi(w_1,z)\cdots\phi(w_n,z)
    \right\rangle\,.
\end{equation}
Here $w=x_3+ix_4$ is the complex coordinate along the defect, and translations in $w$ are generated by the anticommutator of two preserved supercharges,
\begin{equation}\label{eq_app_ac_P}
    P_w=\frac12\{Q_1,\bar Q_{\dot 1}\}\,.
\end{equation}
Since $\bar Q_{\dot 1}$ annihilates the chiral primary, it is then easy to show that $G_n$ is independent of the $w_i$'s. Indeed, using \eqref{eq_app_ac_P}, we can write
\begin{equation}
\begin{split}
    \partial_{w_i}G_n
    &\propto
    \left\langle
    \phi_1\cdots
    [\{Q_1,\bar Q_{\dot 1}\},\phi_i]
    \cdots\phi_n
    \right\rangle
    \\
    &=
    \left\langle
    \phi_1\cdots
    \{\bar Q_{\dot 1},[Q_1,\phi_i]\}
    \cdots\phi_n
    \right\rangle=0\,,
\end{split}
\end{equation}
where in the second line we used the super-Jacobi identity together with the chirality condition $[\bar Q_{\dot 1},\phi_i]=0$, while in the last step we used the supersymmetric Ward identity to move $\bar Q_{\dot 1}$ through the correlator. Its action on all the remaining insertions vanishes by chirality, while its action on the defect vanishes because $\bar Q_{\dot 1}$ is preserved.

This shows that $G_n$ can only depend on the $\bar w_i$'s. However, since the OPE of chiral primaries is nonsingular, this dependence has no singularities at coincident points. Moreover, cluster decomposition implies that the correlator is bounded as the insertions are taken to infinity along the defect. By Liouville's theorem, $G_n$ must therefore be independent of the $\bar w_i$'s as well.

We may therefore take the separations $|w_i-w_j|$ to be arbitrarily large. Cluster decomposition then gives
\begin{equation}
    G_n(w_1,\ldots,w_n;z)
    =
    \prod_{i=1}^n
    \left\langle\phi(w_i,z)\right\rangle
    =
    \left(\frac{\lambda}{z^\Delta}\right)^n\,.
\end{equation}
Since the left-hand side is independent of the $w_i$, this relation holds for arbitrary separated insertion points. Since the OPE of two chiral primaries is nonsingular, we can then take the coincident-point limit, with the standard normalization
\begin{equation}\label{eq_app_nonsingular_chiral_OPE}
    \phi^n(w,z)=
    \lim_{w_i\rightarrow w}
    \phi(w_1,z)\cdots\phi(w_n,z)\,.
\end{equation}
Taking this limit in the previous result gives
\begin{equation}\label{eq:chiral_VEV_factorization}
    \left\langle\phi^n(w,z)\right\rangle
    =
    \left\langle\phi(w,z)\right\rangle^{\,n}
    =
    \frac{\lambda^n}{z^{n\Delta}}\,.
\end{equation}
The argument is the defect analogue of the standard proof of position independence of extremal correlators; see e.g. \cite{Eberhardt:2020cxo}.

The same argument applies to the $3d$ $\mathcal{N}=2$ case. Indeed, the preserved supercharges also obey $\{\mathcal Q,\overline{\mathcal Q}\}\propto P_t$, where $t$ is the coordinate along the defect and $\overline{\mathcal Q}$ annihilates chiral primaries. The supersymmetric Ward identity thus implies that correlators of chiral primaries are independent of their positions along the defect. Using cluster decomposition and the regularity of the OPE between chiral fields as in~\eqref{eq_app_nonsingular_chiral_OPE}, we obtain again the result \eqref{eq:chiral_VEV_factorization}.

\subsection{The supersymmetric tilt and its partner}\label{app:susy_tilt}

Here we derive the statements of section~\ref{subsec_BPS}. We consider line defects in $d=3$, with coordinate $y$ along the line, and comment on surface defects in $d=4$ at the end.

The preserved superconformal algebra $su(1,1|1)$ contains the supercharges $\mathcal Q,\overline{\mathcal Q}$, their superconformal partners, the conformal generators along the defect and the $u(1)$ generator $r$. As in appendix~\ref{app:BPS_condition}, $\overline{\mathcal Q}$ annihilates the bulk chiral primary $\phi$. We normalize
\begin{equation}
    \{\mathcal Q,\overline{\mathcal Q}\}=2P\,,\qquad [r,\overline{\mathcal Q}]=\frac12\overline{\mathcal Q}\,,\qquad [r,\mathcal Q]=-\frac12\mathcal Q\,,
\end{equation}
where $P$ is the translation along the defect. Unitarity requires $\Delta\geq|r|$ for superconformal primaries, which are annihilated by $\mathcal Q$ if and only if $\Delta=-r$, and by $\overline{\mathcal Q}$ if and only if $\Delta=r$ \cite{Agmon:2020pde}.

We are interested in defects which break the $R$-symmetry, yielding a tilt operator as in~\eqref{eq_tilt_def}. We assume that the action of $R$ maps the preserved supercharges and their superconformal partners into themselves, $[R,\mathcal Q]\propto\mathcal Q$ and similarly for the other generators. Consequently, the defect rotated by $R$ preserves the same superconformal algebra as the original one. In particular, the profile~\eqref{eq:gen_bg_bps} remains BPS as we change $\arg\lambda$. This implies that the integrated tilt is invariant under $su(1,1|1)$, hence $\mathcal Q\mathcal T$ and $\overline{\mathcal Q}\mathcal T$ are total derivatives along the defect.\footnote{Formally, let $\mathcal F$ be any preserved fermionic generator and let $[X,\mathcal D]$ denote the variation of the defect under $X$. The super-Jacobi identity gives
\begin{equation}
[\mathcal F,[R,\mathcal D]] =[[\mathcal F,R],\mathcal D]+[R,[\mathcal F,\mathcal D]]=0\,,
\end{equation}
where we used $[\mathcal F,R]\propto\mathcal F$ and $[\mathcal F,\mathcal D]=0$. Since $[R,\mathcal D]$ is the integrated tilt, the latter preserves the same supercharges as the original defect.}

The tilt is a real conformal primary with $\Delta=1$ and, since $[r,R]=0$, with $r=0$. Supersymmetry implies that it must be an F-term or a D-term. The latter would have the form $\mathcal Q\overline{\mathcal Q}\hat\chi$, where $\hat\chi$ is a superconformal primary with $\Delta_{\hat\chi}=r_{\hat\chi}=0$. However, the unitarity bound implies that such a primary is annihilated by both supercharges, and hence the D-term vanishes. The tilt must therefore be an F-term of the form
\begin{equation}
    \mathcal T=\overline{\mathcal Q}\hat\psi+
    \mathcal Q\bar{\hat\psi}\,,\qquad
    \mathcal Q\hat\psi=0\,,\qquad
    \Delta_{\hat\psi}=-r_{\hat\psi}=\frac12\,,
\end{equation}
where $\hat\psi$ is a fermionic superconformal primary belonging to a short multiplet. It follows that the partner 
\begin{equation}
    \mathcal T'=i\left(\overline{\mathcal Q}\hat\psi-
    \mathcal Q\bar{\hat\psi}\right)
\end{equation}
is an independent real conformal primary with $\Delta=1$ and $r=0$, whose integral is also invariant under $su(1,1|1)$.

We now argue that $\mathcal T'$ is exactly marginal to all orders in conformal perturbation theory following~\cite{Leigh:1995ep,Green:2010da}. Consider the complex F-term deformation
\begin{equation}
\delta S_D=\int dy\,\left(\kappa\,\overline{\mathcal Q}\hat\psi+\bar\kappa\,\mathcal Q\bar{\hat\psi}\right)\,,
\end{equation}
whose real and imaginary parts correspond to $\mathcal T$ and $\mathcal T'$, respectively. In a supersymmetric renormalization scheme, the beta functions of F-term couplings are holomorphic functions of these couplings~\cite{Seiberg:1993vc}.\footnote{This follows by promoting the F-term couplings to background chiral superfields.} Since the deformation by $\mathcal T$ generates $R$-rotations, the beta functions vanish along the corresponding exact family of conformal defects. By holomorphy, the vanishing of the F-term beta functions along the exact family generated by $R$-rotations extends to its complexification. In particular, the deformation by $\mathcal T'$, corresponding to the imaginary part of the complex F-term coupling, does not generate any F-term beta function to all orders in perturbation theory. More generally, one can show that F-term couplings are not renormalized in an appropriate supersymmetric scheme, as a consequence of the nonsingularity of chiral-chiral OPEs~\cite{Green:2010da}.

To complete the argument, we must also exclude possible running induced by D-term counterterms. A marginal D-term would be the top component of a multiplet whose superconformal primary has $\Delta=r=0$. As argued above, unitarity implies that such a primary is annihilated by both supercharges, and hence its supersymmetric descendants vanish. There are thus no nontrivial marginal D-terms for line defects in $d=3$, and we conclude that $\mathcal T'$ is exactly marginal to all orders in conformal perturbation theory.

Note that under the deformation $\int dy\,(\kappa\,\overline{\mathcal Q}\hat{\psi}+\bar\kappa\,\mathcal Q\bar{\hat{\psi}})$, the $\overline{\mathcal Q}$ Ward identity and $\overline{\mathcal Q}\phi=0$ give $\partial_\kappa\langle\phi(x)\rangle=-\int dy\,\langle\overline{\mathcal Q}(\phi(x)\,\hat{\psi}(y))\rangle=0$, so that $\lambda$ is a holomorphic function of $\bar\kappa$. Since the deformation by real $\kappa$ rotates the phase of $\lambda$, holomorphy implies that the deformation by imaginary $\kappa$, corresponding to $\mathcal T'$, changes $|\lambda|$ at fixed $\arg\lambda$, at least infinitesimally. This establishes that the magnitude of the one-point function~\eqref{eq:gen_bg_bps} is associated with an exactly marginal deformation.

Let us now comment on how the argument changes for surface defects in $d=4$. In that case the preserved algebra is $su(1,1|1)\oplus su(1,1)$. The same arguments as above, with $\Delta$ replaced by the weight $h$ under $su(1,1|1)$, establish that $\mathcal T'$ is an F-term, and that deforming the defect by it does not generate beta functions for itself or the other F-terms. The absence of D-term contributions, however, requires additional care. Indeed, a marginal D-term can arise from a superconformal primary with weights $(h,\bar h)=(0,1)$, corresponding to an antiholomorphic conserved current on the defect. If the corresponding symmetry is broken by the deformation, the current multiplet can recombine into a long multiplet, generating a nontrivial D-term beta function~\cite{Leigh:1995ep,Green:2010da}. However, as explained for instance in~\cite{Cuomo:2023qvp}, conserved defect currents generate symmetries that act trivially on bulk operators, and hence on all operators appearing in their bulk-to-defect OPE. In particular, the tilt $\mathcal T$ is neutral under such symmetries, and so is its partner $\mathcal T'$, since they appear in the bulk-to-defect OPE of the R-current. Therefore, the deformation cannot break these symmetries, excluding the corresponding D-term running. We conclude that $\mathcal T'$ is exactly marginal also for surface defects in $d=4$.

As a check, consider a free $4d$ $\mathcal N=1$ chiral multiplet, with $R_\Phi=\frac23$ and $\Delta_\Phi=1$. Since $\lambda/z$ is single valued, the vortex is defined by the local defect action $S_D=\int d^2y\,(a\,\partial_z\bar\phi+\bar a\,\partial_{\bar z}\phi)$, which gives $\langle\phi\rangle=\lambda/z$ with $\lambda\propto a$. By \eqref{eq:Q_transfs}, $\partial_{\bar z}\phi=\overline{\mathcal Q}\hat{\psi}_0$ with $\hat{\psi}_0\propto\psi_2$ and $\mathcal Q\hat{\psi}_0=0$ on-shell, so $S_D$ is an F-term. Using that $R_{\Phi}=\frac23$, the tilt and its partner are found to be
\begin{equation}
    \mathcal{T}=\frac{2i}{3}\left(\bar a\,\partial_{\bar z}\phi-a\,\partial_z\bar\phi\right),\qquad
    \mathcal{T}'=-\frac23\left(\bar a\,\partial_{\bar z}\phi+a\,\partial_z\bar\phi\right),
\end{equation}
in agreement with \eqref{eq_susy_tilt} and \eqref{eq_tilt_partner} for $\hat{\psi}=\frac{2i}{3}\bar a\,\hat{\psi}_0$. The tilt rotates the phase of $a$, while $\mathcal{T}'$ rescales $a$; both are exactly marginal, since $S_D$ is linear in the free fields.

As an example of how the above arguments fail when the broken $R$-charge does not map the preserved supercharges to themselves, consider the  $1/2$-BPS Maldacena--Wilson line $W=\operatorname{Tr}\mathcal P\exp\int dy\,(iA_y+\Phi_6)$ in $\mathcal N=4$ SYM. This breaks $SO(6)_R$ to $SO(5)$. In ten-dimensional notation its preserved supercharges are eigenspinors of $\Gamma^y\Gamma^6$, while a broken generator $R_{a6}$ acts on spinors as $\frac12\Gamma^a\Gamma^6$, which anticommutes with $\Gamma^y\Gamma^6$. Hence $[\mathcal Q,R_{a6}]$ is a broken supercharge, and the tilt deformation does not preserve the original supersymmetry. Accordingly, the tilts $\Phi_a$ \cite{Drukker:2022pxk} are superconformal primaries of the displacement multiplet \cite{Liendo:2018ukf}, rather than F-terms, and no marginal partner is required. Indeed, $\Phi_6$, which changes the magnitude of the scalar coupling, is irrelevant on the line \cite{Polchinski:2011im}.

\section{EFT calculation of the Gukov--Witten conformal anomalies}\label{app_GW_WZ}

In this appendix we evaluate explicitly the supersymmetric Wess--Zumino action on the Gukov--Witten background and extract the defect Weyl anomaly $b$, as well as the quantum correction to $d_2$. We first perform the calculation of $b$ on $S^4$ and then reproduce the result directly in flat space. We finally compute the contribution of the Wess-Zumino action to the stress tensor one-point function to extract $d_2$.

On a conformally flat Euclidean background, and in the absence of background $R$-symmetry gauge fields, the complete bosonic $\tau,\chi$ sector of the supersymmetric Wess--Zumino action relevant for our calculation can be written as~\cite{Hellerman:2017sur,Hellerman:2018xpi}
\begin{equation}\label{eq_app_full_WZ}
S_{\rm WZ}^{E}
=
\frac{\Delta a}{16\pi^2}
\int d^4x\sqrt{g}\,
\mathcal{I}(\tau,\chi)\,,
\end{equation}
where
\begin{equation}\label{eq_app_full_WZ_integrand}
\begin{split}
\mathcal{I}(\tau,\chi)
={}&
\tau E_4
+
4\left(
\mathcal{R}_{\mu\nu}
-\frac{1}{2}\mathcal{R}g_{\mu\nu}
\right)
\nabla^\mu\tau\nabla^\nu\tau
-
4(\nabla\tau)^2\nabla^2\tau
+
2(\nabla\tau)^4
\\
&
+
4\left(
\mathcal{R}_{\mu\nu}
-\frac{1}{6}\mathcal{R}g_{\mu\nu}
\right)
\nabla^\mu\chi\nabla^\nu\chi
+
8(\nabla_\mu\nabla_\nu\tau)
\nabla^\mu\chi\nabla^\nu\chi
\\
&
+
4\left[
2(\nabla\tau\cdot\nabla\chi)^2
-
(\nabla\tau)^2(\nabla\chi)^2
\right]
+
2(\nabla\chi)^4\,,
\end{split}
\end{equation}
and
\begin{equation}
E_4
=
\mathcal{R}_{\mu\nu\rho\sigma}^2
-
4\mathcal{R}_{\mu\nu}^2
+
\mathcal{R}^2\,.
\end{equation}

\subsection{Calculation on \texorpdfstring{$S^4$}{S4}}

We use coordinates adapted to a maximal two-sphere,
\begin{equation}\label{eq_app_S4_metric}
ds_{S^4}^2
=
R^2\left(
d\vartheta^2
+
\cos^2\vartheta\,d\Omega_2^2
+
\sin^2\vartheta\,d\psi^2
\right)\,,
\qquad
0\leq\vartheta\leq\frac{\pi}{2}\,,
\end{equation}
where the Gukov--Witten defect is located at $\vartheta=0$. Up to an irrelevant constant normalization, the classical solution yields
\begin{equation}\label{eq_app_tau_chi_sphere}
\tau
=
\log\sin\vartheta+\tau_0\,,
\qquad
\chi
=
\psi+\chi_0\,.
\end{equation}

For the round sphere one has
\begin{equation}
\mathcal{R}_{\mu\nu}
=
\frac{3}{R^2}g_{\mu\nu}\,,
\qquad
\mathcal{R}
=
\frac{12}{R^2}\,,
\qquad
E_4
=
\frac{24}{R^4}\,,
\end{equation}
while the profile~\eqref{eq_app_tau_chi_sphere} gives
\begin{equation}
(\nabla\tau)^2
=
\frac{\cot^2\vartheta}{R^2}\,,
\qquad
\nabla^2\tau
=
-\frac{3}{R^2}\,,
\qquad
(\nabla\chi)^2
=
\frac{1}{R^2\sin^2\vartheta}\,,
\end{equation}
and
\begin{equation}
\nabla\tau\cdot\nabla\chi
=
0\,,
\qquad
(\nabla_\mu\nabla_\nu\tau)
\nabla^\mu\chi\nabla^\nu\chi
=
\frac{\cos^2\vartheta}{R^4\sin^4\vartheta}\,.
\end{equation}
The Euler term is integrable at the defect and has been computed in the main text. In particular, it does not contribute to a logarithm of the cutoff. Substituting the remaining terms in~\eqref{eq_app_full_WZ_integrand}, we find
\begin{equation}
\mathcal{I}_{\rm non\text{-}Euler}
=
\frac{2}{R^4}
\frac{\cos^4\vartheta+3}{\sin^4\vartheta}\,.
\end{equation}

Using that
\begin{equation}
\sqrt{g}=
R^4\cos^2\vartheta\sin\vartheta
\sqrt{g_{S^2}}\,,
\end{equation}
and a cutoff $\vartheta\geq\vartheta_0$ around the defect, the profile-dependent contribution to the action is therefore
\begin{equation}\label{eq_app_WZ_sphere_integral}
S_{\rm WZ,def}^{E}
=
\Delta a
\int_{\vartheta_0}^{\pi/2}
d\vartheta\,
\frac{
\cos^2\vartheta
\left(
\cos^4\vartheta+3
\right)
}{
\sin^3\vartheta
}\,.
\end{equation}
Near the defect the integrand behaves as
\begin{equation}
\frac{
\cos^2\vartheta
\left(
\cos^4\vartheta+3
\right)
}{
\sin^3\vartheta
}
=
\frac{4}{\vartheta^3}
-
\frac{4}{\vartheta}
+
O(\vartheta)\,.
\end{equation}
The proper distance from the defect is
\begin{equation}
\epsilon
=
R\vartheta_0\,,
\end{equation}
and hence
\begin{equation}
S_{\rm WZ,def}^{E}
=
2\Delta a\frac{R^2}{\epsilon^2}
-
4\Delta a
\log\left(\frac{R}{\epsilon}\right)
+
O(1)\,.
\end{equation}
The power divergence can be removed by a cosmological constant counterterm on the defect worldvolume. The logarithmic term in the defect partition function determines the $b$-function~\eqref{eq_GW_b_result}.

\subsection{Calculation in flat space}

The same universal coefficient can be extracted while keeping the ambient metric flat. We consider a spherical defect of radius $R$, located at $x_4=0$ and centered around the origin. It is convenient to introduce toroidal coordinates
\begin{equation}
x_1^2+x_2^2+x_3^2
=
\frac{R^2\sinh^2\rho}{D^2}\,,
\qquad
x_4=
\frac{R\sin\psi}{D}\,,
\end{equation}
where
\begin{equation}
D=\cosh\rho+\cos\psi\,.
\end{equation}
In these coordinates the flat metric takes the form
\begin{equation}\label{eq_app_flat_toroidal_metric}
ds_{\mathds{R}^4}^2
=
\frac{R^2}{D^2}
\left(
d\rho^2
+
\sinh^2\rho\,d\Omega_2^2
+
d\psi^2
\right)\,.
\end{equation}
Note that the above coincides with the AdS$_{d-1}\times S^1$ metric in global coordinates up to the conformal prefactor $R^2/D^2$. We will nonetheless work in flat space. The defect is approached as $\rho\rightarrow\infty$. 

The dilaton and Goldstone profiles are
\begin{equation}\label{eq_app_tau_chi_flat}
\tau
=
-\log D+\tau_0\,,
\qquad
\chi
=
\psi+\chi_0\,.
\end{equation}
In flat space, after integrations by parts, the supersymmetric Wess--Zumino functional can be represented as
\begin{equation}\label{eq_app_flat_WZ}
S_{\rm WZ}^{E}
=
\frac{\Delta a}{8\pi^2}
\int d^4x\sqrt{g}\,
\mathcal{B}(\tau,\chi)\,,
\end{equation}
where
\begin{equation}\label{eq_app_flat_B}
\begin{split}
\mathcal{B}(\tau,\chi)
={}&
(\nabla\tau)^4
-
2(\nabla^2\tau)(\nabla\tau)^2
+
2(\nabla^2\tau)(\nabla\chi)^2
-
4(\nabla\tau\cdot\nabla\chi)\nabla^2\chi
\\
&
-
2(\nabla\tau)^2(\nabla\chi)^2
+
4(\nabla\tau\cdot\nabla\chi)^2
+
(\nabla\chi)^4\,.
\end{split}
\end{equation}
For the profile~\eqref{eq_app_tau_chi_flat}, the required contractions are
\begin{equation}
(\nabla\tau)^2
=
\frac{\sinh^2\rho+\sin^2\psi}{R^2}\,,
\qquad
(\nabla\chi)^2
=
\frac{D^2}{R^2}\,,
\qquad
\nabla\tau\cdot\nabla\chi
=
\frac{D\sin\psi}{R^2}\,,
\end{equation}
and
\begin{equation}
\nabla^2\tau
=
-\frac{2D\cos\psi}{R^2}\,,
\qquad
\nabla^2\chi
=
\frac{2D\sin\psi}{R^2}\,.
\end{equation}
Substituting these expressions in~\eqref{eq_app_flat_B}, all terms combine into
\begin{equation}
\mathcal{B}
=
-\frac{4D^2}{R^4}\,.
\end{equation}
Using the metric~\eqref{eq_app_flat_toroidal_metric}, this gives
\begin{equation}
\sqrt{g}\,\mathcal{B}
=
-4
\frac{\sinh^2\rho}{D^2}
\sqrt{g_{S^2}}\,.
\end{equation}
The angular integral can be performed explicitly,
\begin{equation}
\int_0^{2\pi}
\frac{d\psi}{
\left(
\cosh\rho+\cos\psi
\right)^2
}
=
\frac{2\pi\cosh\rho}{\sinh^3\rho}\,.
\end{equation}
It follows that the contribution from a tubular region $\rho_\ast<\rho<\rho_{\rm max}$ is
\begin{equation}
S_{\rm WZ,tube}^{E}
=
-4\Delta a
\int_{\rho_\ast}^{\rho_{\rm max}}
d\rho\,\coth\rho
=
-4\Delta a
\left[
\log\sinh\rho
\right]_{\rho_\ast}^{\rho_{\rm max}}\,.
\end{equation}

At large $\rho$, the proper distance $s$ from the spherical defect behaves as
\begin{equation}
s
=
2Re^{-\rho}
+
O\left(e^{-2\rho}\right)\,.
\end{equation}
A proper-distance cutoff $s=\epsilon$ therefore corresponds to
\begin{equation}
\rho_{\rm max}
=
\log\left(\frac{2R}{\epsilon}\right)
+
O\left(\frac{\epsilon}{R}\right)\,.
\end{equation}
Consequently,
\begin{equation}
S_{\rm WZ,tube}^{E}
=
-4\Delta a
\log\left(\frac{R}{\epsilon}\right)
+
O(1)\,,
\end{equation}
and hence
\begin{equation}
\log\langle\mathcal{D}\rangle_{\rm WZ}
=
4\Delta a
\log\left(\frac{R}{\epsilon}\right)
+
O(1)\,,
\end{equation}
in agreement with the sphere calculation~\eqref{eq_GW_b_result}.

\subsection{Stress tensor one-point function}

We now compute the contribution of the Wess--Zumino action~\eqref{eq_app_full_WZ} to the stress tensor one-point function of the planar defect. We define $T_{mn}=\frac{2}{\sqrt{g}}\frac{\delta S^E}{\delta g^{mn}}$ and expand the metric around flat space, $g_{mn}=\delta_{mn}+h_{mn}$. The Euler term is quadratic in the curvature and does not contribute at linear order in $h_{mn}$. The remaining terms are evaluated on the profile
\begin{equation}
\tau=\log r+\tau_0\,,\qquad\chi=\theta+\chi_0\,,
\end{equation}
where $z=re^{i\theta}$ is the transverse coordinate. Along the defect, the terms in the first line of~\eqref{eq_app_full_WZ_integrand} give $T_{\mu\nu}=-\frac{\Delta a}{8\pi^2}\frac{\delta_{\mu\nu}}{r^4}$, and those in the second and third lines give $T_{\mu\nu}=-\frac{5\Delta a}{24\pi^2}\frac{\delta_{\mu\nu}}{r^4}$. Altogether
\begin{equation}\label{eq_app_T_WZ}
T^{\rm WZ}_{\mu\nu}=-\frac{\Delta a}{3\pi^2}\frac{\delta_{\mu\nu}}{r^4}\,,\qquad
T^{\rm WZ}_{ij}=-\frac{\Delta a}{3\pi^2}\frac{4n_in_j-3\delta_{ij}}{r^4}\,.
\end{equation}
The result is traceless and has the tensor structure required by conformal invariance, which provides a check of the calculation. Comparing with~\eqref{eq_GW_T}, the Wess--Zumino term contributes $12\Delta a$ to $d_2$. With the same conventions, the improved stress tensor of the scalars on the classical profile is $T_{\mu\nu}=-\frac{1}{3g^2}\delta_{\mu\nu}\operatorname{Tr}(\partial_\rho\phi^i)^2$ along the defect, which gives the classical term in~\eqref{eq_GW_d2}.

\section{Useful formulae}

\subsection{Regulated AdS volume}\label{App_AdS_volume}

In this appendix we review the regulated volume of Euclidean AdS$_{d-1}$ and its physical significance. We introduce embedding coordinates for Euclidean AdS$_{d-1}$ through
\begin{equation}
-X_0^2+\sum_{i=1}^{d-1}X_i^2=-R^2\,.
\end{equation}
Global coordinates are defined by
\begin{equation}
X_0=R\cosh\rho\,,
\qquad
X_i=R\hat n_i\sinh\rho\,,
\qquad
\sum_{i=1}^{d-1}\hat n_i^2=1\,,
\end{equation}
in terms of which the metric reads
\begin{equation}
ds^2=R^2\left(d\rho^2+\sinh^2\rho\,d\Omega_{d-2}^2\right)\,.
\end{equation}
The asymptotic boundary is located at $\rho\rightarrow\infty$.

To regulate the volume, we introduce a cutoff at $\rho=\rho_c\gg 1$. The induced metric at the cutoff surface is that of a $(d-2)$-sphere of radius
\begin{equation}
R_{\rm bdry}=R\sinh\rho_c\simeq\frac{R}{2}e^{\rho_c}\,.
\end{equation}
The regulated AdS volume is therefore
\begin{equation}\label{eq_app_AdS_vol_pre}
\begin{split}
R^{1-d}\operatorname{Vol}(\mathrm{AdS}_{d-1})
&=
\Omega_{d-2}
\int_0^{\rho_c}d\rho\,\sinh^{d-2}\rho=
\Omega_{d-2}
\frac{
\sinh^{d-1}\rho_c\,
{}_2F_1\left(
\frac12,\frac{d-1}{2};
\frac{d+1}{2};
-\sinh^2\rho_c
\right)
}{d-1}
\\
&=
\Omega_{d-2}
\left[
\frac{
\Gamma\left(1-\frac d2\right)
\Gamma\left(\frac{d-1}{2}\right)
}{
2\sqrt{\pi}
}+\frac{2^{2-d}e^{(d-2)\rho_c}}{d-2}-\frac{2^{2-d}(d-2)e^{(d-4)\rho_c}}{d-4}+\ldots\right].
\end{split}
\end{equation}
Here
\begin{equation}
\Omega_d=\frac{2\pi^{\frac{d+1}{2}}}{\Gamma\left(\frac{d+1}{2}\right)}
\end{equation}
denotes the volume of the unit $d$-sphere. When one of the denominators in the asymptotic expansion vanishes, the corresponding power divergence is replaced by a term proportional to $\rho_c$, or equivalently by a logarithmic divergence in $R_{\rm bdry}$.

The first term in the last line of~\eqref{eq_app_AdS_vol_pre} defines the dimensionally regulated volume of AdS$_{d-1}$,
\begin{equation}\label{eq_vol_AdS}
R^{1-d}\operatorname{Vol}(\mathrm{AdS}_{d-1})
\big\vert_{\rm reg}
=\Omega_{d-2}
\frac{\Gamma\left(1-\frac d2\right)\Gamma\left(\frac{d-1}{2}\right)}{2\sqrt{\pi}}
=\pi^{\frac d2-1}\Gamma\left(1-\frac d2\right)\,.
\end{equation}
For odd $d$, this expression is finite and isolates the universal constant term in the regulated volume. For even $d$, it has a pole in dimensional regularization, whose residue captures the coefficient of the universal logarithmic divergence. For instance, setting the AdS radius to one, we have
\begin{align}
\operatorname{Vol}(\mathrm{AdS}_{2})\big\vert_{\rm reg}
&=-2\pi\,,\\
\label{eq_app_Vol_AdS3}
\operatorname{Vol}(\mathrm{AdS}_{3})
&=-2\pi\log\left(R_{\rm bdry}\right)+\ldots\,,\\
\operatorname{Vol}(\mathrm{AdS}_{4})\big\vert_{\rm reg}
&=\frac{4\pi^2}{3}\,.
\end{align}

To understand the physical significance of~\eqref{eq_vol_AdS}, consider for instance the difference between the partition functions of a QFT on AdS$_{d-1}$ with two different boundary conditions. Schematically, this takes the form
\begin{equation}
\log Z
=-\operatorname{Vol}(\mathrm{AdS}_{d-1})F+\text{counterterms}\,,
\end{equation}
where $F$ is the corresponding free-energy density. This quantity is ambiguous because of local counterterms supported on the asymptotic boundary, of the schematic form
\begin{equation}
\int d^{d-2}x\sqrt{g_{\rm bdry}}\,
\mathcal{R}_{\rm bdry}^{\,n}\,,
\qquad n=0,1,\ldots\,,
\end{equation}
for $2n\leq d-2$. Evaluated at the cutoff surface, these terms scale as $R_{\rm bdry}^{d-2-2n}$ and reproduce precisely the power-laws appearing in~\eqref{eq_app_AdS_vol_pre}. The regulated volume~\eqref{eq_vol_AdS} therefore isolates the universal contribution to the free energy: the finite constant term for odd $d$, and the coefficient of the logarithmic divergence for even $d$.

\subsection{One-loop determinants and infinite KK sums}
\label{app:one_loop_determinants}

When computing the defect free energy on $\mathrm{AdS}_{d-1}\times S^1$, the one-loop contribution can often be expressed as an infinite KK sum of scalar and fermionic determinants on hyperbolic space. In this appendix we review the basic formulas and explain the dimensional-regularization prescription that we use to evaluate the resulting KK sums. See e.g.~\cite{Giombi:2021cnr,Giombi:2021uae} for analogous calculations in a similar context.

Consider first a real scalar of mass $M$ on $\mathrm{AdS}_{d-1}$. As well known, it admits two boundary conditions given by the solutions of the following quadratic equation
\begin{equation}
\Delta(\Delta-d+2)= m^2\quad\implies\quad
\Delta_{\pm}=\frac{d-2}{2}\pm\sqrt{\left(\frac{d-2}{2}\right)^2+m^2}\,,
\end{equation}
where the plus sign solution is normally referred to as ``standard'' boundary condition, while the minus sign is the ``alternate'' boundary condition. The latter is admissible for a unitary theory only when $\Delta_{-}>\text{max}\left(\frac{d-4}{2},0\right)$.

With standard boundary conditions, the contribution of a real scalar to the free energy, normalized by the AdS volume, is
\begin{equation}
    f_S(M)
    =
    \int_0^\infty d\nu\,\rho_B(\nu)
    \log\left[
        \nu^2+M^2+\left(\frac{d-2}{2}\right)^2
    \right]\,,
    \label{eq:app_scalar_determinant}
\end{equation}
where
\begin{equation}
    \rho_B(\nu)
    =
    \frac{ \left|
        \Gamma\left(i\nu+\frac{d-2}{2}\right)
        \right|^2}
    {(4\pi)^{\frac{d-1}{2}}\Gamma\left(\frac{d-1}{2}\right) |\Gamma(i\nu)|^2}\,.
    \label{eq:app_scalar_density}
\end{equation}
With alternate boundary conditions instead we find 
\begin{equation}
f^{(-)}_{S}(M)=f_{S}(M)+\delta f_{S}(M)\,,
\end{equation}
where
\begin{equation}\label{eq_dF_scalar}
\delta f_S(M)=\frac{ \Gamma \left(\frac{d}{2}\right)}{\pi^{\frac{d}{2}}\,\Gamma (d-1)}\int_0^{\Delta_{+}-\frac{d-2}{2}}du\,u\sin(\pi u)\Gamma\left(\frac{d-2}{2}+u\right)\Gamma\left(\frac{d-2}{2}-u\right)\,.
\end{equation}

Similarly, a Dirac fermion of mass $M$ in AdS$_{d-1}$ admits both standard and alternate boundary conditions, given by
\begin{equation}
\Delta_{\pm}=\frac{d-2}{2}\pm |M|\,,
\end{equation}
where the alternate is unitary only for $\Delta_{-}\geq\frac{d-3}{2}$. With standard boundary conditions, its free energy density reads
\begin{equation}
    f_D(M)
    =
    -c_{d-1}
    \int_0^\infty d\nu\,\rho_F(\nu)
    \log\left(\nu^2+M^2\right)\,,
    \label{eq:app_fermion_determinant}
\end{equation}
where $c_{d-1}=2^{\lfloor(d-1)/2\rfloor}$ denotes the number of complex components of the Dirac spinor and
\begin{equation}
    \rho_F(\nu)
    =
    \frac{\left|
        \Gamma\left(i\nu+\frac{d-1}{2}\right)
        \right|^2}
    {(4\pi)^{\frac{d-1}{2}}
    \Gamma\left(\frac{d-1}{2}\right)  \left|
        \Gamma\left(i\nu+\frac12\right)
        \right|^2}\,.
    \label{eq:app_fermion_density}
\end{equation}
With alternate boundary its free energy density instead reads
\begin{equation}
f^{(-)}_{D}(M)=f_{D}(M)+\delta f_{D}(M)\,,
\end{equation}
where
\begin{equation}\label{eq_dF_Dirac}
\delta f_{D}(M)=\frac{c_{d-1} \Gamma \left(\frac{d}{2}\right)}{\pi^{\frac{d}{2}}\,\Gamma (d-1)}
\int_0^{\Delta_{+}-\frac{d-2}{2}}du\,\cos(\pi u)\Gamma\left(\frac{d-1}{2}+u\right)\Gamma\left(\frac{d-1}{2}-u\right)\,.
\end{equation}

The spectral densities simplify considerably in four dimensions,
\begin{equation}
    \rho_B(\nu)\Big|_{d=4}
    =
    \frac{\nu^2}{4\pi^2}\,,
    \qquad
    \rho_F(\nu)\Big|_{d=4}
    =
    \frac{\nu^2}{4\pi^2}
    +\frac{1}{16\pi^2}\,.
    \label{eq:app_densities_d4}
\end{equation}
More generally, for calculations to leading nontrivial order in the $\eps$ expansion near four dimensions it is sufficient to replace the exact spectral densities by the first three terms in their large-$\nu$ expansion,
\begin{align}
    \rho_B(\nu)
    &=
    a_d\nu^{d-2}
    +b_{B,d}\nu^{d-4}
    +c_{B,d}\nu^{d-6}
    +\ldots\,,
    \nonumber\\
    \rho_F(\nu)
    &=
    a_d\nu^{d-2}
    +b_{F,d}\nu^{d-4}
    +c_{F,d}\nu^{d-6}
    +\ldots\,.
    \label{eq:app_density_asymptotics}
\end{align}
The coefficients are
\begin{equation}
    a_d
    =
    \frac{2^{1-d}\pi^{\frac{1-d}{2}}}
    {\Gamma\left(\frac{d-1}{2}\right)}\,,
\end{equation}
and
\begin{align}
    b_{B,d}
    &=
    \frac{
        2^{-d-1}(d-4)(d-2)
    }{
        3\pi^{\frac{d-1}{2}}
        \Gamma\left(\frac{d-3}{2}\right)
    }\,,
    &
    b_{F,d}
    &=
    \frac{
        2^{-d-1}(d-2)(d-1)
    }{
        3\pi^{\frac{d-1}{2}}
        \Gamma\left(\frac{d-3}{2}\right)
    }\,,
    \nonumber\\
    c_{B,d}
    &=
    \frac{
        (d-6)(d-4)(d-2)(5d-8)
    }{
        45\,2^{d+4}\pi^{\frac{d-1}{2}}
        \Gamma\left(\frac{d-5}{2}\right)
    }\,,
    &
    c_{F,d}
    &=
    \frac{
        (d-4)(d-2)(d-1)(5d-3)
    }{
        45\,2^{d+4}\pi^{\frac{d-1}{2}}
        \Gamma\left(\frac{d-5}{2}\right)
    }\,.
    \label{eq:app_density_coefficients}
\end{align}
Notice that the last terms in \eqref{eq:app_density_asymptotics} vanish as $O(\eps)$ for $d=4-\eps$. They must nonetheless be retained, since the subsequent KK sum may often generate a $1/\eps$ pole which compensates this suppression. Further subleading terms in the spectral density may instead be safely neglected, as it will become clear below.

The advantage of \eqref{eq:app_density_asymptotics} is that the spectral integrals can be performed explicitly while keeping $d$ arbitrary. Using
\begin{equation}
    \int_0^\infty d\nu\,\nu^q
    \log\left(
        \frac{\nu^2+M_1^2}{\nu^2+M_0^2}
    \right)
    =
    \frac{
        \pi\left(M_1^{q+1}-M_0^{q+1}\right)
    }{
        (q+1)\cos\left(\frac{\pi q}{2}\right)
    }\,,
    \label{eq:app_spectral_integral}
\end{equation}
we may write the scalar determinant, up to an $M$-independent constant and corrections which vanish at the order of interest, as
\begin{equation}
    f_S(M)
    =
    A_B(d)\,\mathcal M^{d-1}
    +B_B(d)\,\mathcal M^{d-3}
    +C_B(d)\,\mathcal M^{d-5}
    +O(\eps)\,,
    \quad
    \mathcal M^2
    =
    M^2+\left(\frac{d-2}{2}\right)^2\,,
    \label{eq:app_scalar_formal_d}
\end{equation}
where
\begin{equation}
    A_B(d)
    =
    -\frac{\pi a_d}
    {(d-1)\cos\left(\frac{\pi d}{2}\right)}\,,
    \qquad
    B_B(d)
    =
    \frac{\pi b_{B,d}}
    {(d-3)\cos\left(\frac{\pi d}{2}\right)}\,,
    \qquad
    C_B(d)
    =
    -\frac{\pi c_{B,d}}
    {(d-5)\cos\left(\frac{\pi d}{2}\right)}\,.
    \label{eq:app_scalar_coefficients}
\end{equation}
Similarly, for a Dirac fermion we obtain
\begin{equation}
    f_D(M)
    =
    A_F(d)\,|M|^{d-1}
    +B_F(d)\,|M|^{d-3}
    +C_F(d)\,|M|^{d-5}
    +O(\eps)\,,
    \label{eq:app_fermion_formal_d}
\end{equation}
with
\begin{equation}
    A_F(d)
    =
    \frac{c_{d-1}\pi a_d}
    {(d-1)\cos\left(\frac{\pi d}{2}\right)}\,,
    \qquad
    B_F(d)
    =
    -\frac{c_{d-1}\pi b_{F,d}}
    {(d-3)\cos\left(\frac{\pi d}{2}\right)}\,,
    \qquad
    C_F(d)
    =
    \frac{c_{d-1}\pi c_{F,d}}
    {(d-5)\cos\left(\frac{\pi d}{2}\right)}\,.
    \label{eq:app_fermion_coefficients}
\end{equation}
These expressions should be understood as dimensional-regularization formulas for differences of determinants, so that $M$-independent constants are immaterial.

As a simple check, setting $d=4$ after performing the spectral integral gives
\begin{equation}
    f_S(M)
    =
    -\frac{1}{12\pi}
    \left[
        M^2+1
    \right]^{3/2}\,,
\end{equation}
while for a two-component Dirac fermion
\begin{equation}
    f_D(M)
    =
    \frac{|M|^3}{6\pi}
    -\frac{|M|}{8\pi}\,,
\end{equation}
again up to mass-independent constants.

In practice, it is important to retain the $d$-dependence in \eqref{eq:app_scalar_formal_d} and \eqref{eq:app_fermion_formal_d}, since we often need to sum over KK masses that grow linearly with the KK momentum, $M_n\propto n$. Note that further subleading powers in \eqref{eq:app_scalar_formal_d} and \eqref{eq:app_fermion_formal_d} do not generate logarithmic divergences when summed over, and hence may be safely neglected. The KK sums can be defined by analytic continuation from sufficiently negative values of $d$. Useful identities are
\begin{equation}
    \sum_{n=0}^{\infty}(n+a)^q
    =
    \zeta(-q,a)\,,
    \label{eq:app_hurwitz_basic}
\end{equation}
and, for $0<a<1$,
\begin{equation}
    \sum_{n\in\mathbb Z}|n+a|^q
    =
    \zeta(-q,a)+\zeta(-q,1-a)\,,
    \label{eq:app_hurwitz_bilateral}
\end{equation}
where $\zeta(x,y)$ is the Hurwitz zeta function. For massive KK towers it is also useful to employ the inhomogeneous Epstein--Hurwitz sum
\begin{equation}
    \sum_{n\in\mathbb Z}
    \frac{1}{\left[(n+a)^2+m^2\right]^s}
    =
    \frac{\sqrt{\pi}}{\Gamma(s)}
    \left[
        \Gamma\left(s-\frac12\right)m^{1-2s}
        +
        4\sum_{k=1}^{\infty}
        \cos(2\pi k a)
        \left(\frac{\pi k}{m}\right)^{s-\frac12}
        K_{s-\frac12}(2\pi k m)
    \right]\,,
    \label{eq:app_epstein_hurwitz}
\end{equation}
which is derived using the Schwinger representation
\begin{equation}
    \frac{1}{\left[(n+a)^2+m^2\right]^s}
    =
    \frac{1}{\Gamma(s)}
    \int_0^\infty dt\,t^{s-1}
    e^{-t\left[(n+a)^2+m^2\right]}\,,
\end{equation}
and exchanging the sum and integration. The representation \eqref{eq:app_epstein_hurwitz} is particularly useful at large $m$, where the first term gives the asymptotic power-law contribution, while the Bessel-function sum is exponentially suppressed.

\section{One-loop free energy of the 5-field WZ model in the \texorpdfstring{$\eps$}{eps}-expansion}\label{App_5_field_eps}

In this appendix we provide details on the one-loop calculation discussed in sec. \ref{sec:5_field_eps}. The Euclidean action of the continuation to $d=4-\eps$ dimensions is~\cite{Cuomo:2024fuy}
\begin{align}
    \mL &=
    |\pd x|^2+|\pd y|^2+\frac12(\pd a)^2
    +m_d^2\left(|x|^2+|y|^2+\frac{a^2}{2}\right) +\frac{g^2}{8}\left(|x|^2-|y|^2\right)^2
    +\frac{g^2}{4}a^2\left(|x|^2+|y|^2\right)
    \nonumber\\
    &+i\bar{\tilde a}\slashed{\nabla}\tilde a
    +i\bar{\tilde x}\slashed{\nabla}\tilde x
    +i\bar{\tilde x}_c\slashed{\nabla}\tilde x_c
    +i\bar{\tilde y}\slashed{\nabla}\tilde y
    +i\bar{\tilde y}_c\slashed{\nabla}\tilde y_c
    -\frac{g}{2}a
    \left(
        \bar{\tilde x}\tilde x+\bar{\tilde x}_c\tilde x_c
        -\bar{\tilde y}\tilde y-\bar{\tilde y}_c\tilde y_c
    \right)
    \nonumber\\
    &-\frac{g}{2}\left[
        \bar{\tilde a}
        \left(
            \tilde x_c x+\tilde x x^*
            -\tilde y_c y-\tilde y y^*
        \right)
        +\text{h.c.}
    \right]\,.
    \label{eq:fiveWZ_app_action}
\end{align}
Here the fermions carry an additional flavor index running from $1$ to $N_D$, which we suppressed. At the end of the calculation we take the formal limit $N_D\rightarrow1/4$ corresponding to the analytic continuation to four dimensions of the $3d$ $\mathcal N=1$ theory.

We expand around the symmetric configuration
\begin{equation}
    x=v e^{-i\alpha\theta}\,,
    \qquad
    y=v e^{-i\alpha\theta}\,,
    \qquad
    a=0\,,
\end{equation}
and define, as in the main text,
\begin{equation}
    m=gvR\,.
\end{equation}

\paragraph{Bosonic spectrum.}

We parametrize the fluctuations as
\begin{equation}
    x=e^{-i\alpha\theta}
    \left(v+\frac{r_x+i\pi_x}{\sqrt2}\right)\,,
    \qquad
    y=e^{-i\alpha\theta}
    \left(v+\frac{r_y+i\pi_y}{\sqrt2}\right)\,.
\end{equation}
Introducing $r_\pm=(r_x\pm r_y)/\sqrt2$ and $\pi_\pm=(\pi_x\pm\pi_y)/\sqrt2$, the $x,y$ sector decomposes into two independent $2\times2$ blocks. For a KK mode proportional to $e^{in\theta}$ they are
\begin{equation}
    R^2\mathcal M_{B,n,+}^2
    =
    \begin{pmatrix}
        n^2+\alpha^2+R^2m_d^2 & -2in\alpha\\
        2in\alpha & n^2+\alpha^2+R^2m_d^2
    \end{pmatrix},
    \label{eq:fiveWZ_app_bos_matrix_plus}
\end{equation}
and
\begin{equation}
    R^2\mathcal M_{B,n,-}^2
    =
    \begin{pmatrix}
        n^2+\alpha^2+R^2m_d^2+m^2 & -2in\alpha\\
        2in\alpha & n^2+\alpha^2+R^2m_d^2
    \end{pmatrix}.
    \label{eq:fiveWZ_app_bos_matrix_minus}
\end{equation}
The fluctuation of $a$ decouples. Diagonalizing, we obtain the five scalar masses
\begin{align}
    R^2M_{B,n,1/2}^2
    &=
    (n\pm\alpha)^2
    -\left(\frac{d-2}{2}\right)^2\,,
    \nonumber\\
    R^2M_{B,n,3/4}^2
    &=
    n^2+\alpha^2+\frac{m^2}{2}
    -\left(\frac{d-2}{2}\right)^2
    \pm
    \sqrt{4\alpha^2n^2+\frac{m^4}{4}}\,,
    \nonumber\\
    R^2M_{B,n,5}^2
    &=
    n^2+m^2
    -\left(\frac{d-2}{2}\right)^2\,.
    \label{eq:fiveWZ_app_bos_spectrum}
\end{align}
On the classical saddle, $\alpha=(d-2)/2$, the first three modes become the three massless degrees of freedom of the moduli-space EFT in the $m\rightarrow\infty$ limit. More explicitly,
\begin{align}
    R^2M_{B,n,1/2}^2
    &=
    n^2\pm2\alpha n\,,
    \nonumber\\
    R^2M_{B,n,3}^2
    &=
    n^2-\frac{4\alpha^2n^2}{m^2}
    +O(m^{-6})\,,
    \nonumber\\
    R^2M_{B,n,4}^2
    &=
    m^2+n^2+\frac{4\alpha^2n^2}{m^2}
    +O(m^{-6})\,,
    \nonumber\\
    R^2M_{B,n,5}^2
    &=
    m^2+n^2-\alpha^2\,.
    \label{eq:fiveWZ_app_bos_large_m}
\end{align}

\paragraph{Fermionic spectrum.}
Each $4d$ Dirac field gives to two $3d$ Dirac fields per KK momentum, thus each of the masses below must be counted twice. It is convenient to define
\begin{equation}
    p=n+\frac12\,,
    \qquad n=0,1,\ldots\,.
\end{equation}
Two of the five fermionic masses are particularly simple,
\begin{equation}
    RM_{F,p,1/2}
    =
    |p\pm\alpha|\,.
    \label{eq:fiveWZ_app_ferm_simple}
\end{equation}
The remaining three masses are the absolute values of the roots of
\begin{equation}
    \lambda^3
    -p\lambda^2
    -\left(p^2+\alpha^2+m^2\right)\lambda
    +p\left(p^2-\alpha^2+m^2\right)
    =0\,.
    \label{eq:fiveWZ_app_ferm_cubic}
\end{equation}
The explicit expressions for the roots are not particularly illuminating, but several limits will be useful below.

For $m\ll1$ and away from $\alpha=2p$, the signed roots of \eqref{eq:fiveWZ_app_ferm_cubic} are
\begin{align}
    \lambda_3
    &=
    -p
    -\frac{2pm^2}{4p^2-\alpha^2}
    +\frac{2pm^4(4p^2+\alpha^2)}
    {(4p^2-\alpha^2)^3}
    +O(m^6)\,,
    \nonumber\\
    \lambda_4
    &=
    p+\alpha
    +\frac{m^2}{2(2p+\alpha)}
    +\frac{(2p-\alpha)m^4}
    {8\alpha(2p+\alpha)^3}
    +O(m^6)\,,
    \nonumber\\
    \lambda_5
    &=
    p-\alpha
    +\frac{m^2}{2(2p-\alpha)}
    -\frac{(2p+\alpha)m^4}
    {8\alpha(2p-\alpha)^3}
    +O(m^6)\,.
    \label{eq:fiveWZ_app_ferm_small_m}
\end{align}
This expansion however is singular for $\alpha=2p$, since $\lambda_3$ and $\lambda_5$ become degenerate. For $(\alpha-2p)^2\sim m^2\ll 1$ these instead behave as
\begin{equation}
    \lambda_\pm
    =
    -p-\frac{\alpha-2p}{2}
    \pm\frac12
    \sqrt{(\alpha-2p)^2+2m^2}
    +\ldots\,.
    \label{eq:fiveWZ_app_ferm_crossing}
\end{equation}

In the opposite limit $m\gg1$, one of the three modes remains light,
\begin{equation}
    RM_{F,p,3}
    =
    p\left[
        1-\frac{2\alpha^2}{m^2}
        +\frac{2\alpha^4}{m^4}
        +O(m^{-6})
    \right]\,.
    \label{eq:fiveWZ_app_ferm_large_m_light}
\end{equation}
Together with \eqref{eq:fiveWZ_app_ferm_simple}, this leaves three light fermionic towers as $m\rightarrow\infty$ as expected from the moduli-space EFT. The other two fermionic KK towers become heavy instead at large $m$:
\begin{equation}
    RM_{F,p,4/5}
    =
    m+\frac{p^2+\alpha^2}{2m}
    \pm\frac{\alpha^2p}{m^2}
    +\frac{-p^4+6\alpha^2p^2-\alpha^4}{8m^3}
    +O(m^{-4})\,.
    \label{eq:fiveWZ_app_ferm_large_m_heavy}
\end{equation}

Finally, for regulating the infinite KK sums it is useful to record the large-$p$ expansion of the three nontrivial modes,
\begin{align}
    RM_{F,p,3}
    &=
    p+\frac{m^2}{2p}
    -\frac{m^2(m^2-\alpha^2)}{8p^3}
    +O(p^{-4})\,,
    \nonumber\\
    RM_{F,p,4/5}
    &=
    p\pm\alpha+\frac{m^2}{4p}
    \pm\frac{m^2(m^2-4\alpha^2)}{32\alpha p^2}
    -\frac{m^2(m^2-\alpha^2)}{16p^3}
    +O(p^{-4})\,.
    \label{eq:fiveWZ_app_ferm_large_p}
\end{align}
The second line assumes fixed nonzero $\alpha$.

\paragraph{One-loop free energy.}

Denoting by $f_S(M)$ and $f_D(M)$ the scalar and Dirac determinants normalized by the $\mathrm{AdS}_{d-1}$ volume, as defined in appendix~\ref{app:one_loop_determinants}, the determinant contribution to the one-loop free energy can be written formally as
\begin{equation}
\begin{split}
    f_0^{(\mathrm{det})}(m,\alpha)
    ={}&
    \sum_{n\in\mathbb Z}
    \sum_{k=1}^5
    f_S(M_{B,n,k})
    +
    4N_D
    \sum_{n=0}^{\infty}
    \sum_{k=1}^5
    f_D(M_{F,n+\frac12,k})
    \\
    &-
    \left.
    \left[
        \sum_{n\in\mathbb Z}
        \sum_{k=1}^5
        f_S(M_{B,n,k})
        +
        4N_D
        \sum_{n=0}^{\infty}
        \sum_{k=1}^5
        f_D(M_{F,n+\frac12,k})
    \right]
    \right|_{m=\alpha=0}\,.
    \label{eq:fiveWZ_app_free_energy_sum}
\end{split}
\end{equation}
The last line fixes the normalization such that the free energy of the trivial defect vanishes. The factor of four in front of the fermionic contribution accounts for the two signs of the half-integer KK momentum and for the fact that each $4d$ field gives rise to two identical KK towers of Dirac fields on AdS$_{3}$.

To evaluate \eqref{eq:fiveWZ_app_free_energy_sum}, we proceed as described in appendix~\ref{app:one_loop_determinants}. We first expand the scalar and fermionic contributions at large KK momentum, keeping the full $d$-dependence of the terms that may contribute through $O(\eps^0)$ after summation. For each KK sum we then write schematically
\begin{equation}
    \sum_n \mathcal F_n(d)
    =
    \sum_n
    \left[
        \mathcal F_n(4)-\mathcal F_n^{\rm as}(4)
    \right]
    +
    \sum_n\mathcal F_n^{\rm as}(d)
    +O(\eps)\,,
    \label{eq:fiveWZ_app_KK_reg}
\end{equation}
where $\mathcal F_n^{\rm as}(4)$ is such that the first sum in \eqref{eq:fiveWZ_app_KK_reg} is convergent and can be evaluated directly in $d=4$, while the second is computed analytically in terms of Riemann and Hurwitz zeta functions. This gives a representation of the one-loop free energy as the sum of an analytic expression and a rapidly convergent infinite series. The explicit result for generic $m$ and $\alpha$ is rather lengthy and we do not report it here.

Specializing to $N_D=1/4$, the divergent part of this expression is
\begin{equation}
    f_0^{(\mathrm{det})}(m,\alpha)\Big|_{\rm div}
    =
    \frac{m^2(\alpha^2-1)}{8\pi\eps}\,.
    \label{eq:fiveWZ_app_free_energy_div}
\end{equation}
This divergence is canceled by the wave-function counterterm. In the minimal subtraction scheme,
\begin{equation}
    \delta Z_X=-\frac{g^2}{32\pi^2\eps}\,,
\end{equation}
which contributes
\begin{equation}
    f_0^{(\mathrm{ct})}(m,\alpha)
    =
    -\frac{m^2}{8\pi\eps}
    \left[
        \alpha^2-\left(\frac{d-2}{2}\right)^2
    \right]\,.
    \label{eq:fiveWZ_app_counterterm}
\end{equation}
Expanding around $d=4-\eps$, the pole in \eqref{eq:fiveWZ_app_counterterm} precisely cancels \eqref{eq:fiveWZ_app_free_energy_div}.

The small-$m$ expansion is conveniently obtained from the convergent representation following \eqref{eq:fiveWZ_app_KK_reg}, by expanding the individual masses at fixed KK momentum before performing the remaining sums. There are two subtleties in taking the limit $\alpha\rightarrow1$. First, the $p=1/2$ fermionic mode lies at the singular point $\alpha=2p$ of the low $m$ expansion; this implies that the two eigenvalues which become degenerate at that point and therefore must be treated together using~\eqref{eq:fiveWZ_app_ferm_crossing}, rather than using the generic expansion \eqref{eq:fiveWZ_app_ferm_small_m}. Second, one of the bosonic modes becomes light in this limit and its determinant must be kept unexpanded until after taking $\alpha\rightarrow1$. Proceeding in this way gives
\begin{equation}
    f_0(m,1)
    =
    \frac{m^2}{8\pi}
    -\frac{1+1/\sqrt2}{12\pi}m^3
    +O(m^4)\,.
    \label{eq:fiveWZ_app_small_m_alpha1}
\end{equation}

The large-$m$ expansion cannot instead be obtained by expanding the summand at fixed KK momentum, since we also need to sum over KK modes with $n\sim m$. We therefore use the matched asymptotic expansion method along the lines of appendix D.2 of~\cite{Cuomo:2021cnb}: the summand is expanded separately in the regions of fixed $n$ and fixed $n/m$, and the two expansions are matched in their common regime of validity. The contribution from the latter region can be organized using the Euler--Maclaurin formula. This yields
\begin{equation}
\begin{split}
    f_0(m,\alpha)
    ={}&
    \frac{1-\alpha^2}{8\pi}m^2\log m
    +\frac{m^2}{16\pi}
    \left[
        -1
        +(1-\alpha^2)
        \left(
            \gamma_E-1-\log(4\pi)
        \right)
    \right]
    \\
    &-\frac{\alpha^4}{16\pi}\log m
    +C(\alpha)
    +O\left(\frac{\log m}{m^2}\right)\,,
    \label{eq:fiveWZ_app_large_m}
\end{split}
\end{equation}
where $C(\alpha)$ is independent of $m$ and reads
\begin{equation}
\begin{split}
    C(\alpha)={}&
    \frac{1}{64\pi}
    +\frac{\alpha^2}{3\pi}
    +\frac{\alpha^4}{16\pi}
    \left(1+\log 2-\gamma_E\right)
    -\frac{\alpha^3}{4\pi}
    -\frac{1}{6\pi}
    \sum_{n=1}^{\infty}
    \left[
        \left(n^2+\alpha^2\right)^{3/2}
        -n^3
        -\frac{3}{2}\alpha^2 n
        -\frac{3\alpha^4}{8n}
    \right]
    \\
    &+\Theta\left(\alpha-\frac12\right)
    \left[
        \frac{\left(\alpha-\frac12\right)^3}{3\pi}
        -\frac{\alpha-\frac12}{4\pi}
    \right]\,.
\end{split}
\end{equation}
In particular, setting $\alpha=1$ at the order relevant for the calculation in the main text,
\begin{equation}
    f_0(m,1)
    =
    -\frac{m^2}{16\pi}
    -\frac{1}{16\pi}\log m
    +C(1)
    +O\left(\frac{\log m}{m^2}\right)\,.
    \label{eq:fiveWZ_app_large_m_alpha1}
\end{equation}

\section{Details on large-\texorpdfstring{$N$}{N} computations}\label{app:large_N}

In this appendix we provide more details for the calculations of section \ref{sec:large_N}.

\subsection{Effective action}\label{app:large_N_EFT}

The theory has superpotential
\begin{equation}
    W=A(X_i^2-Y_i^2)\;.
\end{equation}

In addition to the usual kinetic terms, which include $-\frac12F^2$ for each auxiliary field $F$, the potential terms in components follow from $F\,\partial W-\frac12\,\partial^2W\,\psi\psi$ (keeping the auxiliary fields explicit):
\begin{gather}
    V = 
    2 a (x_i F^X_i - y_i F^Y_i) - a (\psi^X_i \psi^X_i - \psi^Y_i \psi^Y_i) + F_A(x_i^2 - y_i^2) - 2 \eta(x_i \psi^X_i - y_i \psi^Y_i)\;.
\end{gather}
Here the components of the superfield $A$ are $a,\eta,F_A$.
We will also be working on conformally flat spaces, which requires adding a conformal mass term for the scalars:
\begin{equation}
    V_m=\frac12m_d^2(a^2+x_i^2+y_i^2)\;.
\end{equation}
We will be working on $\mathrm{AdS}_2\times S^1$ and we will set both radii to 1 so that $m_d^2=-1/4$. The general result for any ratio of radii can be recovered easily.

Note that at large $N$, $F_A$ naturally plays the role of the Hubbard-Stratonovich field, and its term $-\frac12F_A^2$ can be dropped, since it is irrelevant in the IR, where $F_A$ has dimension 2. Next we assume some background for $x_i,y_i$ and integrate out the fluctuations around these fields at large $N$. Specifically, we can use the $O(N)\times O(N)$ symmetry to ensure that only $x\equiv (x_1+ix_2)/\sqrt2$ and $y\equiv (y_1+iy_2)/\sqrt2$ have nontrivial vevs, and integrate out fluctuations around this background.\footnote{In principle we can ensure that only $x_1,y_1\neq 0$ but since we will soon discuss monodromies it will be more useful to keep a $U(1)$ symmetry manifest.} Defining $F_X\equiv(F^X_1+iF^X_2)/\sqrt2$, $F_Y\equiv(F^Y_1+iF^Y_2)/\sqrt2$ and rescaling $F_A\to F_A/2$, $a\to a/2$, the result at large $N$ is
\begin{equation}
    \begin{split}
         S_{\rm eff} &= \frac{N}{2} \operatorname{Tr}\log \left[-\Delta + F_A + a^2 + m_d^2\right] + \frac{N}{2} \operatorname{Tr}\log \left[-\Delta - F_A + a^2 + m_d^2 \right] \\
    &- \frac{N}{2} \operatorname{Tr}\log\left[i \gamma^\mu \partial_\mu + a \right]  - \frac{N}{2} \operatorname{Tr}\log\left[i \gamma^\mu \partial_\mu - a \right] \\
    &+\int d^3x\sqrt{g}\left[\left(F_A +m_d^2\right) |x|^2 + \left(-F_A +  m_d^2\right) |y|^2  \right.\\
    &\hspace{5.3em}\left.+ 2a\,\mathrm{Re}\left(\bar x F_X - \bar y F_Y\right) - |F_{X}|^2 - |F_{Y}|^2\right] \;,
    \end{split}
\end{equation}
with $\Delta$ the Laplacian. 

Next we integrate out $F_X,F_Y$ to get
\begin{equation}
    \begin{split}
     S_{\rm eff} &= \frac{N}{2} \operatorname{Tr}\log \left[-\Delta + F_A + a^2 + m_d^2\right] + \frac{N}{2} \operatorname{Tr}\log \left[-\Delta - F_A + a^2 + m_d^2 \right] \\
    &- \frac{N}{2} \operatorname{Tr}\log\left[i \gamma^\mu \partial_\mu + a \right]  - \frac{N}{2} \operatorname{Tr}\log\left[i \gamma^\mu \partial_\mu - a \right] \\
    &+\int d^3x\sqrt{g}\left[\left(F_A+ m_d^2+a^2\right) |x|^2 + \left(-F_A + m_d^2+a^2\right) |y|^2\right] \,.
    \end{split}
\end{equation}
This is the large-$N$ effective action we are interested in.

\subsection{Computing the partition function}\label{app:large_N_I}

At $a=0$ the fermionic determinants in \eqref{eq_largeN_I} are independent of $F_A$, so only the bosonic part enters the saddle equations; their contribution to the $g$-function is discussed in section~\ref{sec:large_N}. The bosonic part is
\begin{equation}
    \mathcal{I}_{\rm B}=\frac{N}{2}\operatorname{Tr}\log(-\Delta_x+F_A+m_d^2)+\frac{N}{2}\operatorname{Tr}\log(-\Delta_y-F_A+m_d^2)\;,
\end{equation}
on $\mathrm{AdS}_2\times S^1$ (with $m_d^2=-1/4$), where $\Delta_x$ and $\Delta_y$ are the Laplacians built from the twisted derivatives $D^x_\mu$ and $D^y_\mu$ of section~\ref{sec:large_N}, which include the monodromies $\alpha_x$ and $\alpha_y$. The result is \cite{Giombi:2021uae}
\begin{equation}\label{eq:sum_I}
\begin{split}
    \mathcal{I}_{\rm B}=\frac{N}{2}\frac{\operatorname{Vol}\bigl(\mathrm{AdS}_{d-1}\bigr)}
     {(4\pi)^{\frac{d-1}{2}}\;\Gamma\!\bigl(\tfrac{d-1}{2}\bigr)}
\int_{-\infty}^{\infty}d\nu\,
\frac{\bigl|\Gamma\!\bigl(i\nu+\tfrac{d-2}{2}\bigr)\bigr|^2}
     {\bigl|\Gamma(i\nu)\bigr|^2}&
\left[\sum_{n\in\mathbb Z}
\log\left(\nu^2+(n+\alpha_x)^2+F_A\right)\right.\\
&\left.+\sum_{n\in\mathbb Z}\log\left(\nu^2+(n+\alpha_y)^2-F_A\right)\right]\,.
\end{split}
\end{equation}
We can simplify this a bit more since we are directly in 3d by using
\begin{equation}
\frac{\bigl|\Gamma\!\bigl(i\nu+\tfrac{d-2}{2}\bigr)\bigr|^2}
     {\bigl|\Gamma(i\nu)\bigr|^2}|_{d=3}=\nu\tanh(\pi\nu)\,.
     \end{equation}

Next we use
\begin{equation}
   \sum_{n\in\mathbb{Z}}
\log(\nu^2+(n+\alpha)^2+F_A)=\log(2\cosh(2\pi \sqrt{\nu^2+F_A})-2\cos(2 \pi \alpha))\,.
\end{equation}
For $0\leq \alpha<1$, reality requires all the eigenvalues in the sum to be non-negative. Thus the $x$ and $y$ determinants require, respectively, $F_A\geq-\min\{\alpha_x^2,(1-\alpha_x)^2\}$ and $F_A\leq\min\{\alpha_y^2,(1-\alpha_y)^2\}$. Note that this is the same condition as for the large-$N$ monodromy defect in the usual $O(N)$ model \cite[sec.~4]{Giombi:2021uae}. On the locus \eqref{eq:largeN_alphas} we have $F_A+\alpha_x^2=-F_A+\alpha_y^2=1/4$, $F_A+(1-\alpha_x)^2=5/4-2\alpha_x$ and $-F_A+(1-\alpha_y)^2=5/4-2\alpha_y$, so the conditions reduce to $\alpha_{x,y}\leq5/8$, i.e.\ $|F_A|\leq 9/64$, and are saturated by the $n=-1$ modes.

The free energy is then
\begin{equation}
\begin{split}
    \mathcal{I}_{\rm B}(F_A,\alpha_x,\alpha_y)=&\frac{N\,\operatorname{Vol}(\mathrm{AdS}_2)}{8\pi}
\int_{-\infty}^{\infty}d\nu\,
\nu\tanh(\pi\nu)
\left[\log(2\cosh(2\pi \sqrt{\nu^2+F_A})-2\cos(2\pi\alpha_x))\right.\\
&\hspace{4.9cm}\left.+\log(2\cosh(2\pi \sqrt{\nu^2-F_A})-2\cos(2\pi\alpha_y))\right].
\end{split}  
\end{equation}
This integral is divergent, but can be made finite by subtracting the result for the trivial defect, so that
\begin{equation}
\mathcal{I}_{{\rm B},\text{reg}}=\mathcal{I}_{\rm B}(F_A,\alpha_x,\alpha_y)-\mathcal{I}_{\rm B}(0,0,0)\;.
\end{equation}
Taking a derivative, the second equation in \eqref{eq:largeN_eqs} gives
\begin{equation}\label{eq:largeN_diff}
    \frac{|x|^2-|y|^2}{N}=-\frac{1}{16\pi^2}\int_{-\infty}^{\infty}d\nu\,\nu\tanh(\pi\nu)\left[G(\nu,F_A,\alpha_x)-G(\nu,-F_A,\alpha_y)\right]\,,
\end{equation}
where
\begin{equation}
    G(\nu,F_A,\alpha)=\frac{\pi\sinh\left(2\pi\sqrt{\nu^2+F_A}\right)}{\sqrt{\nu^2+F_A}\left[\cosh\left(2\pi\sqrt{\nu^2+F_A}\right)-\cos(2\pi\alpha)\right]}\,.
\end{equation}
This integral is convergent, and it remains finite when a mode saturates the BF bound, since $\nu\tanh(\pi\nu)$ vanishes at $\nu=0$.

\bibliography{refs}
	\bibliographystyle{JHEP}

\end{document}